\documentclass[12pt]{article}
\pdfoutput=1 

\usepackage{amsmath}
\usepackage{amssymb}
\usepackage{amsfonts}
\usepackage{graphicx}           
\usepackage[utf8]{inputenc}     
\usepackage{aas_macros}				  
\usepackage{bm}                 
\usepackage{amsthm}
\newtheoremstyle{mystyle}
{1em}
{1em}
{\itshape}
{}
{\bfseries}
{.}
{ }
{}

\theoremstyle{mystyle}
\newtheorem{thumb}{Rule of Thumb}
\newtheorem{corollary}{Corollary of Thumb}

\usepackage{slashed}				
\usepackage{mathrsfs}			 	
\usepackage{bbm}					  
\usepackage{cancel}		
\usepackage[normalem]{ulem} 		

\usepackage{lineno}

\usepackage[dvipsnames]{xcolor}
\usepackage[hang,flushmargin]{footmisc} 

\usepackage{fancyhdr}		
\usepackage{lipsum}			
\usepackage{framed}			
\usepackage{subcaption}	
\usepackage{cite}			  
\usepackage{xspace}			

\usepackage{booktabs}		
\usepackage{nicefrac}		

\usepackage[font=small]{caption} 
\usepackage{float}         

\usepackage[margin=2cm]{geometry}   
\graphicspath{{figures/}}			
\numberwithin{equation}{section}    

\usepackage{multicol}
\usepackage{etoolbox}
\usepackage{relsize}
\patchcmd{\thebibliography}
{\list}
{\begin{multicols}{2}\smaller\list}
	{}
	{}
	\appto{\endthebibliography}{\end{multicols}}

\let\oldenumerate\enumerate
\renewcommand{\enumerate}{
	\oldenumerate
	\setlength{\itemsep}{2pt}
	\setlength{\parskip}{0pt}
	\setlength{\parsep}{0pt}
}

\let\olditemize\itemize
\renewcommand{\itemize}{
	\olditemize
	\setlength{\itemsep}{1pt}
	\setlength{\parskip}{0pt}
	\setlength{\parsep}{0pt}
}

\newcommand{\acro}[1]{\textsc{\MakeLowercase{#1}}}    

\renewcommand{\tilde}{\widetilde}   
\renewcommand{\vec}[1]{\mathbf{#1}} 

	\definecolor{nirmal}{rgb}{1,0,1}
	\definecolor{nirmalcomment}{rgb}{.4,.5,1}

	\newcommand{\CM}{\text{\rm \tiny{CM}}}
	\newcommand{\Mol}{\text{\rm \tiny{M\o l}}}
	\newcommand{\esc}{\text{\rm {esc}}}
	\newcommand{\bbeta}{\boldsymbol{\beta}}
	\newcommand{\kcm}{\vec{k}_\CM}

	\newcommand{\email}[1]{\href{mailto:#1}{#1}}
	\newenvironment{institutions}[1][2em]{\begin{list}{}{\setlength\leftmargin{#1}\setlength\rightmargin{#1}}\item[]}{\end{list}}

	\usepackage[
	colorlinks=true,
	citecolor=green!50!black,
	linkcolor=NavyBlue!75!black,
	urlcolor=green!50!black,
	hypertexnames=false]{hyperref}

	\fancypagestyle{firststyle}{
		\rhead{\footnotesize%
			\texttt{UCR-TR-2019-FLIP-NCC-1701-C}%
		}}
		
		\usepackage{etoc}

\begin{document}
  \thispagestyle{firststyle}          

\begin{center} 

{\large \bf Dark Kinetic Heating of Neutron Stars from Contact Interactions with Relativistic Targets}


    \vskip .7cm

   { \bf 
    Aniket Joglekar$^{a}$, 
    Nirmal Raj$^{b}$,
    Philip Tanedo$^{a}$,
    and 
    Hai-Bo Yu$^{a}$
    } 
   \\ 
   \vspace{-.2em}
   { \tt \footnotesize
      \email{aniket@ucr.edu},
      \email{nraj@triumf.ca},
      \email{flip.tanedo@ucr.edu},
      \email{haiboyu@ucr.edu}
   }
  
   \vspace{-.2cm}

   \begin{institutions}[2.25cm]
   \footnotesize
   $^{a}$ 
   {\it 
      Department of Physics \& Astronomy, 
      University of  California, Riverside, 
      {CA} 92521      
   }    
  \\ 
  \vspace*{0.05cm}   
  $^{b}$ 
  {\it 
       \acro{TRIUMF}, 4004 Westbrook Mall, Vancouver, BC V6T 2A3, Canada
  }
  \end{institutions}

\end{center}


\begin{abstract}
\noindent 
Dark matter can capture in neutron stars from scattering off ultra-relativistic electrons. 
We present a method to calculate the capture rate on degenerate targets with ultra-relativistic momenta in a compact astronomical object. Our treatment accounts for the target momentum and the Fermi degeneracy of the system.
We derive scaling relations for scattering with relativistic targets and confirm consistency with the non-relativistic limit and Lorentz invariance.
The potential observation of kinetic heating of neutron stars has a larger discovery reach for dark matter--lepton interactions than conventional terrestrial direct detection experiments. We map this reach onto a set of bosonic and fermionic effective contact interactions between dark matter and leptons as well as nucleons.
%
We show the results for the contact operators up to dimension 6 for spin-0 and spin-1/2 dark matter interactions with relativistic as well as non-relativistic Standard Model fermions. Highlights of this program in the case of vector mediated interactions are presented in a companion letter~\cite{Joglekar:2019vzy}.
Our method is generalizable to dark matter scattering in any degenerate medium where the Pauli exclusion principle leads to relativistic targets with a constrained phase space for scattering.
\end{abstract}

\small
\setcounter{tocdepth}{2}
\tableofcontents
\normalsize
\clearpage



\section{Introduction}

Astronomical data unambiguously establishes the existence of dark matter. 
Interactions between dark matter and visible matter are predicted by many models to set the cosmological abundance of dark matter.
This motivates experimental search strategies such as direct detection, which looks for the recoils of ordinary matter particles scattered by dark matter. 
To date, however, terrestrial experiments have only set limits on on the strength of dark matter--visible matter interactions.

An innovative extension of this program is the proposal that neutron stars can be used as
dark matter laboratories in space. 
Dark matter falls into a neutron star's steep gravitational potential at semi-relativistic speeds and scatters with visible matter. The recoil energy of stellar constituents heats the neutron star.
A sufficiently old neutron star is expected to be cold enough that this \emph{kinetic heating} is an observable signature of dark matter scattering on ordinary matter~\cite{Baryakhtar:2017dbj,Raj:2017wrv}. 
If radio telescopes such as \acro{FAST}~\cite{FAST}, {\acro{SKA}~\cite{SKA}, and \acro{CHIME}~\cite{CHIME}} detect the radio pulses of a $\mathcal O(10^9~\text{year})$-old pulsar, then upcoming infrared telescopes like the 
\acro{JWST}~\cite{Gardner:2006ky},  
\acro{TMT}~\cite{Skidmore:2015lga}, and \acro{ELT}~\cite{andersen2003euro50}
would be able to detect kinetic heating with $\mathcal O(10^4~\text{sec})$ integration time.
The discovery reach of such an observation is favorable compared to terrestrial direct detection experiments~\cite{Baryakhtar:2017dbj,Raj:2017wrv}.

Recent efforts in this program focus on dark matter that interacts primarily with leptons~\cite{Joglekar:2019vzy}.
Early work on this subject assumes scattering with non-relativistic targets that are at rest in the neutron star frame~\cite{Garani:2019fpa, Bell:2019pyc}. However, this treatment fails for the electron targets. The matter in a neutron star is degenerate. The Pauli exclusion principle forces the neutron star constituents to have non-zero momenta and restricts the phase space for scattering by blocking degenerate final states. Neutrons, protons and muons in a neutron star have a Fermi energy well below their mass and thus remain non-relativistic. Electrons in a neutron star, on the other hand, are \emph{ultra}-relativistic, except in the outermost layers of the crust. 
In a companion paper, we showed that this can change capture kinematics significantly, causing the non-relativistic approximation to underestimate the capture rate in some parametric regions by five orders of magnitude~\cite{Joglekar:2019vzy}.

The present paper extends that analysis to a complete set of contact operators up to dimension-6 between spin-0 or spin-$1/2$ dark matter and nucleonic or leptonic targets. 
We give a physical interpretation for the features of relativistic capture that cause the non-relativistic approximation to grossly underestimate the capture rate in some regions.
We derive how the phase space for this process scales as a function of the target Fermi momentum, target mass, and dark matter mass. 
Our results are consistent with both the non-relativistic limit and Lorentz invariance.
This study opens a new frontier for the capture of dark matter on neutron stars, a subject that began thirty years ago in studies of black hole formation~\cite{Goldman:1989nd, Gould:1989gw}.
It is part of a larger multi-messenger frontier to study the potential to understand dark matter from its capture in compact stars~\cite{%
  Kouvaris:2007ay, 
  Bertone:2007ae, 
  McCullough:2010ai, 
  Kouvaris:2010jy,  
  deLavallaz:2010wp, 
  Kouvaris:2010vv, 
  Kouvaris:2011fi, 
  McDermott:2011jp, 
  Guver:2012ba, 
  Bertoni:2013bsa, 
  Bramante:2013hn, 
  Bramante:2013nma, 
  Bell:2013xk, 
  Perez-Garcia:2014dra,
  Graham:2015apa,
  Bramante:2015cua,
  Cermeno:2016olb, 
  Baryakhtar:2017dbj,
  Raj:2017wrv,
  Krall:2017xij, 
  Graham:2018efk, 
  Garani:2018kkd, 
  Chen:2018ohx, 
  Bell:2018pkk, 
  McKeen:2018xwc,
  Acevedo:2019agu, 
  Acevedo:2019gre, 
  Janish:2019nkk, 
  Hamaguchi:2019oev,
  Camargo:2019wou,
  Dasgupta:2019juq,
  Bell:2020jou,
  Dasgupta:2020dik,
  Garani:2020wge
}. 

The organizational structure of this paper is manifest in the table of contents.  Sections \ref{sec:model:and:conventions} and \ref{sec:Review:Heating} are a self-contained introduction to standard formalism of kinetic heating. Sections \ref{sec:formalism} and \ref{sec:RelScat} introduce our revised formalism for ultra-relativistic targets and summarizes the qualitative physical properties. The discovery reach with respect to contact operators is presented in the figures of Section~\ref{sec:bounds:on:contact:ops}. We list and estimate sources of uncertainty in Section~\ref{sec:uncertainties}.
The appendices include detailed derivations of key results. Some of these are known results that may not be obvious and others are technical calculations that confirm the qualitative discussions in the paper.

\section{Neutron Star Model and Conventions}
\label{sec:model:and:conventions}

\begin{table}
  \centering
  \begin{tabular}{lllll}  
  \toprule
  Component 
  & $Y_\text{T}$ \quad\;
  & $\langle n_\text{T} \rangle~[\text{cm}^{-3}]$ \qquad
  & $p_\text{F}$~[MeV]
  & $E_\text{F}$~[MeV]
  \\
  \midrule
  $e^-$
  & 0.06
  & $1.27\times10^{37}$
  & 146
  & 146
  \\
  $\mu^-$
  & 0.02
  & $4.23\times10^{36}$
  & 50
  & 118
  \\
  $p^+$
  & 0.07
  & $1.48\times10^{37}$
  & 160
  & 951
  \\
  $n$
  & 0.93
  & $1.97\times10^{38}$
  & 373
  & 1011
  \\
  \bottomrule
  \end{tabular}
  \caption{
  Components of a neutron star core for star mass of $M_\star=1.5~\text{M}_\odot$ as computed in Ref.~\cite{Bell:2019pyc} with the Brussels--Montreal unified equation of state BSk24~\cite{Pearson:2018tkr}.
  For each component, we list the volume-averaged abundance $Y_\text{T}$ normalized to the nucleon abundance, average number density $\langle n_\text{T}\rangle$, volume-averaged Fermi three-momentum $p_\text{F}$ and Fermi energy $E_\text{F}$.
  }
  \label{tab:neutron:star:components}
\end{table}

We use a simple model of the neutron star that assumes neutrons, protons, electrons, and muons are the sole stellar constituents in the core. We relate the volume-averaged abundance of a target species $Y_\text{T}$ to its average number density $\langle  n_\text{T}\rangle$,
\begin{align}
\langle  n_\text{T}\rangle
&=
Y_T\frac{M_\star}{m_\text{n}} \left(\frac{4}{3}\pi R_\star^3\right)^{-1},
\label{eq:average:n:T}
\end{align}
where we use the following benchmark values for the neutron star mass, radius, and ambient dark matter density~\cite{Pato:2015dua}:
\begin{align}
  M_\star&=1.5~\text{M}_\odot
  &
  R_\star&=12.6~\text{km} 
  &
  \rho_\chi &= 0.4~\text{GeV}/\text{cm}^3
  \ .
\end{align}

We assume that the neutron star is effectively at rest relative to the dark matter halo. In the stellar core, we have used a volume-averaged target number per nucleon, $Y_\text{T}$, and a volume-averaged target Fermi momentum, $p_\text{F}$, as computed in~\cite{Bell:2019pyc} using the BSk24 unified equation of state~\cite{Pearson:2018tkr} at $M_\star=1.5~\text{M}_\odot$ and $R_\star=12.6~\text{km}$.
The target particles in the neutron star are degenerate: their chemical potentials, $\mu_\text{T} \sim \mathcal O(100~\text{MeV})$, are all much greater than the neutron star temperature, $T_\star \sim \mathcal O(\text{eV})$. We thus assume that the energy levels for each target are filled to its Fermi energy, $E_\text{F}$. These properties are summarized in Table~\ref{tab:neutron:star:components}. We also neglect the effects of interactions among neutron star constituents.

The relativistic treatment of capture requires relating kinematic quantities that are naturally defined in different frames. In order to simplify notation, we assume that quantities are defined in the neutron star frame unless otherwise identified with a subscript, e.g.~$(d\sigma)_\CM$ is a cross section in the center of momentum frame. We explain additional conventions in Appendix~\ref{app:conventions}; these are mostly for the technical work in the appendices.


\section{Review of Dark Kinetic Heating}\label{sec:Review:Heating}

We summarize the relevant background material for the kinetic heating of neutron stars from dark matter capture~\cite{Baryakhtar:2017dbj}.

\begin{figure}[tb]
  \centering
  \includegraphics[width=.7\textwidth]{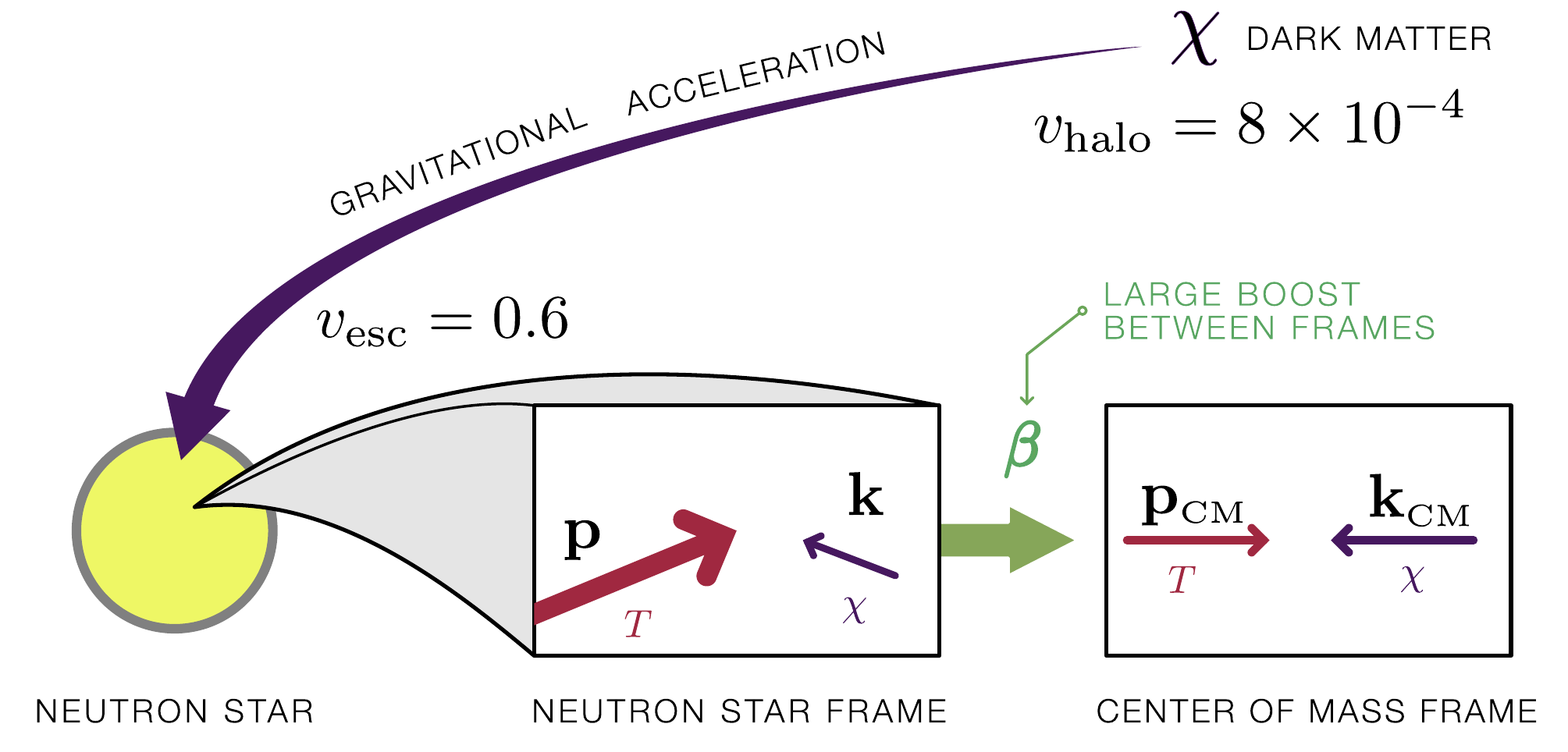}
  \caption{%
  Graphical definition of the kinematic quantities in this paper. The dark matter velocity asymptotically far from the star, $v_\textnormal{halo}$, and at the point of impact, $v_\text{esc}$, are defined in the neutron star frame. Frame-dependent with no subscripts are assumed to be in the neutron star frame, whereas center-of-momentum frame quantities are labeled with a subscript \acro{CM}.
  }
  \label{fig:NS:Boosts} 
\end{figure}

\subsection{Acceleration of Dark Matter}

A dark matter particle $\chi$ in the halo is gravitationally accelerated toward a neutron star. At the star’s surface, $\chi$ has an effectively radially inward trajectory with total energy, boost, and velocity
\begin{align}
  \gamma_\esc m_\chi &= m_\chi + \frac{2GM_\star m_\chi}{R_\star} 
  &
  \gamma_\esc &= 1.24
  &
  v_\esc &= 0.6
  \ ,
  \label{eq:gamma:chi:v:chi}
\end{align}
where $m_\chi$ is the dark matter mass.
In this estimate we ignore the dark matter’s velocity in the halo, which is a negligible contribution its energy at the star’s surface. We use the fact that the escape velocity in a Schwarzschild background is identical to the Newtonian escape velocity; this accounts for the factor of two in the gravitational potential term of $\gamma_\esc m_\chi$~\cite[ex.~9.1]{hartle2003gravity}.
We depict these velocities and our conventions for the initial scattering state in Figure~\ref{fig:NS:Boosts}.

\subsection{Kinetic Heating}
\label{sec:Kinetic:Heating}

The flux of dark matter onto the neutron star depends on the maximum impact parameter for incident dark matter to intersect the star~\cite{Goldman:1989nd},
\begin{align}
  b_\text{max} &= 
  \frac{R_\star}{v_\text{halo}} 
  \sqrt{\frac{2GM_\star}{R_\star}}
  \left(1-\frac{2GM_\star}{R_\star}\right)^{-1/2} 
    = R_\star \frac{v_{\rm esc}}{v_{\rm halo}} \gamma_{\rm esc}
    \ ,
  \label{eq:bmax}
\end{align}
where $v_\text{halo}=8\times 10^{-4}$ is the dark matter velocity in the halo, asymptotically far from the star.
The expression for $b_\text{max}$ follows from conservation of energy and angular momentum in a Schwarschild metric, see Appendix~\ref{app:bmax}. The total dark matter mass passing through the neutron star per unit time is then
\begin{align}
  \dot{M}_\chi 
  =
  \pi b_\text{max}^2 
  v_\esc
  \rho_\chi 
  \approx 
  3.1 \times 10^{25}~\frac{\text{GeV}}{\text{s}}
  \approx 55~\frac{\text{g}}{\text{s}}
  \ .
\end{align}
Over the age of the universe, this accreted dark matter density is negligible compared to the visible matter species in a neutron star. However, this dark matter deposits a constant flux of kinetic energy onto the neutron star that is converted into heat:
\begin{align}
  \dot K 
  &\;=\; 
  \left(
    \gamma_\esc
    - 1
  \right)
  \dot M
  f
  \;\approx\; 
  6.5 \times 10^{24}\,f\text{ GeV s}^{-1},
\end{align}
where $f$ is the dark matter \emph{capture efficiency}.  The central result of this paper is to calculate $f$ for dark matter scattering on degenerate, relativistic targets.
The energy deposited in the neutron star is converted into a kinetic heating of the apparent blackbody temperature 
\begin{align}
  T_\text{kin} &=  1600~f^{1/4} \, \text{K}\ .
  \label{eq:T:kinetic:f:1:4}
\end{align}
A $\mathcal O(10^9~\text{year})$-old neutron star is expected to cool to $\mathcal O(100~\text{K})$~\cite{Yakovlev:2004iq,Page:2004fy}. 
The measurement of kinetic heating above the expected neutron star {luminosity} by an infrared telescope is a smoking gun signature for dark matter--visible matter interactions.

\subsection{Dark Matter Capture}

The possibility that dark matter may capture on celestial objects has long been an opportunity for the \emph{indirect detection} of dark matter’s annihilation products~\cite{Press:1985ug, Silk:1985ax}, see e.g.~\cite{Gould:1987ir} for a detailed treatment.
In contrast, our treatment of kinetic heating is an extension of the \emph{direct detection} process intrinsic to dark matter capture.
In this paper, we focus only on heating from dark matter that is captured in the neutron star. With this assumption, all of the dark matter's kinetic energy is converted into heat.
This simplification is conservative: some dark matter may deposit energy by scattering but not capture in the star. Further, it is possible that captured dark matter may subsequently annihilate within the neutron star, converting its mass energy into additional heating~\cite{Baryakhtar:2017dbj,Raj:2017wrv,Acevedo:2019agu}. This effect enhances the proposed kinetic heating signal.

There are two conditions for dark matter to capture in a celestial object like a neutron star: 
\begin{enumerate}

\item {The dark matter--target scattering cross section is large enough for a transiting dark matter particle to interact with the target particles.}
This means that over the transit time $\Delta t$ across the star, there are a sufficient number of interactions, $d\sigma\, v_\text{rel} \langle n_\text{T}\rangle \Delta t$. 

\item {The dark matter particle loses enough energy from scattering that it is unable to escape the gravitational potential of the capturing object.}
This means that by the time it exits the star, the dark matter has lost its asymptotic initial kinetic energy. Effectively this requires that its radial velocity is less than the star’s escape velocity at some point of its transit.
\end{enumerate}
In the non-relativistic treatment of dark matter capture in neutron stars, the first condition is diagnosed by comparing the dark matter--target cross section to a \emph{threshold (saturation) cross section}. 
Increasing the cross section beyond this threshold value does not increase probability of capture.
The second condition is determined by the fixed-target kinematics wherein gravitationally accelerated dark matter hits a stationary target in the neutron star rest frame.

\begin{figure}
\centering
\includegraphics[width=\textwidth]{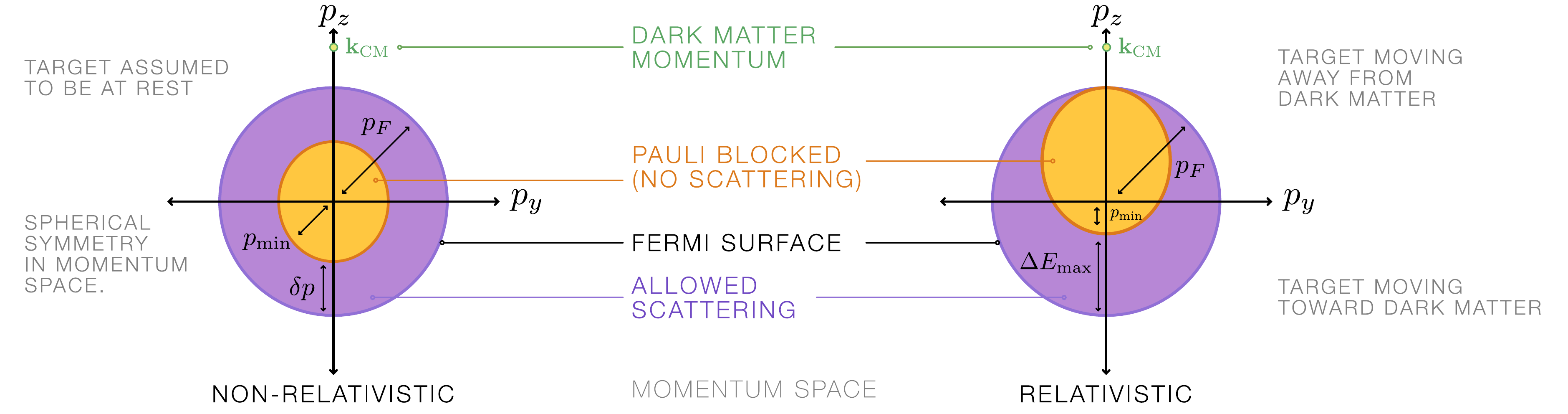}
\caption{
The Pauli exclusion principle blocks the available final state phase space for degenerate targets. We sketch this effect for non-relativistic (left) and relativistic (right) targets in momentum space. Dark matter is assumed to enter along the positive $z$-direction with some kinetic energy. The purple regions represents targets that are allowed to scatter because the final state will be outside the Fermi surface. Orange regions, on the other hand, are \emph{Pauli blocked} and are forbidden from scattering. 
}
\label{fig:pauli:block:compare}
\end{figure}

These conditions are more nuanced for relativistic, degenerate targets like electrons in a neutron star. Because the targets are relativistic, they are not at rest in the neutron star frame where energy loss is calculated. Instead, the dark matter--target scattering events have an ensemble of center-of-momentum kinematic configurations against which the differential cross section must be integrated. 

The Pauli exclusion principle introduces an additional condition on the energy transfer: the outgoing target particle must scatter into phase space that is not already filled by Fermi-degenerate states. In other words, chunks of phase space are \emph{Pauli blocked}; see Figure~\ref{fig:pauli:block:compare}. Pauli blocking occurs for non-relativistic targets as well; for example, neutrons in the neutron star are degenerate, but have a Fermi momentum below their mass. However, relativistic targets have the additional nuance that the final state kinematics in the center of momentum frame are not sufficient to determine if the given scattering is Pauli blocked since there is a non-trivial boost to the neutron star frame.

\subsection{Non-Relativistic Targets: Threshold Cross Section}
\label{sec:NR:target:threshold:xsec}

For non-relativistic targets, the threshold cross section is simply the geometric cross section of the neutron star, $\pi R_\star^2$, divided by the total number of target particles $\approx M_\star/m_\text{T}$:
\begin{align}
  \sigma_\text{thres} 
  &
  =
  \frac{\text{geometric cross section}}{\text{number of targets}} 
  =
  \pi R_\star^2 \frac{m_\text{T}}{M_\star} 
  &
  \text{GeV} \lesssim m_\chi \lesssim 10^6~\text{GeV}
  \ .
  \label{eq:non:relativisticthreshold:cross:section}
\end{align}
This expression is valid in a range of dark matter masses between the target  Fermi energy and a maximum mass above which {multiple scatters per transit are required for successful capture}.

Pauli blocking limits the available phase space for dark matter masses below $\mathcal O(\text{GeV})$: the incident low-mass dark matter does not have enough kinetic energy to scatter a target particle out of its degenerate Fermi surface. In this case only a fraction of the targets near a `skin’ of the Fermi surface are accessible for scattering. This fraction is $\delta p/p_\text{F}$ where $\delta p \approx (\gamma_\chi-1)m_\chi v_\chi$ is the maximum kinetic energy of the incident dark matter. Thus for this case
\begin{align}
  \sigma_\text{thres}^\text{Pauli} 
  &= 
  \frac 13 \frac{\delta p}{p_F}
  \sigma_\text{thres}
  \approx 
  \frac{\text{GeV}}{m_\chi}
  \sigma_\text{thres}
  &
  m_\chi &\lesssim \text{GeV} \ .
\end{align}

Conversely, for very heavy dark matter, the amount of energy transferred $\Delta E$ saturates to a fixed value independent of its mass; see Appendix~\ref{app:Delta:E:Non:Relativistic}. On the other hand, the dark matter's asymptotic kinetic energy, $\frac{1}{2}mv_\text{halo}^2$, scales linearly with its mass. Thus in the heavy dark matter limit, a single scatter is insufficient to transfer the dark matter’s total kinetic energy to a target. In this case, one requires multiple scatters to capture dark matter. The scaling yields 
\begin{align}
  \sigma_\text{thres}^\text{multi} 
  &
  \approx 
  \frac{m_\chi}{10^6~\text{GeV}}
  \sigma_\text{thres}
  &
  m_\chi &\gtrsim 10^6~\text{GeV} \ .
\end{align}

In this paper, we examine the `phase space' of dark matter capture on neutron stars from \emph{relativistic} targets as a function of the dark matter mass. We demonstrate the principles that lead to analogous low mass, intermediate mass, and high-mass scaling regimes.
Relativistic targets, however, require a completely different formalism. For non-relativistic targets one may simply compare the \emph{total} dark matter--target cross section to a threshold cross section; this is a notion that is well defined because the targets are all at rest relative to the neutron star. 
This assumption breaks down for the case of relativistic targets.

\section{Formalism for Relativistic Targets}
\label{sec:formalism}

We systematically develop the relativistic formula for dark matter capture on a compact object with degenerate targets. Our expression is general and matches non-relativistic results in that limit. All quantities are assumed to be in the neutron star frame unless explicitly otherwise indicated by a subscript.

\subsection{Breakdown of the Non-Relativistic Treatment}

The concept of a threshold cross section in Section~\ref{sec:NR:target:threshold:xsec} breaks down when the neutron star targets are relativistic. For example, the electron Fermi energy in a neutron star is much greater than its mass. The target particles are thus typically traveling near the speed of light. The amount of energy transferred from the dark matter particle to the target depends on this initial state kinetic energy. Thus the capture rate depends on the relative orientation of the target and dark matter three-momenta.

Further, it is not possible to define a total cross section $\sigma$ that one may compare to any meaningful threshold, as is standard for non-relativistic targets, c.f.~Section~\ref{sec:NR:target:threshold:xsec}. Instead, one must calculate the differential cross section $d\sigma$ with respect to the `initial state phase space’ of targets. 
What more, the center-of-momentum scattering cross section must be boosted into the neutron star frame. This boost is not collinear with the collision axis so that the cross section is non-trivially length-contracted. This can be a significant effect given the potential magnitude of the boost between these frames.

One of the limitations of the non-relativistic approximation is seen in the expression for the number of captured dark matter particles, which depends on a ratio of cross sections. This is not Lorentz invariant as required. 
Our main result in this section is a careful derivation of a fully relativistic capture efficiency. In Ref.~\cite{Joglekar:2019vzy} we showed  that the fully relativistic capture efficiency is many orders of magnitude larger than the non-relativistic approximation.

\subsection{Capture Probability}
\label{sec:capture:efficiency}

Given some infinitesimal piece of the initial state and final state phase space, the differential \textbf{capture efficiency} for an incident dark matter particle, $df$, is the total number scatters $d\nu$ with the neutron star targets divided by the total number of incident dark matter particles $dN_\chi$ subject to the kinematic conditions that the scatter leads to capture: 
\begin{align}
  df = \left.\frac{d\nu}{dN_\chi}\right|_\text{capture} \ .
  \label{eq:diff:capture:eff:initial}
\end{align}
This capture efficiency replaces the threshold cross section in Section~\ref{sec:NR:target:threshold:xsec}.
The integrated capture efficiency is an expected number of scatters satisfying the capture conditions. When this number is less than one, it can be interpreted as a \emph{capture probability per dark matter particle}. Each of the quantities $df$, $d\nu$, and $dN_\chi$ are Lorentz invariant.

The number of scatters $d\nu$ of dark matter with density $dn_\chi$ on targets with density $dn_\text{T}$ and relative velocity $v_\text{rel}$ is
  \begin{align}
  d\nu
  &= 
  d\sigma \, v_\text{rel} \, dn_\text{T} \, dn_\chi \, \Delta V  \, \Delta t \ .
  \label{eq:dnu:dsig:vrel:dnt:dnchi:Dv:Dt}
\end{align}
Here $\Delta t \approx 2~R_\star$ is the approximate transit time of a dark matter particle through the star.
The neutron star volume is $\Delta V$ and relates the dark matter number density to the total number of dark matter particles available to scatter in the star, $dN_\chi = dn_\chi \Delta V$. We thus write the differential capture efficiency \eqref{eq:diff:capture:eff:initial} as
\begin{align}
  df =&   
  \left. 
    d\sigma \, v_\text{rel} \, dn_\text{T}  \, \Delta t 
    \,
  \right |_\text{capture}
  \ .
  \label{eq:df:dsig:v:dnt:dt}
\end{align}
The capture conditions are restrictions on phase space based on the energy transfer, $\Delta E$. 
We explicitly define these conditions in Section~\ref{sec:energy:transfer:conditions}.

\subsection{Connecting Factors in Different Frames}

While $df$ in \eqref{eq:df:dsig:v:dnt:dt} is Lorentz invariant and can be computed in any frame, its covariant factors are most naturally defined in different frames.
Specifically, capture conditions and the target number density are most simply stated in the neutron star frame where the Fermi surface is spherical and the densities are uniform. In contrast, the cross section $d\sigma$ is typically calculated in the center of momentum frame. 
The large boost between these frames are the origin of the dramatic results of our fully relativistic treatment of dark matter capture on relativistic targets compared to a non-relativistic approximation.

We calculate the right-hand side of \eqref{eq:df:dsig:v:dnt:dt} in the neutron star frame. 
The quantity $d\sigma\, v_\text{rel}$ in this frame is readily expressed with respect to the differential cross section in the center of momentum frame.
The M\" oller velocity, $v_\Mol$, relates the cross section in any frame $d\sigma$ to the cross section in the center of momentum frame, $d\sigma_\CM$:
\begin{align}
        d\sigma \, v_\text{rel} &= d\sigma_\CM \, v_\Mol 
        &
        v_\Mol &= \frac{\sqrt{(p\cdot k)^2 - m_\text{T}^2 m_\chi^2}}{E_p E_k}
        \ .
        \label{eq:dsig:vrel:dsig:CM:vMol}
\end{align} 
This result is well known from the calculation of dark matter annihilation~\cite{Gondolo:1990dk}; 
for completeness we derive it in Appendix~\ref{sec:Moller}.
Thus we write $df$ in the neutron star frame with respect to the center-of-momentum cross section d$\sigma_\CM$,
  \begin{align}
  df =&   
  \left. 
    d\sigma_\CM \, v_\Mol \, dn_\text{T}  \, \Delta t 
    \,
  \right |_\text{capture}
  \ .
  \label{eq:df:dsig:CM:v:dnt:dt}
  \end{align}
To integrate $df$, one requires explicit factors enforcing the energy transfer conditions for capture.

\subsection{Energy Transfer Conditions}
\label{sec:energy:transfer:conditions}

The explicit factors in \eqref{eq:df:dsig:CM:v:dnt:dt} determine the probability of scattering. The contribution of these factors are subject to the condition that the scattering dark matter is captured by the neutron star.
The differential capture efficiency, $df$, should only be non-zero if the energy transferred $\Delta E$ from the dark matter to the target ($i$) overcomes the Fermi degeneracy (Pauli blocking) of outgoing target states and ($ii$) depletes the dark matter’s asymptotic kinetic energy so that it cannot escape the star’s gravitational potential. These conditions are applied as step functions in the capture efficiency:
\begin{align}
  \Theta(x) 
  = 
  \begin{cases}
  1 & \text{if }x > 0
  \\
  0 & \text{otherwise}
  \end{cases} \ .
\end{align}

\subsubsection{Degenerate Targets}
\label{sec:degenerate:targets:EF:pauli:blocking}

Because of the Fermi degeneracy of the neutron star, the outgoing target particle must have momentum greater than the Fermi momentum, $p_\text{F}$, or else the Pauli exclusion principle \emph{blocks} the interaction. 
This restricts the phase space to have a minimum energy transfer from the dark matter to the target, $\Delta E$, in the neutron star frame where the Fermi surface is spherically symmetric:
  \begin{align}
    \Delta E + E_p - E_\text{F} &> 0 
    &
    df \sim 
    \left.
      \Theta\left(\Delta E + E_p - E_\text{F}\right)
    \right|_\textnormal{KE}
    \ ,
    \label{eq:DeltaE:Ep:EF}
  \end{align}
where $E_p$ is the energy of the initial state target particle and $E_\text{F}$ is the Fermi momentum for the target species. 
Figure~\ref{fig:pauli:block:compare} demonstrates why this treatment is necessary compared to the non-relativistic limit.
The subscript \acro{KE} is a reminder that one must still impose the second capture condition on the dark matter's kinetic energy.

\subsubsection{Depleting the Kinetic Energy: Single Scatter Case}

The second capture condition is that the dark matter must transfer enough of its kinetic energy in the scattering event. This is a matter of whether the outgoing dark matter particle has less than it's escape velocity at the point of scattering. This amounts to losing the kinetic energy it had asymptotically far from the star: 
\begin{align}
    \Delta E - \Delta E_\text{min} &> 0
    &
    \Delta E_\text{min} &= E_\text{halo} = \frac 12 m_\chi v_\text{halo}^2 
    \ .
    \label{eq:DeltaE:Ehalo}
  \end{align}
We assume that dark matter--target scattering is elastic. Restricting to the case where dark matter scatters only \emph{once} in the neutron star, this tells us that the single-scatter capture efficiency is
\begin{align}
  df_{1} =&   
    d\sigma_\CM \, v_\Mol \, dn_\text{T}  \, \Delta t 
    \;
  \left.
  \Theta\left(\Delta E - \Delta E_\text{min}\right)
  \right|_{\textnormal{Pauli}}
  \ ,
  \label{eq:df:dsig:CM:v:dnt:dt:1Hit}
  \end{align}
  where the subscript `Pauli' is a reminder that one must still impose the second capture condition on the target final state.

\subsubsection{Multiple Scattering}
\label{sec:multiple:scatter:Nhit}

The energy transfer condition \eqref{eq:DeltaE:Ehalo} is modified when the transiting dark matter particle can scatters more than once in the target volume of the neutron star. In that case it is sufficient for dark matter to lose its asymptotic kinetic energy $\Delta E_\text{min} = E_\text{halo}$ over multiple interactions over the course of its entire transit through the star. The generalization of the condition \eqref{eq:DeltaE:Ehalo} for $N_\text{hit}$ scatters is
  \begin{align}
    \langle\Delta E\rangle - \frac{\Delta E_\text{min}}{N_\text{hit}} > 0
    \ ,
    \label{eq:DeltaE:Ehalo:Nhit}
  \end{align}
where $\langle \Delta E\rangle$ is the average energy transfer over all of the dark matter scatters.
For a dark matter particle that requires $N_\text{hit}$ scatters to capture, each scatter must occur in a fraction of the total transit time,
  \begin{align}
    \Delta t_{N_\text{hit}} = \frac{\Delta t}{N_\text{hit}}  \ .
    \label{eq:Delta:t:Nhit}
  \end{align}
We assume that the dark matter takes a straight line path through the star with no significant deflection. 

A full treatment of the capture including multiple scatters is computationally demanding. In order to impose \eqref{eq:DeltaE:Ehalo:Nhit} one must keep track of the transiting dark matter particle’s scattering history. Further, one must take an appropriately weighted sum over possible number of scatters $N_\text{hit}$.
To make the problem tractable, we make a conservative simplification and replace \eqref{eq:DeltaE:Ehalo:Nhit} with the stronger condition that \emph{each} scatter must have at least the minimum average energy to capture:
      \begin{align}
      \Delta E - \frac{\Delta E_\text{min}}{N_\text{hit}} > 0
      \ .
      \label{eq:DeltaE:Ehalo:Nhit:no:average}
    \end{align}
One may then sum over the number of hits required to scatter:
  \begin{align}
    df &= 
    \sum_{N_\text{hit}} 
    \;
    d\sigma_\CM 
    \, v_\Mol \, dn_\text{T}  \, \frac{\Delta t}{N_\text{hit}} 
      \,
    \left.
      \Theta
      \left(
        \Delta E - \frac{\Delta E_\text{min}}{N_\text{hit}}  
      \right)
      \Theta
      \left(
        \frac{\Delta E_\text{min}}{N_\text{hit}-1}   - \Delta E
      \right)
    \right|_{\textnormal{Pauli}}
    \ .
    \label{eq:df:dsig:CM:v:dnt:dt:total:sum:Nhit}
  \end{align}
A phase space region that captures after $N$ hits is only counted in the term of the sum where $N_\text{hit} = N$; this is imposed by the two step functions. 
See Appendix~\ref{app:multiple:scatter:capture} for a detailed discussion.

\subsection{Capture Probability Formula}

\begin{figure}
\centering
\includegraphics[width=.9\textwidth]{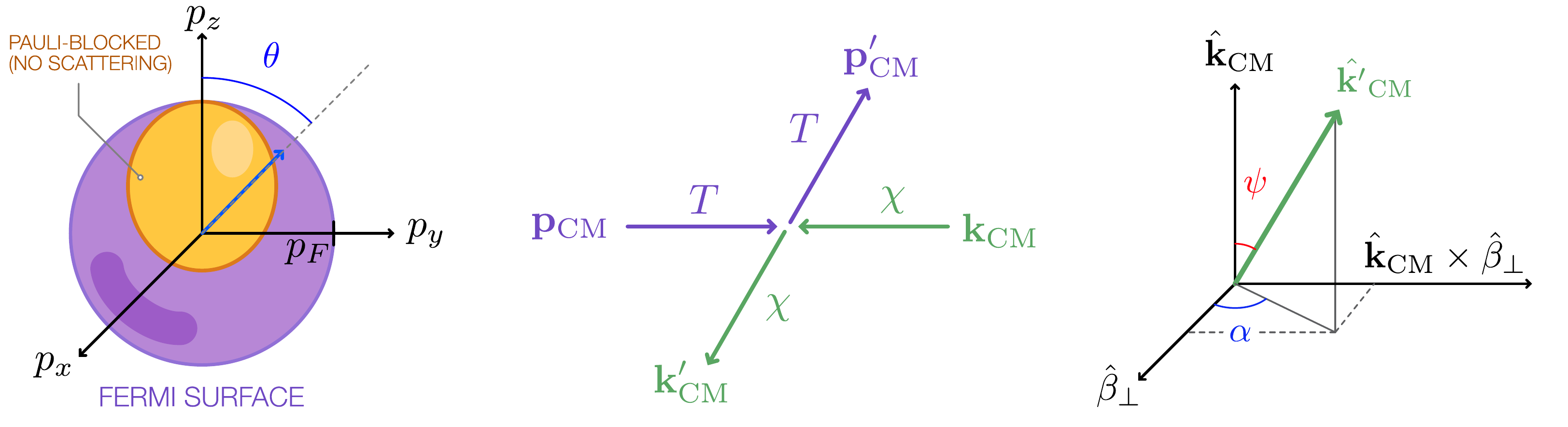}
\caption{Variables and angles. Left: target momentum space. Center: center of momentum kinematic variables between the target, $T$, and dark mater, $\chi$. Right: Center of momentum frame scattered dark matter direction, $\hat{\vec{k}}'_\text{CM}$ with respect to a coordinate system of the initial dark matter direction, $\hat{\vec{k}}'_\text{CM}$, the component of the boost between the two frames perpendicular to the initial direction, $\hat{\vec{\beta}}_\perp$, and the orthogonal direction.}
\label{fig:coordinates}
\end{figure}

The full expression for the differential capture rate combines the base expression for $df$ \eqref{eq:df:dsig:v:dnt:dt} that enforces Pauli blocking \eqref{eq:DeltaE:Ep:EF} and the sum and over multiple scatters \eqref{eq:df:dsig:CM:v:dnt:dt:total:sum:Nhit}:
\begin{align}
  df &= 
  \sum_{N_\text{hit}} 
  \;
  d\sigma_\CM 
  \, v_\Mol \, dn_\text{T}  \, \frac{\Delta t}{N_\text{hit}} 
    \,
  \Theta
  \left(
    \Delta E - \frac{E_\text{halo}}{N_\text{hit}}  
  \right)
    \Theta
    \left(
      \frac{\Delta E_\text{halo}}{N_\text{hit}-1}   - \Delta E
    \right)
  \Theta
  \left(
    \Delta E + E_p - E_\text{F}
  \right)
  \ .
  \label{eq:df:dsig:CM:v:dnt:dt:total:step 1}
  \end{align}
It is convenient to explicitly write the center-of-momentum cross section with respect to the kinematics in that frame
\begin{align}
  d\sigma_\CM 
  &=
  \frac{d\sigma_\CM}{d\Omega_\CM}
  d\Omega_{\CM}
  &
  d\Omega_{\CM} &= d\alpha \; d(\cos \, \psi),
\end{align}
where $\psi$ and $\alpha$ are the polar and azimuthal angles of scattering respectively in the center-of-momentum frame; see Figure~\ref{fig:coordinates}. The differential volume in the target momentum space with respect to its Fermi sphere is
\begin{align}
  dn_\text{T}
  &= 
  \langle n_\text{T}\rangle \frac{p^2 dp\, \Omega_{\text{F}}}{V_\text{F}}
  &
  V_\text{F} &= \frac{4}{3}\pi p_\text{F}^3 
  &
  d\Omega_{\text{F}} &= d\varphi \; d(\cos \, \theta)
  \ ,
  \label{eq:NS:angular:omega:F:phi:theta}
\end{align}
where $\langle n_\text{T}\rangle$ is the average target density in \eqref{eq:average:n:T} and we write $p = |\vec{p}|$ to be the magnitude of the target three-momentum. This is integrated up to $p_\text{F}$, the Fermi three-momentum. The $d\varphi$ integral is trivial.
The final expression is
\begin{align}
  f &= 
  \sum_{N_\text{hit}} 
  \;
  \frac{\langle n_\text{T}\rangle\Delta t}{N_\text{hit}} 
  \int d\Omega_\text{F}
  \int_0^{p_\text{F}}
  \frac{p^2 dp }{V_\text{F}}
  \int d\Omega_\CM
    \frac{d\sigma_\CM }{d\Omega_\CM}
  \, v_\Mol \,  
    \,
  \Theta^3(\Delta E)
  \ ,
  \label{eq:f:full}
  \end{align}
  where we use the shorthand notation $\Theta^3(\Delta E)$ to indicate the step functions from Pauli blocking and multiple scatters,
  \begin{align}
    \Theta^3(\Delta E)
    &\equiv
    \Theta
    \left(
    \Delta E - \frac{E_\text{halo}}{N_\text{hit}}  
    \right)
    \Theta
    \left(
      \frac{\Delta E_\text{halo}}{N_\text{hit}-1}   - \Delta E
    \right)
    \Theta
    \left(
      \Delta E + E_p - E_\text{F}
    \right)
  \ .
  \end{align}

The capture efficiency, $f$, is defined to be the weighted number of scatters for a dark matter particle that captures. When this number is greater than one, we may assume that the dark matter particle is captured. When the number is less than one, then the probability for a given dark matter particle to capture is $f$. In other words, the capture probability of a given dark matter particle in some initial phase space volume $dn_\chi$ is $P_\text{capture} = \min\left(f,1\right)$.

\subsection{Non-Relativistic Target Limit}
\label{sec:non:relativistic:target:limit}

We confirm our primary result \eqref{eq:f:full} by verifying that it reduces to earlier results in the limit of stationary (non-relativistic) target particles~\cite{Baryakhtar:2017dbj,Bell:2019pyc,Raj:2017wrv}. 
In this limit, the initial target three-momentum is zero $\vec{p}=0$, which trivializes the integration over initial target momenta. This means that the step functions in \eqref{eq:DeltaE:Ep:EF} and \eqref{eq:DeltaE:Ehalo:Nhit} no longer impose kinematic constraints on the phase space integrals and may be factored out and treated following the discussion of the non-relativistic case in Section~\ref{sec:NR:target:threshold:xsec}. This approximation was previously applied to neutron star dark kinetic heating from interactions with muons in Ref.~\cite{Garani:2019fpa} and leptophilic interactions in Ref.~\cite{Bell:2019pyc}.%

Define $d \hat f = d\nu/dN_\chi$ to be the differential capture efficiency \emph{without} capture conditions imposed. To show consistency with the non-relativistic limit, it is sufficient to show that $\hat f = \sigma/\sigma_\text{thres}$, where the threshold cross section $\sigma_\text{thres}$ is simply the geometric cross section of a target in the neutron star.
In the $\vec{p}\to 0$ limit, the M\o ller velocity reduces to the dark matter velocity in the neutron star frame, $v_\Mol \to k/E_k = v_\esc$.
The capture efficiency reduces to
\begin{align}
  \hat f 
  &= 
  v_\esc
  \langle n_\text{T}\rangle \Delta t
  \int \frac{p^2 dp \, d\Omega_\text{F}}{V_\text{F}}
  \int d\sigma \ .
\end{align}
Observe that $\int d\sigma = \sigma$ is the total dark matter--target cross section; a quantity which we argued is not well defined for an ensemble of relativistic, degenerate targets. The $d^3\vec{p}$ integral over initial momenta is also trivially equal to unity. One may technically write this by imposing a $\delta^{(3)}(\vec{p})$ distribution forcing the targets to be stationary, or alternatively by taking $p_\text{F}\to 0$. Finally we observe that
\begin{align}
  \hat f 
  &= 
  v_\esc
  \, \Delta t \langle n_\text{T} \rangle \, \sigma
  = \frac{\sigma}{\sigma_\text{thres}}
  = \text{number of dark matter--target scatters} \ ,
\end{align}
so that this indeed recovers the standard non-relativistic treatment in Section~\ref{sec:NR:target:threshold:xsec}.

\subsection{Numerical Methodology}

The results in this paper are based on numerically integrating \eqref{eq:f:full}.
We use Python 3 on a personal computer with the \texttt{NumPy} numerical methods module and the \texttt{Vegas} module for Monte Carlo integration.
The estimated run time for evaluating $f$ at a given dark matter mass depends on the kinematic regimes in Section~\ref{sec:kinematic:regimes}. On a single 2.3~GHz core: a point in the light dark matter regime is evaluated in $\mathcal{O}(\text{minute})$. For heavy and very heavy dark matter, it takes $\mathcal{O}(\text{second})$ and $\mathcal{O}(\text{hour})$, respectively. Together, it takes a few hours to produce a scan over dark matter masses from 10~eV to 10~PeV for a given target and interaction operator.


\section{Scaling of Scattering}
\label{sec:RelScat}

\begin{figure}[tb]
  \centering
  \includegraphics[width=.7\textwidth]{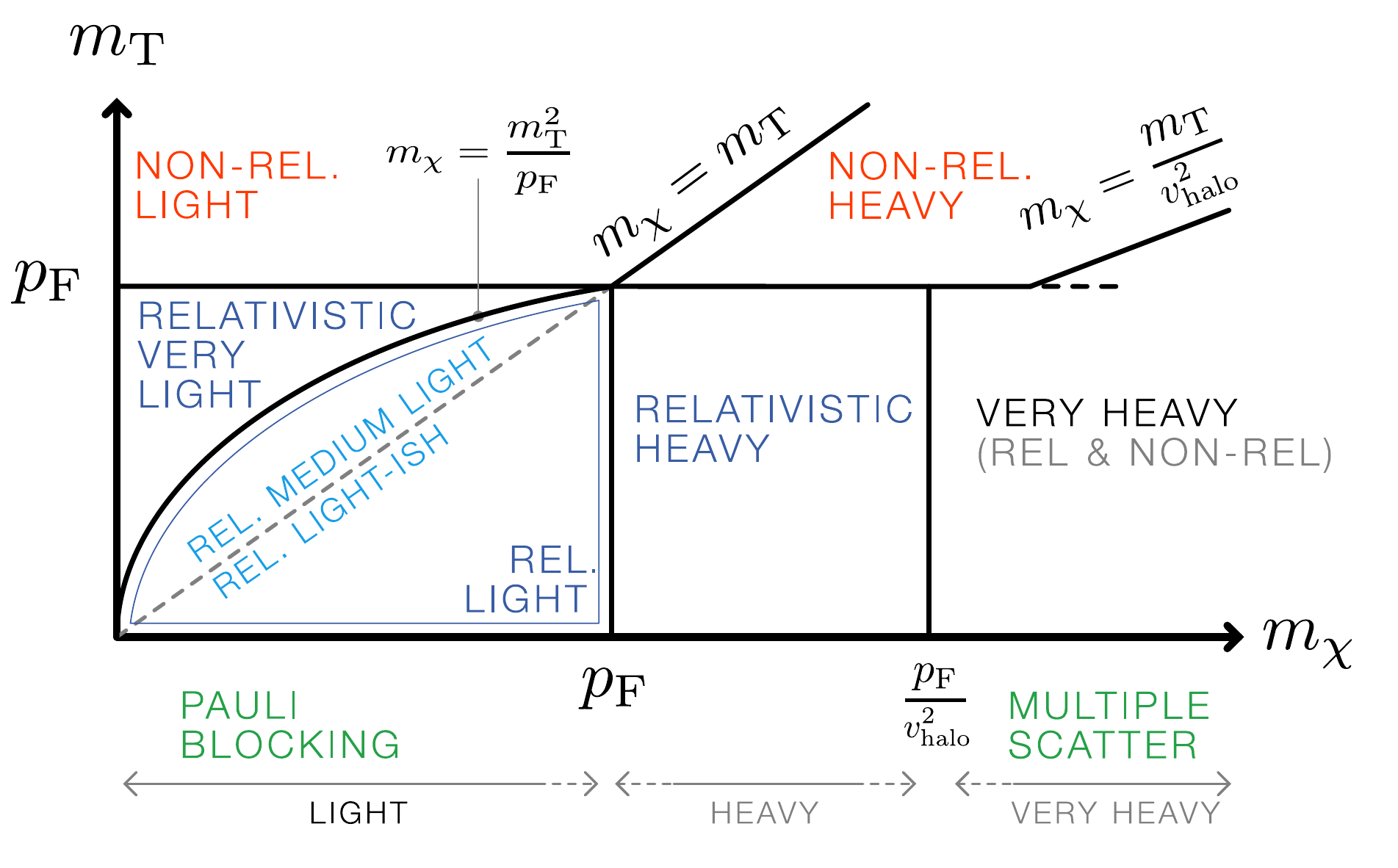}
  \caption{The phase space of scattering kinematics for neutron star kinetic heating. The target mass relative to its Fermi momentum determines whether it is \emph{relativistic} or \emph{non-relativistic}. The dark matter mass relative to the target mass and Fermi momentum determines whether it is \emph{heavy} or \emph{light}. The subdivisions of relativistic dark matter are described in the main text.
  }
  \label{fig:app:phaseblock:phasespace}
\end{figure}

We present the general behavior of dark matter capture on relativistic targets.
The kinematics of dark matter scattering off relativistic targets depends on the dark matter mass relative to the other mass scales in the system: the target mass, $m_\text{T}$, and the Fermi momentum, $p_\text{F}$. 
For relativistic targets, $m_\text{T} \ll p_\text{F}$ so that the Fermi energy and Fermi momentum are effectively equal, $E_\text{F}\approx p_\text{F}$.
Figure~\ref{fig:app:phaseblock:phasespace} shows the regimes of qualitatively different kinematics and sketches the discovery reach with respect to these regimes. This behavior is in contrast to the qualitative behavior of non-relativistic kinematics reviewed in Section~\ref{sec:NR:target:threshold:xsec}.
This section explains the origin of this general behavior with respect to the dark matter mass.

\begin{figure}[tb]
  \centering
  \includegraphics[width=.7\textwidth]{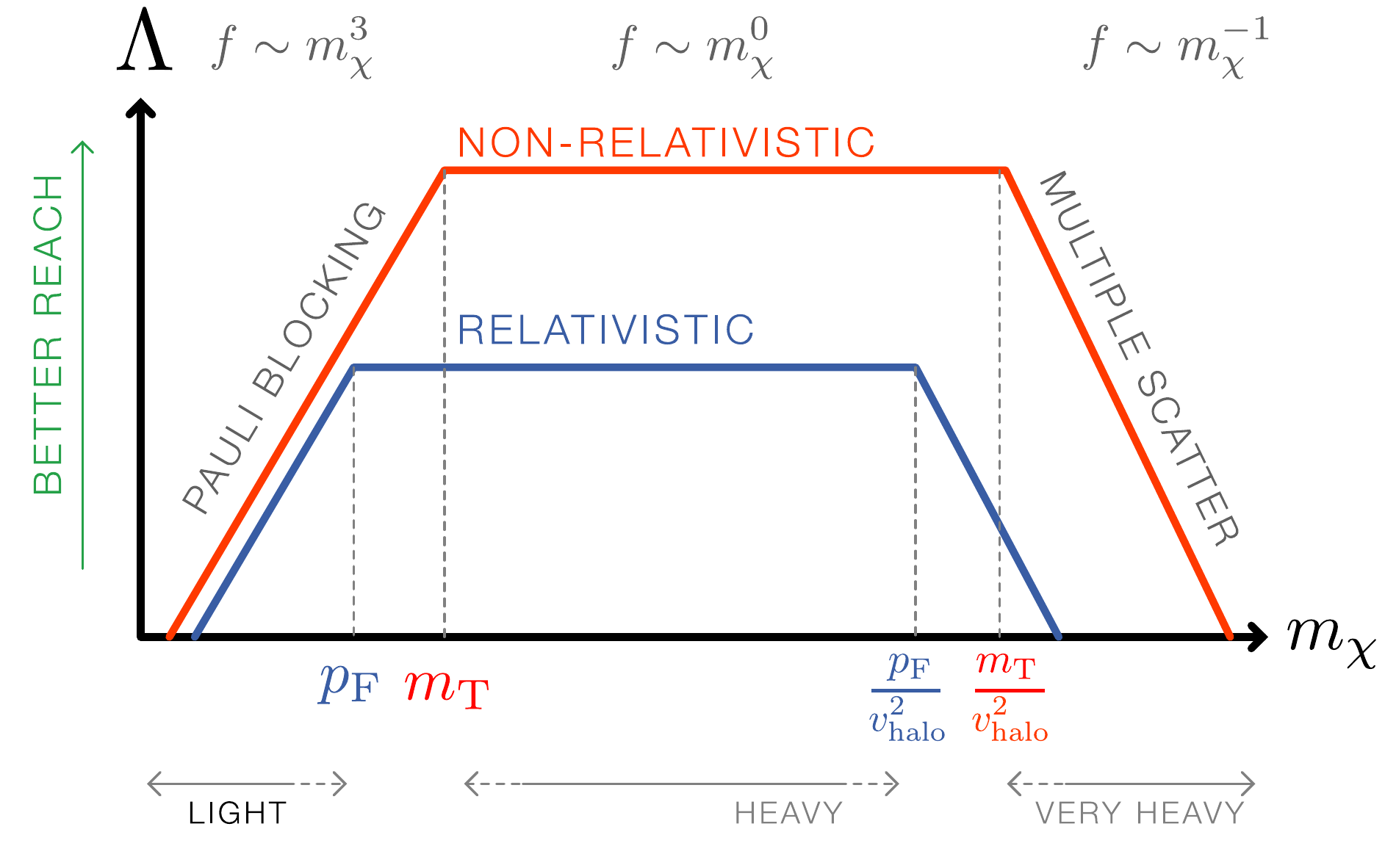}
  \caption{%
  Sketch of the reach of kinetic heating observations for dark matter scattering on relativistic (blue) or non-relativistic (red) targets through contact operators with a characteristic cutoff scale $\Lambda$. The scaling with the dark matter mass, $m_\chi$, corresponds to the phases in Figure~\ref{fig:app:phaseblock:phasespace}.
  }
  \label{fig:app:phaseblock:phasespace:mesa}
\end{figure}

\subsection{Kinematic Regimes}
\label{sec:kinematic:regimes}

A new result in this paper is a classification of the kinematic regimes for dark matter capture in compact objects according to the target and dark matter masses relative to the target Fermi momentum. 
We divide the `phase space’ of kinematic regimes according to
\begin{enumerate}     
  \item \emph{Whether the target is {relativistic} or {non-relativistic}.}\\
  If the target mass $m_\text{T}$ is lighter than its Fermi momentum, $p_\text{F}$, it is \textbf{relativistic}. 
  Otherwise, it is \textbf{non-relativistic}. Muons, for which, $m_\text{T} \approx p_\text{F}$ are marginal.

  \item \emph{Whether the dark matter is {heavy} or {light}.}\\
  If the dark matter is heavier than both $m_\text{T}$ and $p_\text{F}$, then it is \textbf{heavy}. 
  Otherwise, it is \textbf{light}. For relativistic targets, light means $m_\chi \ll p_\text{F}$, whereas for non-relativistic targets this means $m_\chi \ll m_\text{T}$. In fact, there are cases for additional subdivision:
  \begin{itemize}
  
    \item \textbf{Very heavy} dark matter requires multiple scatters to capture. 
    For relativistic targets the threshold for this is the $m_\chi \gg  p_\text{F}/v_\textnormal{halo}^2$, whereas for non-relativistic targets the threshold is $m_\chi \gg  m_\text{T}/v_\textnormal{halo}^2$.

    \item For \emph{relativistic} targets, one must also distinguish \emph{very light} dark matter from \emph{light} dark matter:
    \begin{itemize}
      \item \textbf{Very light} dark matter is lighter than $m_\text{T}^2/p_\text{F}$. The scattering cross section and phase space scales differently in this regime, as seen in \eqref{eq:qualitative:s:scaling} and Appendix~\ref{app:phase:block:rel:light}.

      \item \textbf{Light} dark matter can be further divided into \textbf{light-ish} dark matter (heavier than $m_\text{T}$ but lighter than $p_\text{F}$) and \textbf{medium light} dark matter (heavier than $m_\text{T}^2/p_\text{F}$ but lighter than $m_\text{T}$). 
      These follow the same kinematics, but their squared amplitudes scale differently. These distinctions are explored further in Appendix~\ref{sec:DominantContact} and not discussed further in the main text.
    \end{itemize}
  \end{itemize}
\end{enumerate} 
These phases are sketched in Figure~\ref{fig:app:phaseblock:phasespace}. 

The distinction between light versus heavy dark matter is manifest in the scattering kinematics. In particular, capture on relativistic targets with light dark matter prefers relatively modest momentum transfer. This is because there is a region of scattering kinematics where an energetic target transfers energy to the dark matter rather than vice versa. In this case, the center of momentum frame scattering angle is rather small and corresponds to a very forward scattering. Note that these forward scatters map onto $\mathcal O(1)$ scattering angles in the neutron star frame. This behavior is lost in the non-relativistic treatment of relativistic targets.

\subsection{Cross Section Scaling}

Contact operators parameterize the strength of short-distance interactions between pairs of dark matter particles and pairs of visible matter particles by a cutoff scale, $\Lambda$. Larger values of $\Lambda$ correspond to smaller scattering amplitudes.
Most of the contact operators, that we consider in the next section, produce squared amplitudes that scale as
\begin{align}
	| \mathcal M |^2 &\propto \frac{m_\chi^2 E_p^2}{\Lambda^4} 
	&
	\left(\frac{d\sigma}{d\Omega}\right)_\CM
	  & \propto 
	  \frac{|\mathcal M|^2}{s}
	  \approx
	  \frac{m_\chi^2 E_p^2}{s\Lambda^4}
	  \approx
	  \frac{m_\chi^2 m_\text{T}^2}{s\Lambda^4}
	  \left(1+\frac{p_\text{F}^2}{m_\text{T}^2}\right)
	\ .
  \label{eq:app:dominant:terms:baseline:M2}
\end{align}
We use this scaling relation to establish the baseline behavior of the capture efficiency in different dark matter mass regimes.
Appendix~\ref{sec:DominantContact} motivates this behavior and classifies the exceptional cases. In the non-relativistic target limit, $p_\text{F}^2/m_\text{T}^2\ll 1$ and $s=(m_\chi+m_\text{T})^2$, the above expression reduces to a standard expression familiar from dark matter direct detection:
\begin{align}
  \left(\frac{d\sigma}{d\Omega}\right)_\CM
  \propto 
  \frac{m_\chi^2 m_\text{T}^2}{(m_\chi+m_\text{T})^2\Lambda^4}  \ .
  \label{eq:qualitative:dsig:dOmega:scaling}
 \end{align}
The squared center of momentum energy, $s=E_\CM^2$, depends on the ratios of the dimensionful quantities in this expression:
\begin{align}
  s &= m_\chi^2 
  + m_\text{T}^2 
  + 2\gamma_\esc m_\chi E_p 
  \left( 1 -\frac{pv_\esc}{E_p}\cos\theta \right)
  &
  s & \approx 
  \begin{cases}
    m_\text{T}^2  & m_\chi \ll m_\text{T}^2/E_\text{F}
    \\
    m_\chi E_p    & m_\text{T}^2/E_\text{F} \ll m_\chi \ll E_\text{F}
    \\
    m_\chi^2      & E_\text{F} \ll m_\chi
  \end{cases}
  \ .
  \label{eq:qualitative:s:scaling}
\end{align}

\subsection{Characteristic Features}\label{sec:CharFeatures}

\begin{figure}[tb]
  \centering
  \includegraphics[width=\textwidth]{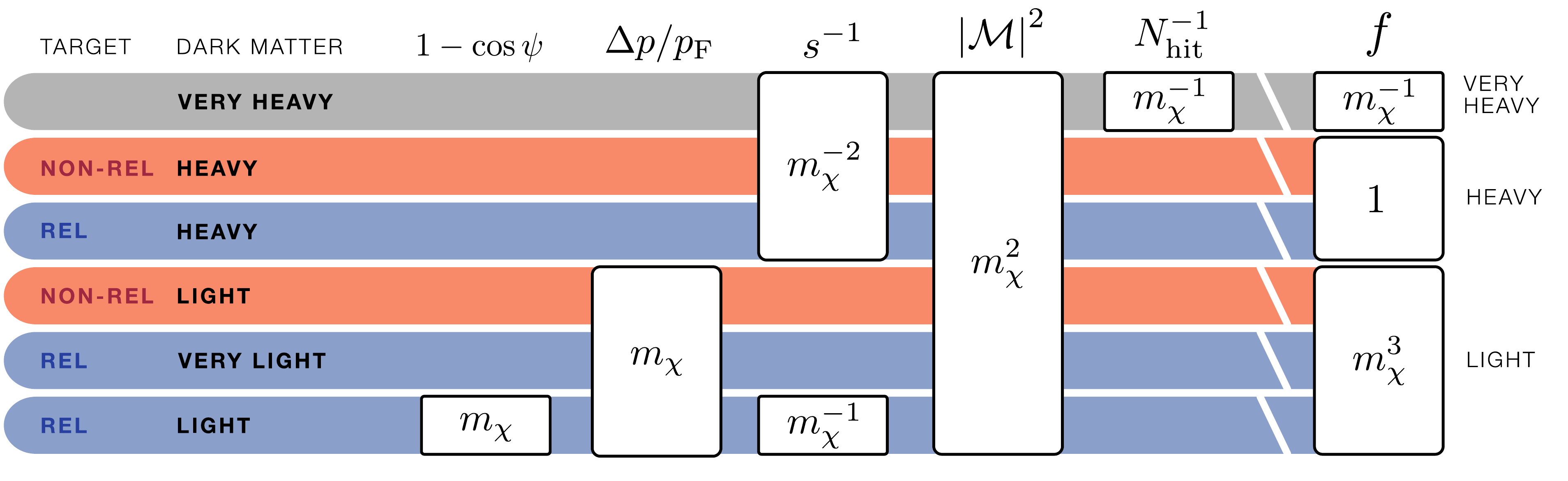}
  \caption{Chart showing the origin of the $m_\chi$ scaling for each of the  regions in~Figure~\ref{fig:app:phaseblock:phasespace}. The `plateau' behavior in Figure~\ref{fig:app:phaseblock:phasespace:mesa} is demonstrated for both relativistic (blue) and non-relativistic (red) targets.  
  }
  \label{fig:app:phaseblock:short:flow:chart}
\end{figure}

Figures~\ref{fig:app:phaseblock:phasespace:mesa} and
\ref{fig:app:phaseblock:short:flow:chart} depict the scaling of the capture efficiency $f$ with the dark matter mass $m_\chi$ and the origin of the scaling in each phase space regime. These features are realized in the numerical results in Section~\ref{sec:bounds:on:contact:ops}.  
To understand this behavior, we identify which factors in the capture efficiency, \eqref{eq:f:full}, scale with $m_\chi$:
\begin{align}
  f \sim 
  \frac{1}{N_\text{hit}} 
  \int_{\cos\psi_\text{max}}^1
  d\cos\psi
  \int_{p_\text{min}}^{p_\text{F}} \frac{p^2 dp}{p_\text{F}^3}
  \,
  \frac{| \mathcal M|^2}{s} \ .
  \label{eq:f:dependence:on:mx}
\end{align}
The scaling factors come from
($i$) a factor accounting for \emph{multiple scattering}, $N_\text{hit}^{-1}$, and
($ii$) the \emph{phase space} integrals $d\cos\psi\, p^2 dp$, and ($iii$) the \emph{differential cross section} $d\sigma/d\Omega_\CM$.

The M\o ller velocity, \eqref{eq:dsig:vrel:dsig:CM:vMol}, reduces to $v_\esc$ in the non-relativistic target limit and to $\left(1-v_\esc \cos\theta\right)$ in the relativistic target limit. In either case it does not contribute to the $m_\chi$ scaling of $f$.
The phase space factors in \eqref{eq:f:dependence:on:mx} neglect the target initial angle $\theta$ and the azimuthal scattering angle $\alpha$.
We show in Appendix~\ref{app:PhaseBlock} that these play a role in understanding the total phase space scaling with $m_\chi$, but their phase space volumes do not themselves scale with $m_\chi$. 
The Pauli blocking step functions in \eqref{eq:f:full} are converted into limits for the phase space integrations. 
The expression for the cross section depends on the details of the interaction between dark matter and the target. In the present study, we use the scaling behavior in 
\eqref{eq:app:dominant:terms:baseline:M2}
which describes most of the contact operators. 

In this analysis and the extended phase space analysis in Appendix~\ref{app:PhaseBlock}, we make the simplifying assumption that all of the $m_\chi$-dependent factors are independent of one another. Thus we treat each phase space integral as having a trivial integrand so that they are purely unweighted volume integrals. In actuality, $| \mathcal M |^2$ depends on both the center of momentum polar angle $\psi$ and target three-momentum $\vec{p}$, but our analysis is sufficient to understand the scaling of the capture efficiency with $m_\chi$. In summary, for each regime the scaling of the capture efficiency $f$ with the dark matter mass $m_\chi$ is determined by following three questions corresponding to the factors in \eqref{eq:f:dependence:on:mx}:
\begin{enumerate} 
  \item How does the differential cross section scale with $m_\chi$?
  \item Is the phase space suppressed with $m_\chi$?
  \item Does capture require multiple scatters?
\end{enumerate}

\subsubsection*{Heavy Dark Matter Regimes}

The heavy dark matter regimes are characterized by large momentum transfer so that Pauli blocking is negligible. 

\paragraph{Heavy, but not very heavy, dark matter.} This corresponds to 
$p_F \ll m_\chi \ll p_F/v_\text{halo}^2$ for relativistic targets and $ m_\text{T} \ll m_\chi \ll m_\text{T}/v_\text{halo}^2$ for non-relativistic targets.
In this limit, the cross section is independent of $m_\chi$, the phase space is unsuppressed, and dark matter captures after a single scatter. Thus, $f$ is independent of $m_\chi$.

Heavy dark matter transfers enough kinetic energy to capture in a neutron star. This is true even for relativistic targets: where the transferred energy to the targets is enough to overwhelm Pauli blocking.
The full expression for the energy transferred from the dark matter to the target, $\Delta E$, is presented in Appendix~\ref{sec:DeltaE}. Because the gravitational acceleration from the neutron star is proportional to the dark matter mass, heavier dark matter has a larger three momentum upon scattering, $\vec{k}$.
With the heavy dark matter scaling $s\sim m_\chi^2$ in \eqref{eq:qualitative:s:scaling}, the cross section $d\sigma/d\Omega$ in \eqref{eq:qualitative:dsig:dOmega:scaling} is independent of $m_\chi$.
Thus in this regime \emph{the capture efficiency is independent of the dark matter mass}, as shown by the plateau feature in Figure~\ref{fig:app:phaseblock:phasespace}.

Observe that the $(p_\text{F}/m_\text{T})^2$ term in the fully relativistic cross section 
\eqref{eq:app:dominant:terms:baseline:M2}
is not present in  the non-relativistic limit. While this is negligible for non-relativistic targets, this factor is on the order of $10^5$ for ultrarelativistic targets like electrons. Thus this is a gross underestimation when using the non-relativistic formulation for relativistic targets.

\paragraph{Very heavy dark matter.}
This corresponds to 
$m_\chi \gg p_F/v_\text{halo}^2$ for relativistic targets and $m_\chi \gg m_\text{T}/v_\text{halo}^2$ for non-relativistic targets.
This behaves like heavy dark matter, except multiple scatters are required to capture. We find that $f$ scales like $m_\chi^{-1}$.

The heavy dark matter behavior above breaks down for dark matter masses in the \emph{very heavy} regime. In this case the energy transfer $\Delta E$ as a function of dark matter mass saturates. However, the required kinetic energy loss scales linearly with the dark matter mass. In Appendix~\ref{sec:app:max:DeltaE:wrt:psi:alpha} we confirm that the maximum energy transfer from a single scatter in the heavy dark matter regime is approximately the target energy, $\Delta E_\text{max} \approx E_p$, which is independent of the dark matter mass. The very heavy dark matter threshold is when this maximum energy transfer is smaller than the minimum required energy transfer for capture $\Delta E_\text{min} = m_\chi v_\text{halo}^2/2$, \eqref{eq:DeltaE:Ehalo}. Since $v_\text{halo}^2 \sim 10^{-6}$, the threshold mass above which multiple scatters is required is
\begin{align}
  \text{very heavy:}\quad
  m_{\chi}
  \gg
  \begin{cases}
      \displaystyle
      \frac{p_\text{F}}{v_\textnormal{halo}^2} \approx 10^6 p_\text{F} 
      & \text{relativistic}
      \\[2em]
      \displaystyle
      \frac{m_\text{T}}{v_\textnormal{halo}^2} \approx 10^6 m_\text{T}
      & \text{non-relativistic}
  \end{cases}
  \ .
  \label{eq:qualitative:very:heavy:threshold}
\end{align}
The number of scatters required to transfer a total energy of $\Delta E_\text{min}$ thus scales like the dark matter mass, $N_\text{hit} \sim m_\chi$. 
Section~\ref{sec:multiple:scatter:Nhit} shows that the capture efficiency, $f$, goes like $N_\text{hit}^{-1}$ and hence in the very heavy regime $f\sim m_\chi^{-1}$.
In this regime \emph{heavier dark matter is a less effective kinetic heat source}, as shown by the falling feature to the right of Figure~\ref{fig:app:phaseblock:phasespace}.

Observe that the onset of the characteristic behavior differs for relativistic versus non-relativistic targets because the thresholds to the very heavy regime are different, \eqref{eq:qualitative:very:heavy:threshold}.
The Fermi energy of electrons is roughly an order of magnitude lower than the mass of nucleons, $p_\text{F}^e \approx 10^{-1} m_N$. Indeed, the downward slope on the reach plots begins at $\sim 10^5$~GeV for electrons versus $\sim 10^6$~GeV for neutrons and protons.

\subsubsection*{Light Dark Matter Regimes}

Light dark matter does not always transfer enough energy to excite the target to an unoccupied momentum state. The Fermi exclusion principle prevents the scattering from occurring and the phase space is {Pauli blocked}. This is depicted on the right-hand side of in Figure~\ref{fig:pauli:block:compare}. Though the blocked volume is not strictly spherical, it is sufficient to treat it as such to determine the scaling with $m_\chi$. 

The momentum space volume of available targets is approximately the volume enclosed by the Fermi sphere.
However, only a fraction of these targets can scatter to an allowed state outside the Fermi sphere. The precise shape of these allowed targets, sketched in Figure~\ref{fig:pauli:block:compare}, is not necessary to track how this volume scales the dark matter mass.
The fraction of the targets which are not blocked scales like
\begin{align}
  \int_{p_\text{min}}^{p_\text{F}}
  \frac{p^2 dp}{ \frac 13 p_\text{F}^3} 
  =
  \frac{p_\text{F}^3-p_\text{min}^3}{p_\text{F}^3}
  \approx
  \frac{3(p_\text{F}-p_\text{min})}{p_\text{F}}
  \approx
  \frac{3\Delta p}{p_\text{F}} 
  \sim
    \frac{3E_\text{F}\Delta E_\text{max}}{p_\text{F}^2} 
  \ ,
\end{align}
where in the last approximation we replace the energy transfer $\Delta E$ with its maximum value in order make the $m_\chi$ scaling manifest. The full expression for the energy transfer is derived in Appendix~\ref{sec:DeltaE}. The relevant limits of $\Delta E_\text{max}$ are shown in Appendix~\ref{sec:app:max:DeltaE:wrt:psi:alpha}. For all of the cases with light dark matter, including very light dark matter, the key result is that 
\begin{align}
  \text{light dark matter:}
  \qquad
  \Delta E_\text{max} \sim m_\chi
  \ .
\end{align}
This factor leads to a reduced (blocked) phase space for lighter dark matter in this regime. As shown in Figure~\ref{fig:app:phaseblock:short:flow:chart}, the specific light dark matter scenarios each carry different factors of $m_\chi$, but they all combine to give $f\sim m_\chi^3$.

\paragraph{Non-relativistic target, light dark matter.} 
This corresponds to 
$m_\chi \ll m_\text{T}$.
In this limit, the cross section scales like $m_\chi^2$, the phase space is Pauli blocked (factor of $m_\chi$), and dark matter captures after a single scatter. Thus, $f$ scales like $m_\chi^3$.

In this regime dark matter is light and the Fermi energy is negligible, thus $s \approx m_\text{T}^2$ and is independent of $m_\chi$. The benchmark cross section \eqref{eq:qualitative:dsig:dOmega:scaling} thus scales like $m_\chi^2$. Combined with the Pauli blocking factor, this gives $f\sim m_\chi^3$.

\paragraph{Relativistic target; light dark matter.}
This corresponds to 
$m_\text{T}^2/p_\text{F} \ll m_\chi \ll p_\text{F}$.
In this limit, the cross section scales like $m_\chi$, the phase space is both Pauli blocked (factor of $m_\chi$) and has constrained $\cos\psi$ space (factor of $m_\chi$), and dark matter captures after a single scatter. Thus, $f$ scales like $m_\chi^3$.

For relativistic targets, the Fermi energy and the target mass combine to introduce a scale that separates light and very light dark matter. In the light dark matter regime $s\sim m_\chi p_\text{F}$, \eqref{eq:qualitative:s:scaling}. The benchmark cross section \eqref{eq:qualitative:dsig:dOmega:scaling} thus scales as $m_\chi$. In addition to the Pauli blocking factor of $m_\chi$, this regime’s phase space is also suppressed in the $\cos\psi$ integration. This phenomenon is detailed in Appendix~\ref{app:phase:block:rel:light}, where we show that
\begin{align}
   \int_{\cos\psi_\text{max}}^1
  d\cos\psi = 1-\cos\psi_\text{max}
  &<
  \frac{v_\esc^2 \sin^2\theta \cos^2\alpha}{\left(1-v_\esc \cos\theta\right)^2}
  \left[
    \frac{m_\text{T}^2}{p^2} 
    + 
    \frac{m_\chi}{p}\left(\cdots\right)
    +
    \mathcal O\left( \frac{m_\chi^2}{p^2}, \frac{m_\text{T}^2m_\chi}{p^3} \right)
  \right]\ ,
  \label{eq:qualitative:psi:phase:space}
 \end{align}
where $(\cdots)$ are terms independent of $m_\chi$ and $m_\text{T}$, see \eqref{eq:app:phasespace:psi:tan2psi:2:upper:limit:approx}. In the light-but-not-too-light dark matter regime, the $\mathcal O(m_\chi/p)$ term dominates the bound and the phase space is suppressed proportionally to the dark matter mass. 
The combination of the scaling factors gives $f\sim m_\chi^3$.

\paragraph{Relativistic target; very light dark matter.}
This corresponds to 
$m_\chi \ll m_\text{T}^2/p_\text{F}$.
In this limit, the cross section scales like $m_\chi^2$, the phase space is Pauli blocked (factor of $m_\chi$), and dark matter captures after a single scatter. Thus, $f$ scales like $m_\chi^3$.

This regime follows the behavior of a non-relativistic target, light dark matter scenario. In both cases the dark matter is lighter than any other scale in the problem. There is no suppression of the $\psi$ phase space in \eqref{eq:qualitative:psi:phase:space} because the upper limit is dominated by the $\mathcal O(m_\text{T}^2/p^2)$ term that is independent of the dark matter mass.


\section{Discovery Reach for Effective Contact Operators}
\label{sec:bounds:on:contact:ops}

We present the discovery reach for spin-0 and spin-1 dark matter interacting with Standard Model fermions through an effective contact operator. Our primary focus is the reach for kinetic heating from ultra-relativistic electrons in a neutron star; we compare this to terrestrial bounds on dark matter--electron scattering, the non-relativistic approximation for kinetic heating off electrons, and the analogous reach for muons and nucleons.

\subsection{Effective Theory}

Effective contact operators are model-independent parameterizations of dark matter--visible matter interactions in the limit of small momentum transfer compared to the mass scale of the dynamics that generate interaction. 
The coupling of these operators are an inverse power of a cutoff scale, $\Lambda$, that is approximately the scale at which the effective description breaks down. In the simplest ultraviolet completion, the cutoff scale is a combination of heavy mediator masses and couplings.


Table~\ref{tab:amplitudes12} presents our basis of contact operators and the squared amplitudes from each operator.
We write our effective operators in the form $\mathcal O_\chi \mathcal O_\xi$ where each $\mathcal O_{\chi}$ is a bilinear of fermionic or scalar dark matter and $\xi$ is a Standard Model fermion with a bilinear $\mathcal O_{\xi}$. 
The effective coupling is $\Lambda^{4-N}$ where $N$ is the dimension of the combined operator. Most of the operators are dimension-6 and carry a coupling $\Lambda^{-2}$, while $\mathcal O^\text{S}_{1,2}$ are dimension-5 with coupling $\Lambda^{-1}$.

\begin{table} 
  \centering
  \begin{tabular}{lll}
  	\toprule
  	Name 
  	& Operator 
  	& $\Lambda^4|\mathcal{M}|^2$
  	\; ($m_\chi^2 E_p^2$ term dominates when present)
  	\\
  	\midrule
  	$\mathcal{O}^\textnormal{F}_1$ 
  	& $
  	\left(\bar{\chi}\chi\right) \left(\bar{\xi}\xi\right)$ 
  	& $
  	\left(4m_\chi^2-t\right)\left(4m_T^2-t\right)$
  	\\
  	$\mathcal{O}^\textnormal{F}_2$ 
  	& $
  	\left(\bar{\chi} i\gamma^5\chi\right)\left(\bar{\xi}\xi\right)$ 
  	& $
  	t\left(t-4m_T^2\right)$
  	\\
  	$\mathcal{O}^\textnormal{F}_3$ 
  	& $
  	\left(\bar{\chi}\chi\right)\left(\bar{\xi} i\gamma^5\xi\right)$ 
  	& $
  	t\left(t-4m_\chi^2\right)$
  	\\
  	$\mathcal{O}^\textnormal{F}_4$ 
  	& $
  	\left(\bar{\chi} i\gamma^5\chi\right)\left(\bar{\xi} i\gamma^5\xi\right)$ 
  	& $
  	t^2$
  	\\
  	$\mathcal{O}^\textnormal{F}_5$ 
  	& $\left(\bar{\chi}\gamma^\mu\chi\right)\left(\bar{\xi}\gamma_\mu\xi\right)$ 
  	& 
  	$4\left(m_T^2+m_\chi^2\right)^2
  	-8s\left(m_T^2+m_\chi^2\right)
  	+4s^2
  	+4st
  	+2t^2
  	$
  	\\
  	$\mathcal{O}^\textnormal{F}_6$ 
  	& $\left(\bar{\chi}\gamma^\mu\gamma^5\chi\right)\left(\bar{\xi}\gamma_\mu\xi\right)$ 
  	& $
  	4\left(m_T^2-m_\chi^2\right)^2
  	-8s\left(m_T^2+m_\chi^2\right)
  	-8\,t\,m_\chi^2
  	+4s^2
  	+4st
  	+2t^2
  	$
  	\\
  	$\mathcal{O}^\textnormal{F}_7$ 
  	& $\left(\bar{\chi}\gamma^\mu\chi\right)\left(\bar{\xi}\gamma_\mu\gamma^5\xi\right)$ 
  	& $
  	4\left(m_T^2-m_\chi^2\right)^2
  	-8s\left(m_T^2+m_\chi^2\right)
  	-8\,t\,m_T^2
  	+4s^2
  	+4st
  	+2t^2
  	$
  	\\
  	$\mathcal{O}^\textnormal{F}_8$ 
  	& $\left(\bar{\chi}\gamma^\mu\gamma^5\chi\right)\left(\bar{\xi}\gamma_\mu\gamma^5\xi\right)$ 
  	& $
  	4\left(m_T^4+10m_T^2m_\chi^2+m_\chi^4\right)
  	-8(s+t)\left(m_T^2+m_\chi^2\right)
  	+4s^2
  	+4st
  	+2t^2
  	$
  	\\
  	$\mathcal{O}^\textnormal{F}_9$ 
  	& $
  	\left(\bar{\chi}\sigma^{\mu\nu}\chi\right)\left(\bar{\xi}\sigma_{\mu\nu}\xi\right)$ 
  	& 
  	$
  	8
  	\left[
  	4\left(m_T^4+4m_T^2m_\chi^2+m_\chi^4\right)
  	-2(4s+t)\left(m_T^2+m_\chi^2\right)
  	+(2s+t)^2
  	\right]
  	$
  	\\
  	$\mathcal{O}^\textnormal{F}_{10}$ 
  	& $
  	\left(\bar{\chi}\sigma^{\mu\nu}i\gamma^5\chi\right)\left(\bar{\xi}\sigma_{\mu\nu}\xi\right)$ 
  	& 
  	$
  	8
  	\left[
  	4\left(m_T^2+m_\chi^2\right)^2
  	-2(4s+t)\left(m_T^2+m_\chi^2\right)
  	+(2s+t)^2
  	\right]
  	$
  	\\
  	%
  	%
  	\midrule
  	$\mathcal{O}^\textnormal{S}_1$ 
  	& 
  	$
  	\left(\chi^\dag\chi\right)
  	\left(\bar\xi\xi\right)\Lambda$
  	& 
  	$
  	\Lambda^2\left(4m_{\rm T}^2-t\right)
  	$
  	\\
  	$\mathcal{O}^\textnormal{S}_2$ 
  	& 
  	$
  	\left(\chi^\dag\chi\right)
  	\left( \bar\xi i\gamma^5 \xi \right) \Lambda$
  	& $
  	\Lambda^2\left(-t\right)
  	$
  	\\
  	$\mathcal{O}^\textnormal{S}_3$ 
  	& 
  	$
  	\left( \chi^\dag i\partial_\mu \chi \right)
  	\left( \bar\xi\gamma^\mu \xi \right)$
  	& $
  	4\left[\left(m_T^2+m_\chi^2\right)^2
  	-2s\left(m_T^2+m_\chi^2\right)
  	+s^2
  	+st
  	-m_{\rm T}^2t\right]
  	$
  	\\
  	$\mathcal{O}^\textnormal{S}_4$ 
  	& 
  	$\left(\chi^\dag i \partial_\mu\chi\right)
  	\left( \bar\xi\gamma^\mu\gamma^5 \xi \right)$
  	& $
  	4\left[\left(m_T^2-m_\chi^2\right)^2
  	-2s\left(m_T^2+m_\chi^2\right)
  	+s^2
  	+st\right]
  	$
  	\\
  	\bottomrule
  	\bottomrule
  \end{tabular}   
  \caption{Effective contact operators and their squared tree-level scattering amplitudes for pairwise interactions of dark matter, $\chi$, with Standard Model fermionic targets, $\xi$. Squared amplitudes are written with respect to the Mandelstam $s$ and $t$ variables and are rescaled by a power of the cutoff $\Lambda^4$ for brevity.
  Operators with superscript F(S) correspond to fermionic (scalar) dark matter.
  }
  \label{tab:amplitudes12}
\end{table}

One must account for additional factors when comparing the contact operators in Table~\ref{tab:amplitudes12} to a full theory. 
The operators with spin-0 or spin-2 Standard Model bilinears---$\mathcal O^\text{F}_{1-4}$, $\mathcal O^\text{F}_{9-10}$, and $\mathcal O^\text{S}_{1-2}$---connect fermions of different chirality. 
Gauge invariance requires that the ultraviolet physics that generates the operator must include an order parameter for electroweak symmetry breaking. 
Furthermore, these operators may introduce tightly constrained flavor violating observables at loop level.
As such, one may complete operators with gauge-invariant contact operators consistent with minimal flavor violation~\cite{DAmbrosio:2002vsn}. The simplest choice is to use the Higgs vacuum expectation value as the order parameter for chiral symmetry breaking and the Standard Model Yukawa couplings as flavor spurions:
\begin{align}
  \frac{1}{\Lambda^{N-4}} \mathcal O_\chi\mathcal O_\xi 
  \equiv 
  \frac{1}{\tilde\Lambda^{N-3}} 
  y^\xi_{IJ}
  \mathcal O_\chi \, \langle H \rangle\cdot \mathcal O_\xi^{IJ} \ ,
\end{align}
where $H\cdot\mathcal O_\xi$ is a Standard Model gauge singlet and $y^\xi_{IJ} O_\xi^{IJ}$ is a flavor singlet. In this way the kinetic heating discovery reach in $\Lambda$ may be mapped onto a bound on the corresponding cutoff $\tilde\Lambda$ of a gauge-invariant, minimal flavor violating contact operator. The completion above is consistent with what one would expect from a heavy scalar mediator mixing with the Standard Model Higgs.

\paragraph{Methodology.} 

We assume that dark matter--visible matter interactions are dominated by a single contact operator with a single target species. 
We numerically integrate \eqref{eq:f:full} to determine the projected reach on the cutoff $\Lambda$ as a function of the dark matter mass $m_\chi$.
We plot the discovery reach for a range of dark matter masses,
\begin{align}
  10~\text{eV}
  \;
  \text{(evaporation)}
  \;
  <
  \;
  m_\chi 
  \;
  <
  \;
  10~\text{PeV}
  \;
  \text{(no new features)} \ .
\end{align}
The lower limit corresponds to the mass at which one must account for the evaporation of dark matter from the neutron star~\cite{Garani:2018kkd}. The upper limit is the scale beyond which there are no new features; the capture efficiency scales as $f\sim m_\chi^{-1}$ from the requirement of multiple scattering, see Section~\ref{sec:kinematic:regimes}.

\paragraph{Limits of the Effective Contact Operator Description.}
If the contact operators are generated by a heavy mediator with $\mathcal O(1)$ couplings, then one may roughly interpret $\Lambda$ to be the mediator mass.
The dark matter--target interactions for kinetic heating are then $t$-channel interactions. In this case, the contact description breaks down when the Mandelstam $t$-variable dominates in the mediator propagator, where $-t$ is the square of the transferred four-momentum. 
For our purposes, the characteristic scale momentum transfer is determined by the kinematics of kinetic heating. This, in turn, defines the condition at which the effective theory may be replaced by its \acro{UV} completion:
\begin{align}
  q^2 \sim 
    \begin{cases}
      \displaystyle
      p_F^2 
      \quad&\Rightarrow \quad \text{breakdown when}\; \Lambda \ll p_F
      \quad\;\text{(heavy dark matter)}
      \\
      \displaystyle
      m_\chi^2
      \quad&\Rightarrow\quad \text{breakdown when}\; \Lambda \ll m_\chi
      \quad\text{(light dark matter)}
  \end{cases}
  \ .
  \label{eq:eft:breakdown}
\end{align}
Assuming $\mathcal O(1)$ couplings, this condition is relevant for the fermionic $\mathcal{O}_{2-4}^F$ operators so that one should use caution when interpreting these plots with respect to the domain of validity.
The scaling of $t$ that leads to \eqref{eq:eft:breakdown} is presented in Table~\ref{tab:sandt} of Appendix~\ref{sec:DominantContact}.


\subsection{Results}

\begin{figure*}
  \centering
  \includegraphics[width=0.48\textwidth]{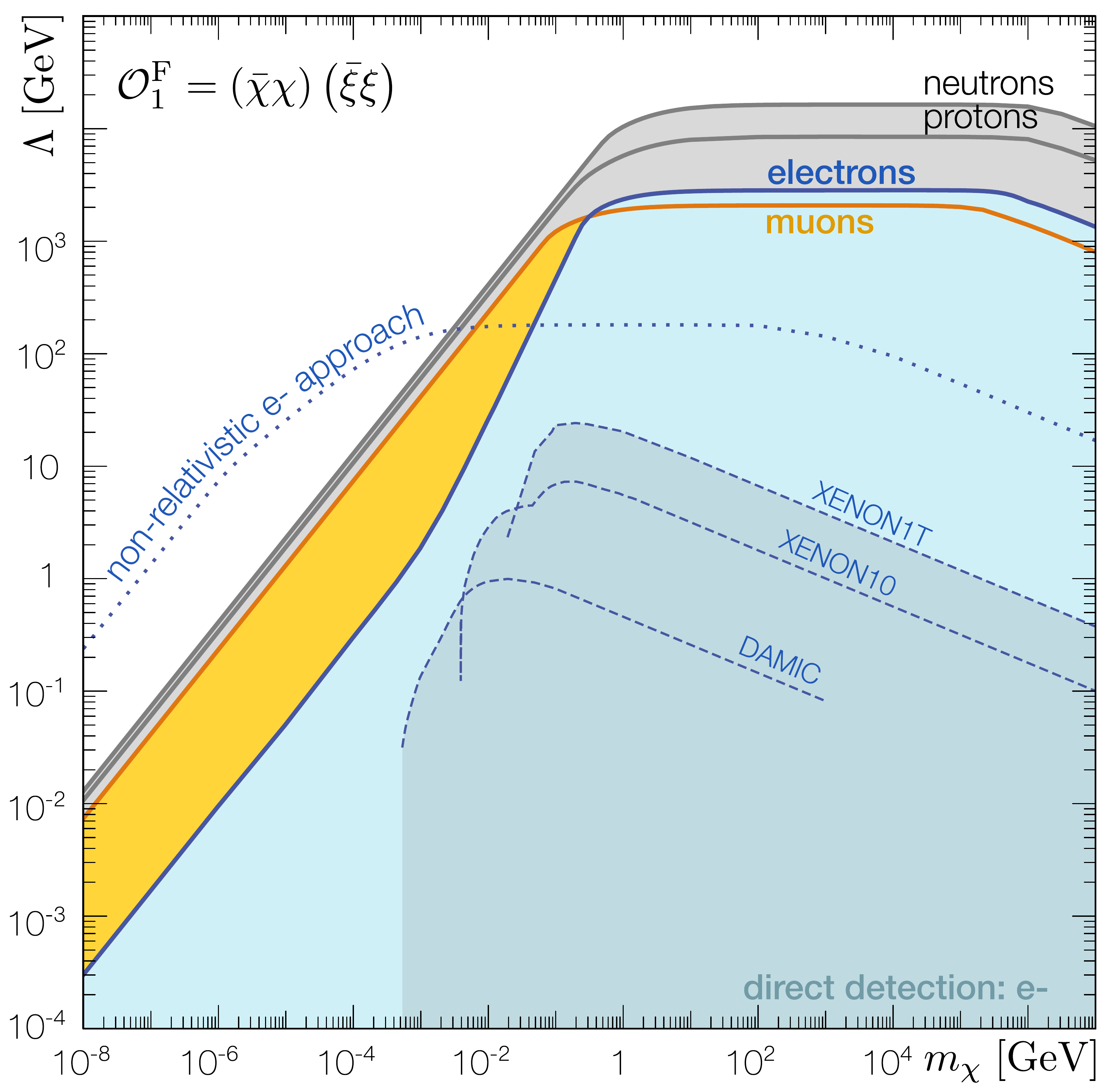}
  \includegraphics[width=0.48\textwidth]{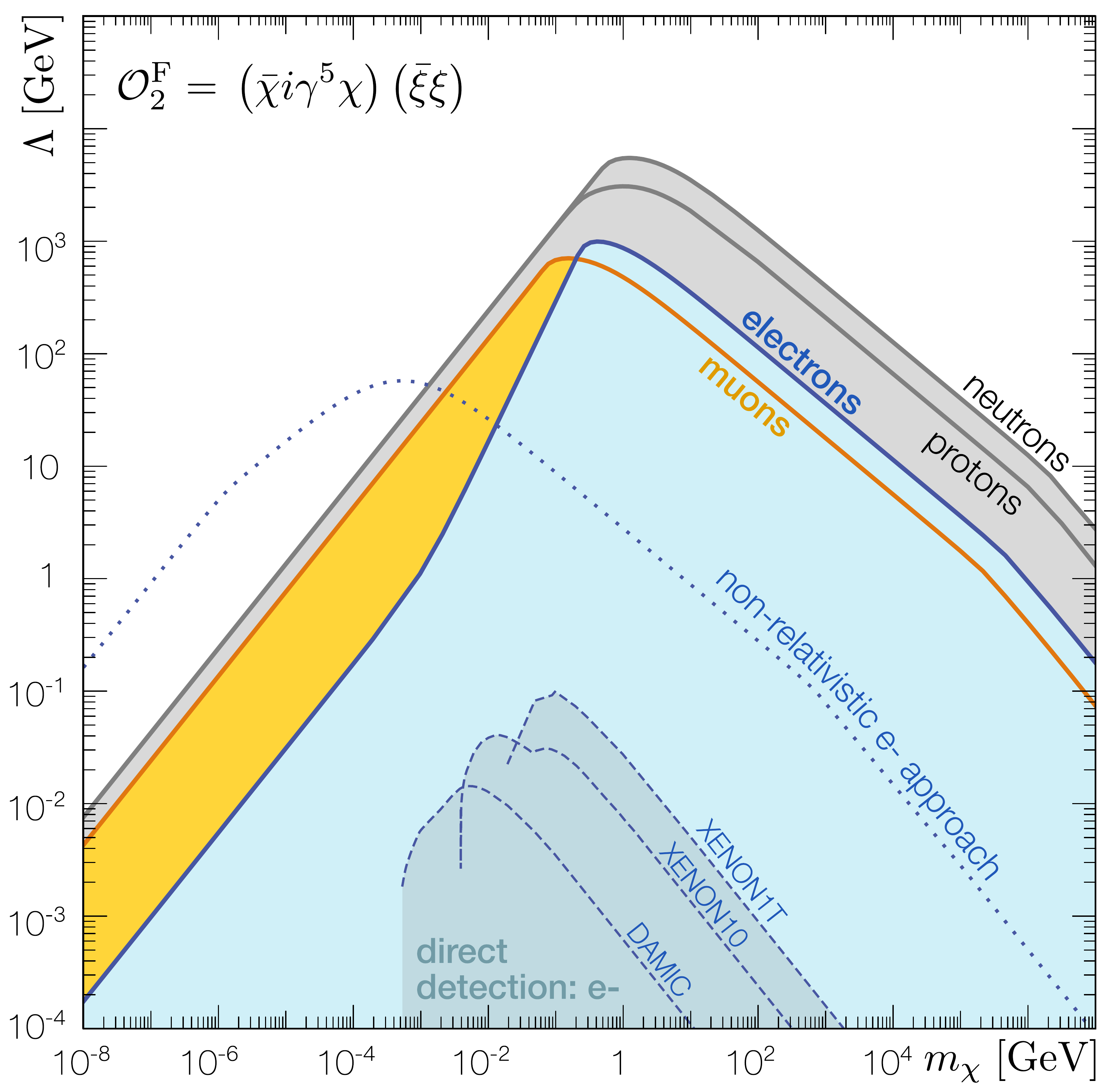}\\
  \includegraphics[width=0.48\textwidth]{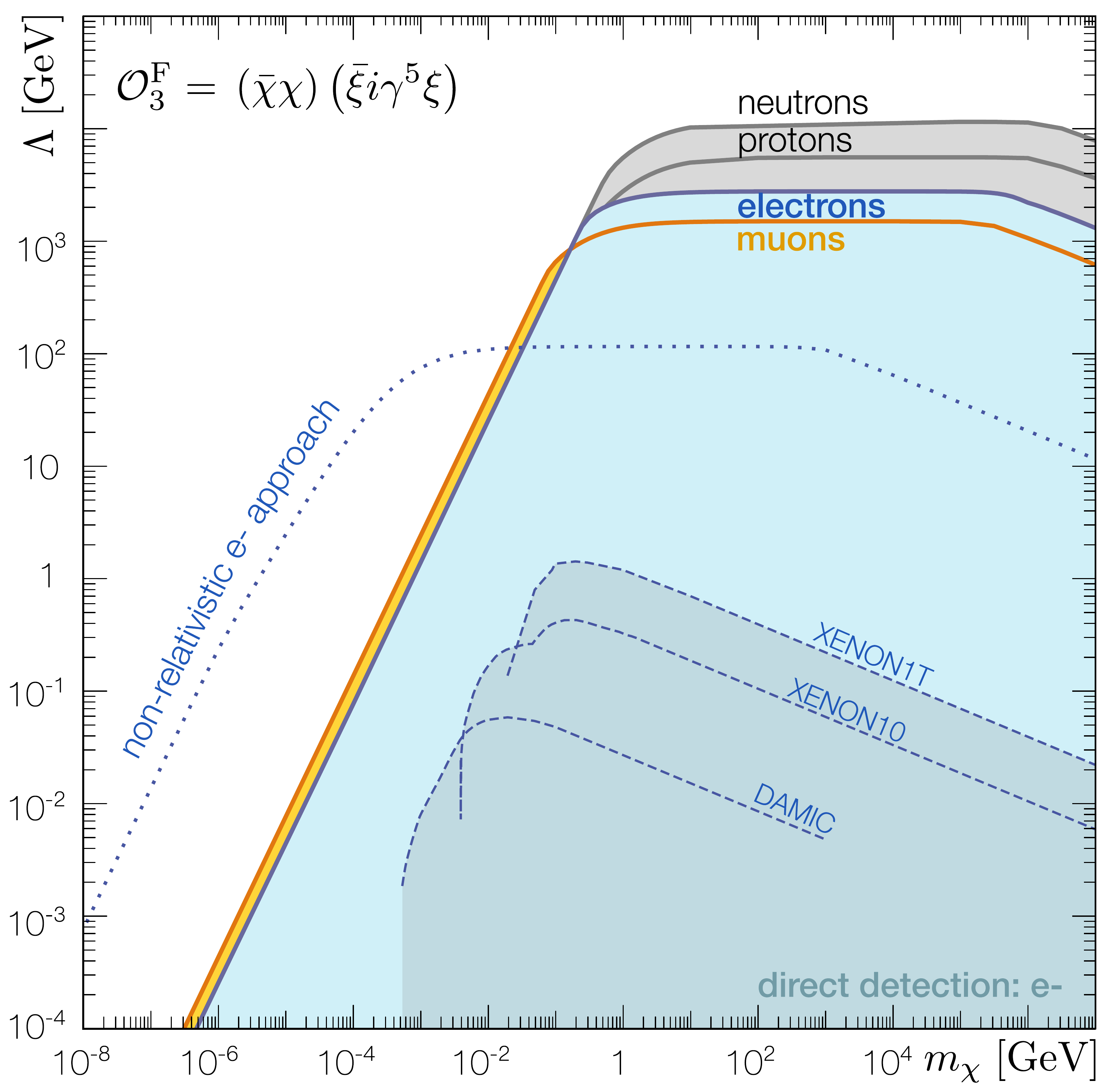}
  \includegraphics[width=0.48\textwidth]{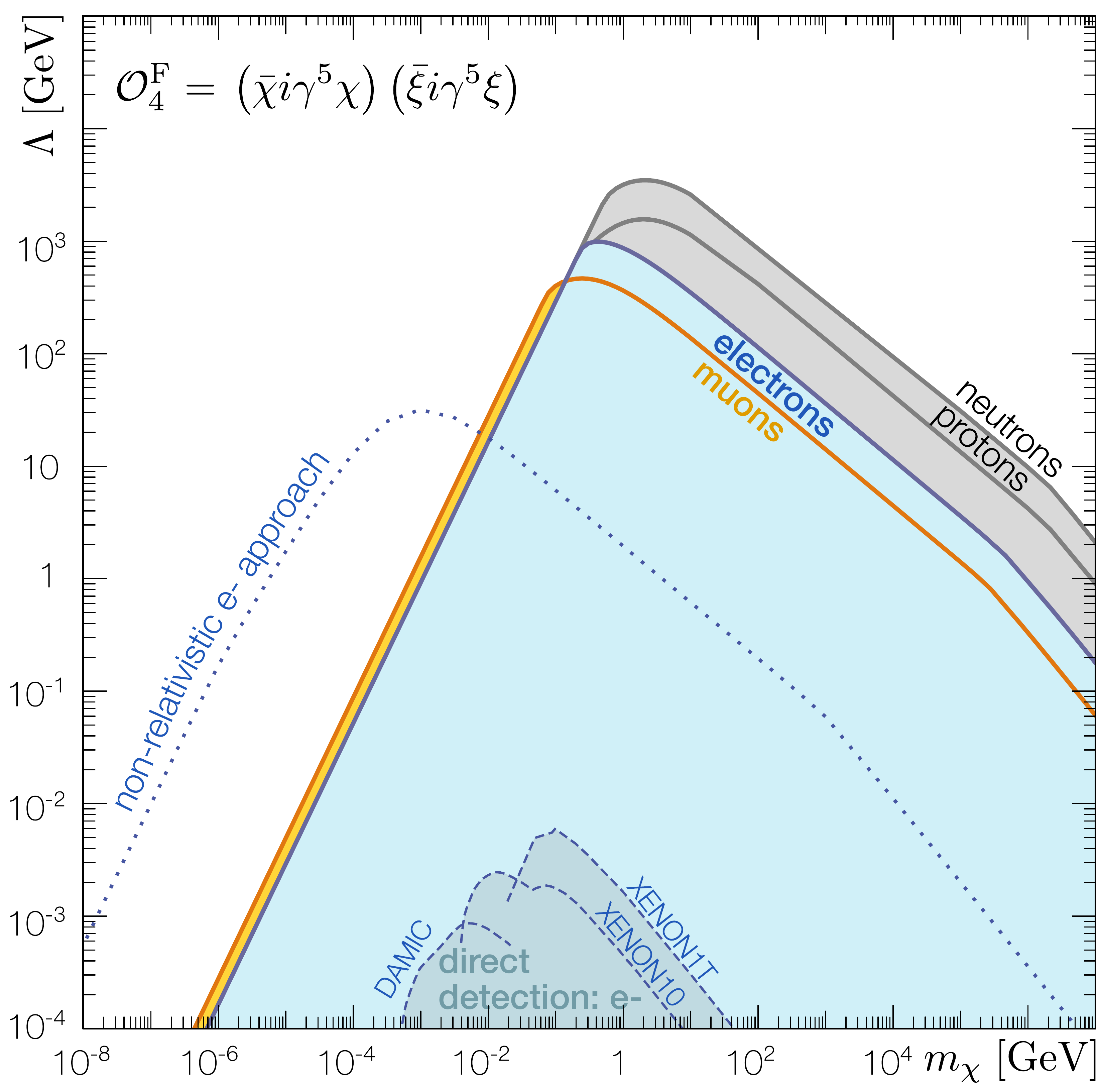}
  \caption{%
    \label{fig:FContactScalar}%
    Projected kinetic heating discovery reach for fermionic dark matter interacting through the spin-0 contact operators $\mathcal{O}_{1-4}^\text{F}$ in Table~\ref{tab:amplitudes12} for different targets: electrons (cyan), muons (yellow), nucleons (gray).
    The corresponding shaded regions are accessible for the benchmark scenario of a neutron star temperature $T_\star = 1600~\text{K}$ ($f=1$). 
    The dotted blue line shows the approximation of non-relativistic electrons and is contrasted with the full relativistic calculation undertaken in this work (solid blue line).
    Dashed blue lines with the corresponding shaded region show the reach of electron recoil direct detection searches~\cite{Essig:2012yx, Agnes:2018ves, Aprile:2019xxb,Aguilar-Arevalo:2019wdi}. 
    Note that for $\mathcal{O}_{2-4}^\text{F}$, the validity of the effective theory may break down when the ultraviolet theory is a heavy mediator with $\mathcal O(1)$ couplings, see \eqref{eq:eft:breakdown}.
    }
\end{figure*}

\begin{figure*}
  \centering
  \includegraphics[width=0.48\textwidth]{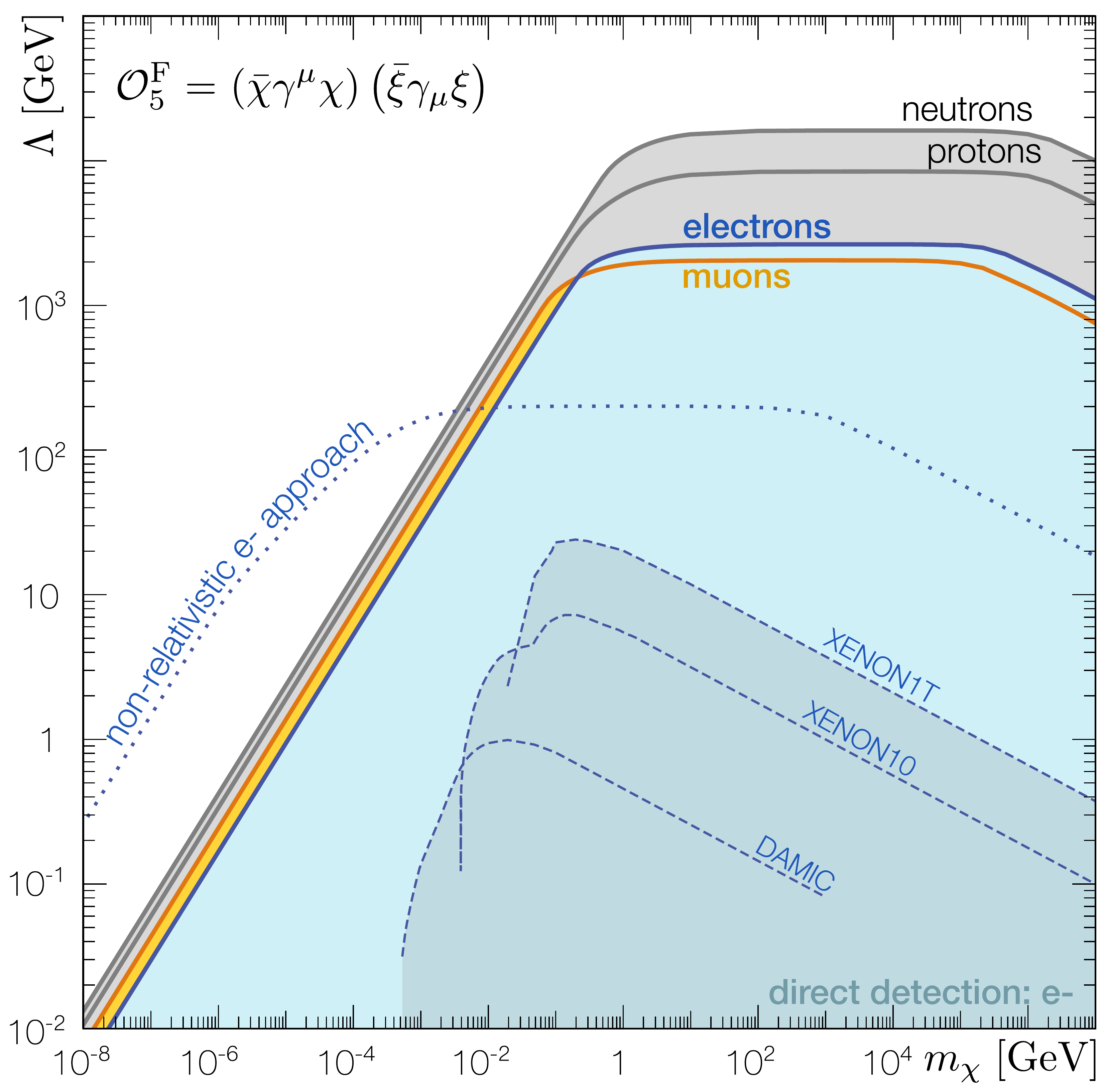}
  \includegraphics[width=0.48\textwidth]{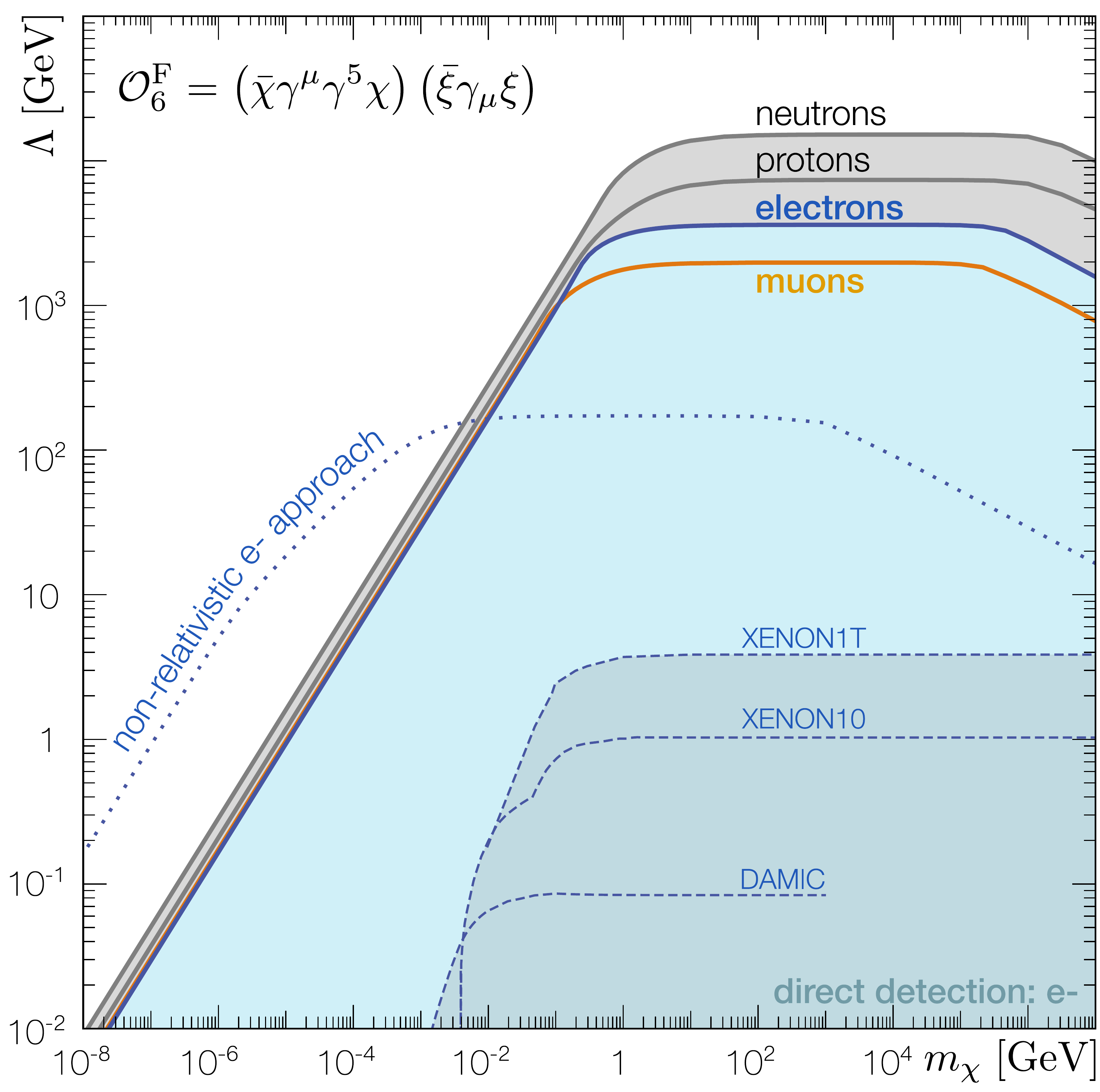}\\
  \includegraphics[width=0.48\textwidth]{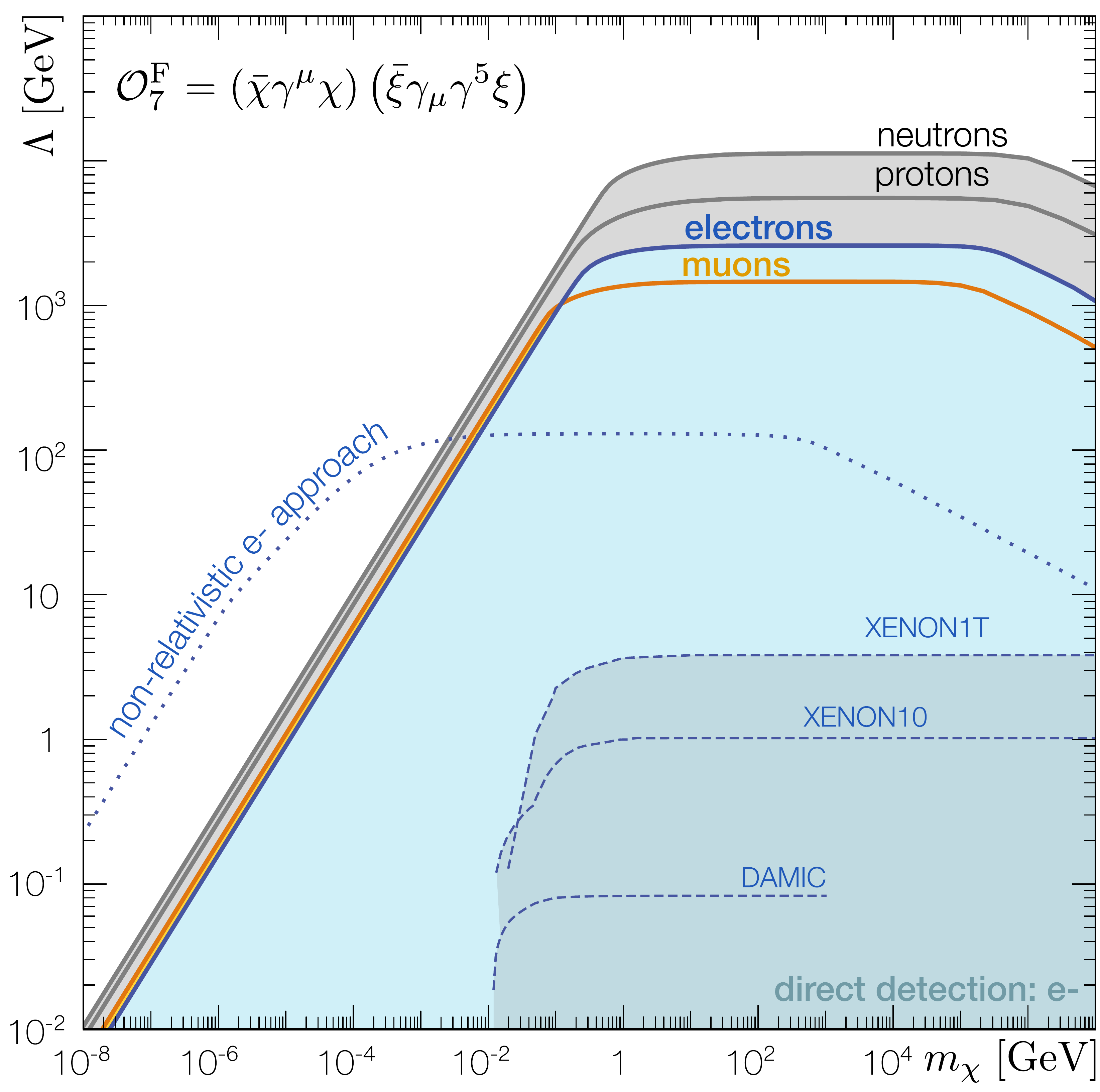}
  \includegraphics[width=0.48\textwidth]{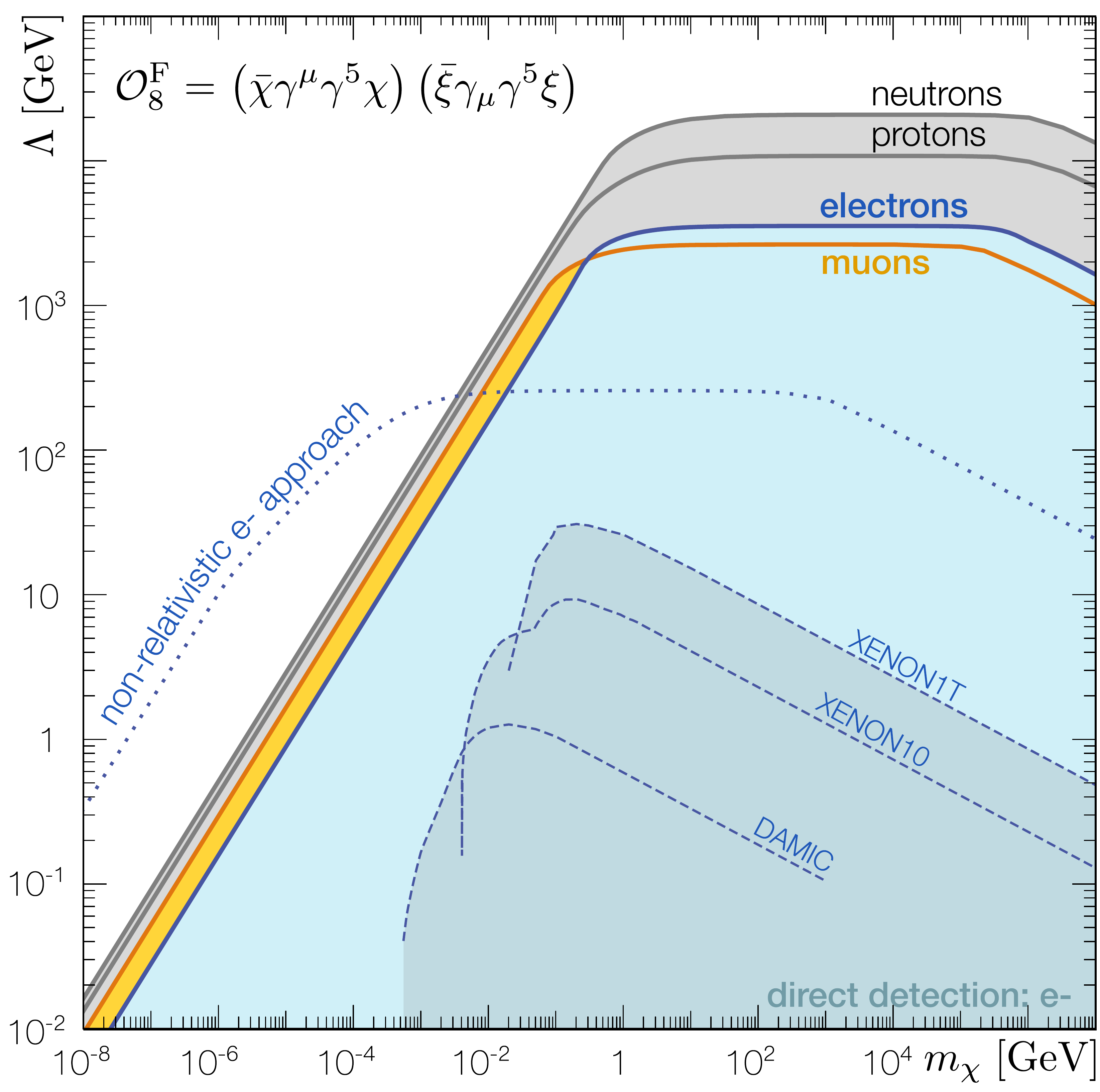}
  \caption{%
  Projected kinetic heating discovery reach for fermionic dark matter interacting through the spin-1 contact operators $\mathcal{O}_{5-8}^\text{F}$ in Table~\ref{tab:amplitudes12} for different targets: electrons (cyan), muons (yellow), nucleons (gray). 
  Other features follow the caption in Figure~\ref{fig:FContactScalar}.
  }
  \label{fig:FContactVector}
\end{figure*}

\begin{figure*}
  \centering
  \includegraphics[width=0.48\textwidth]{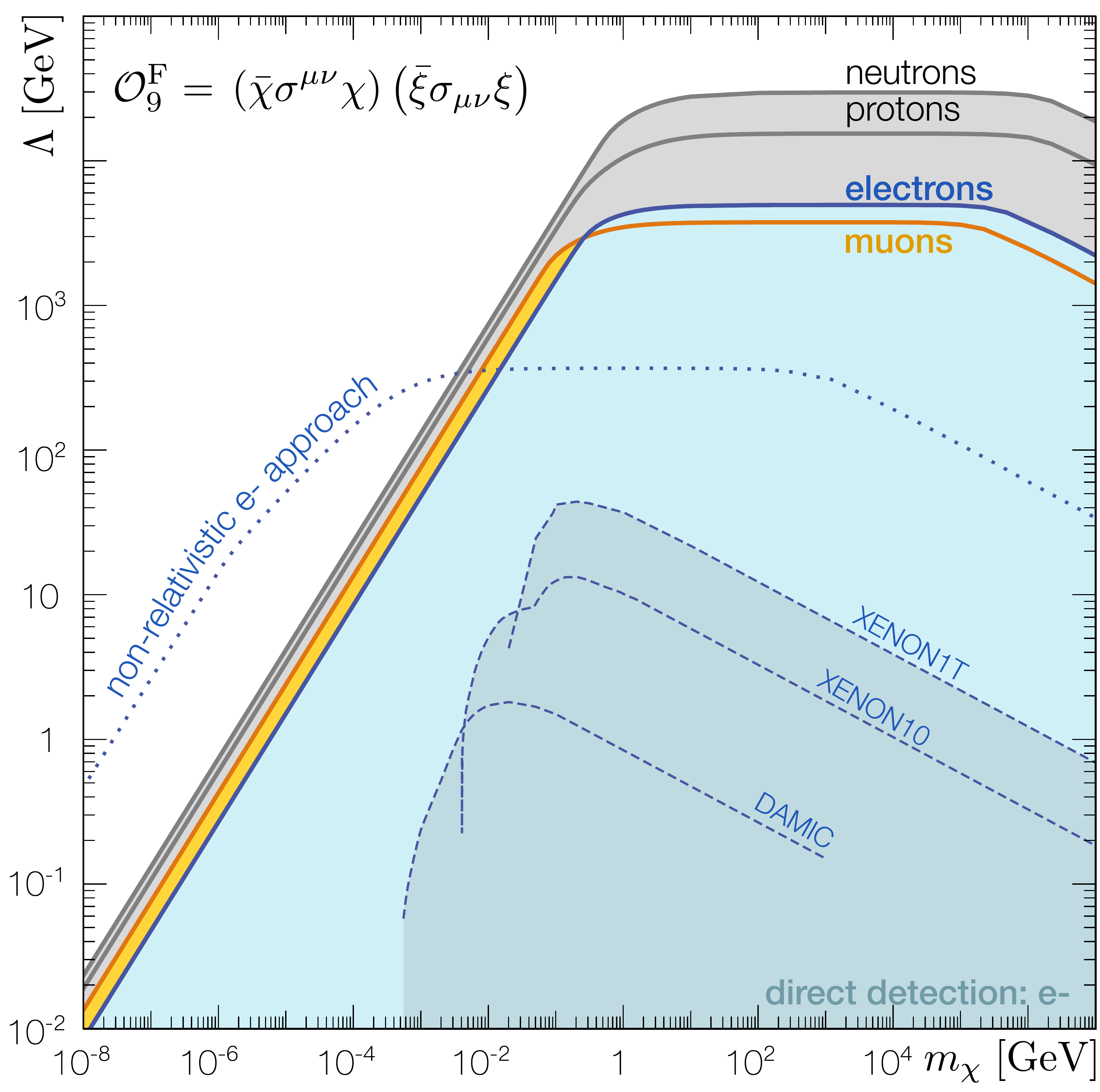}
  \includegraphics[width=0.48\textwidth]{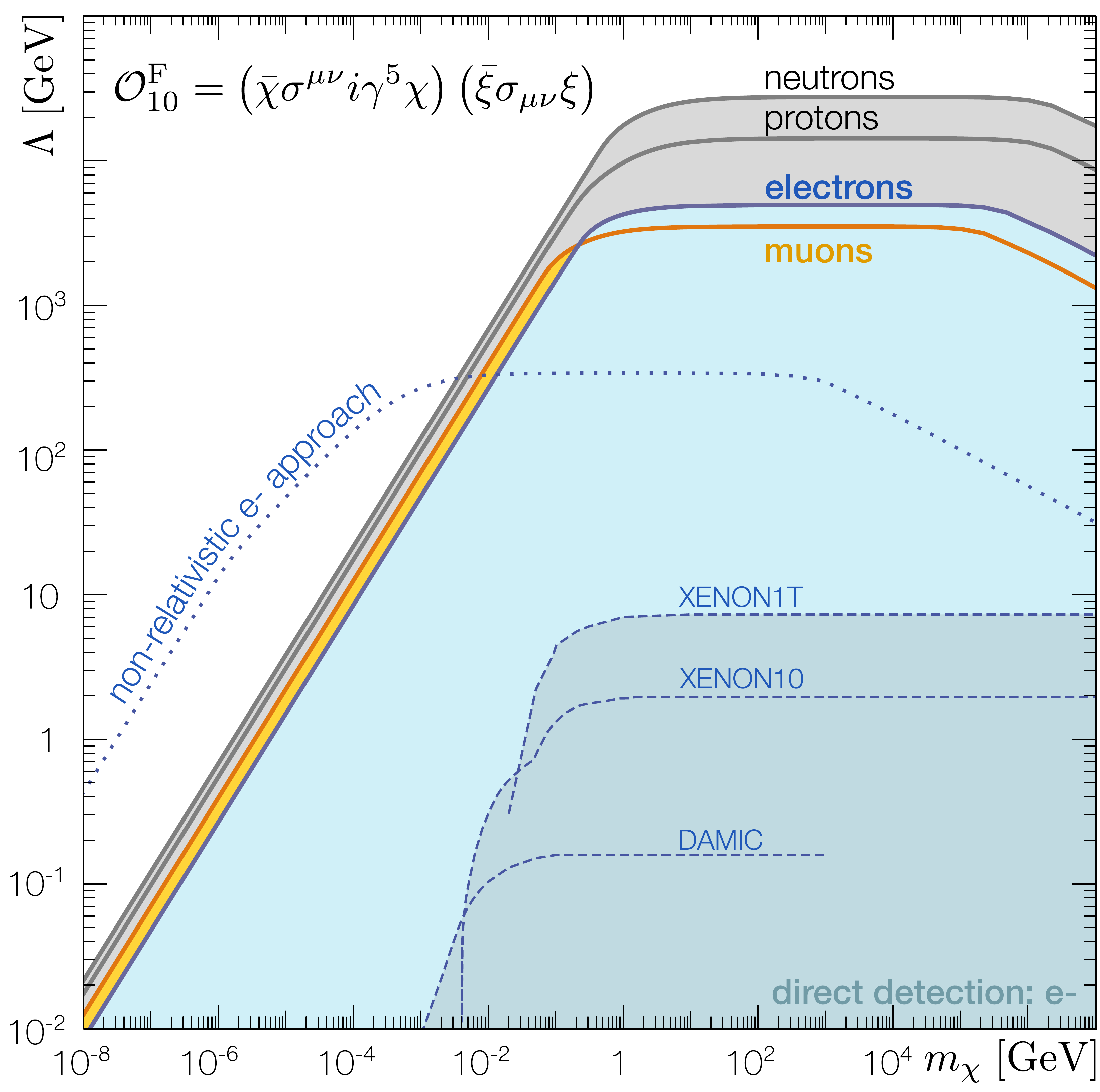}
  \caption{
  Projected kinetic heating discovery reach for fermionic dark matter interacting through the spin-2 contact operators $\mathcal{O}_{9,10}^\text{F}$ in Table~\ref{tab:amplitudes12} for different targets: electrons (cyan), muons (yellow), nucleons (gray).
  Other features follow the caption in Figure~\ref{fig:FContactScalar}.
  }
  \label{fig:FContactTensor}
\end{figure*}

\begin{figure*}
  \centering
  \includegraphics[width=0.48\textwidth]{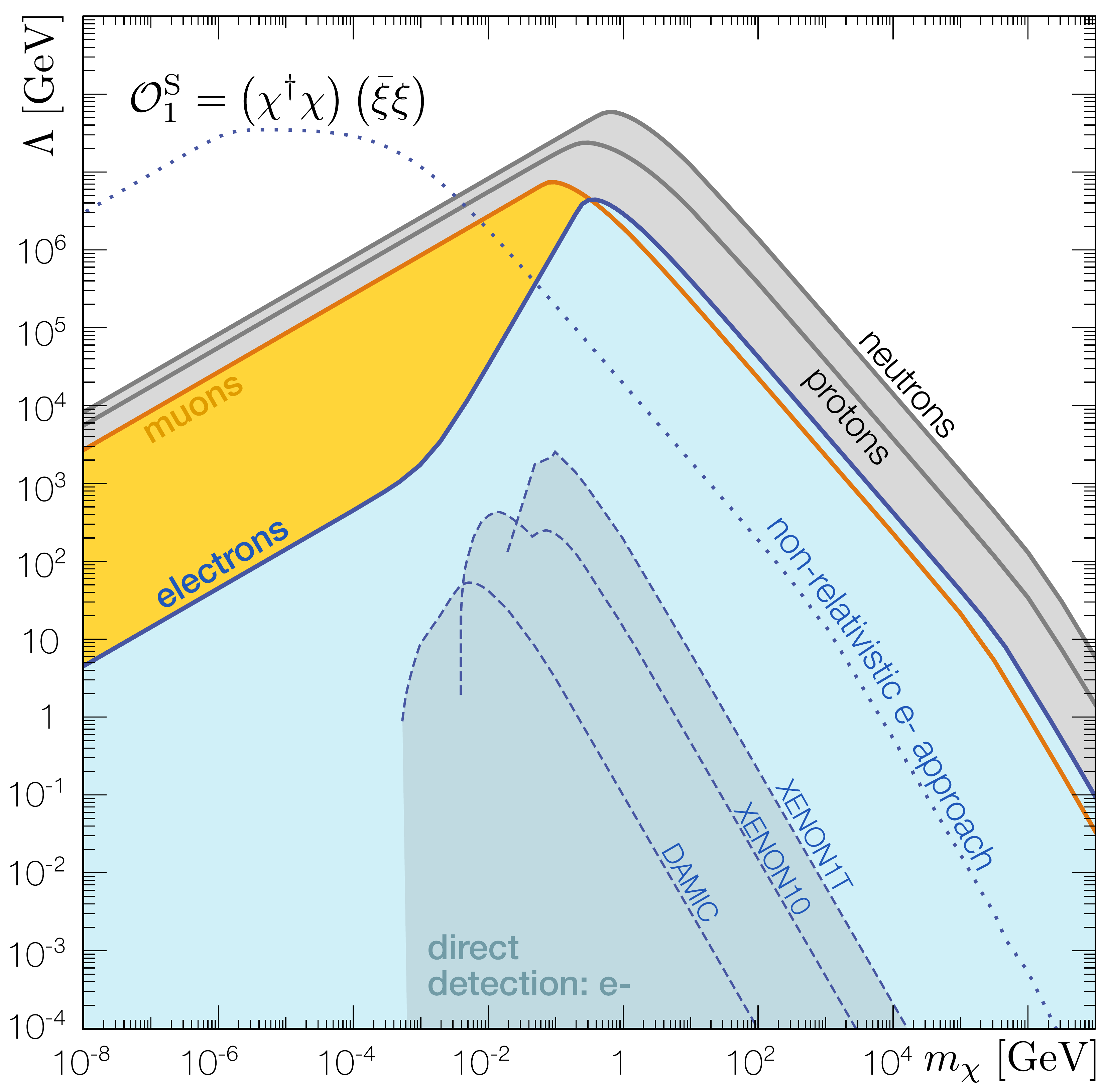}
  \includegraphics[width=0.48\textwidth]{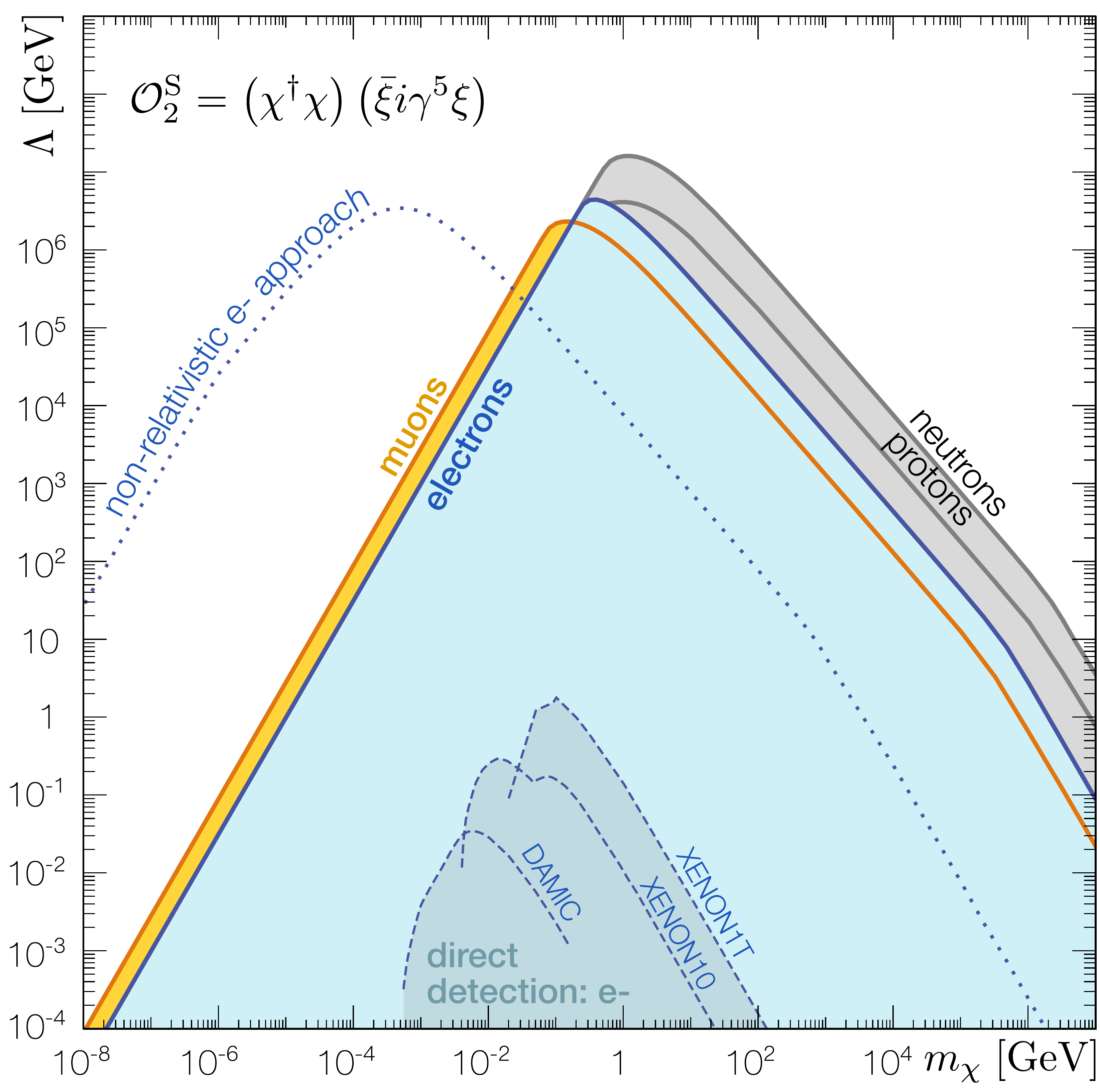}\\
  \includegraphics[width=0.48\textwidth]{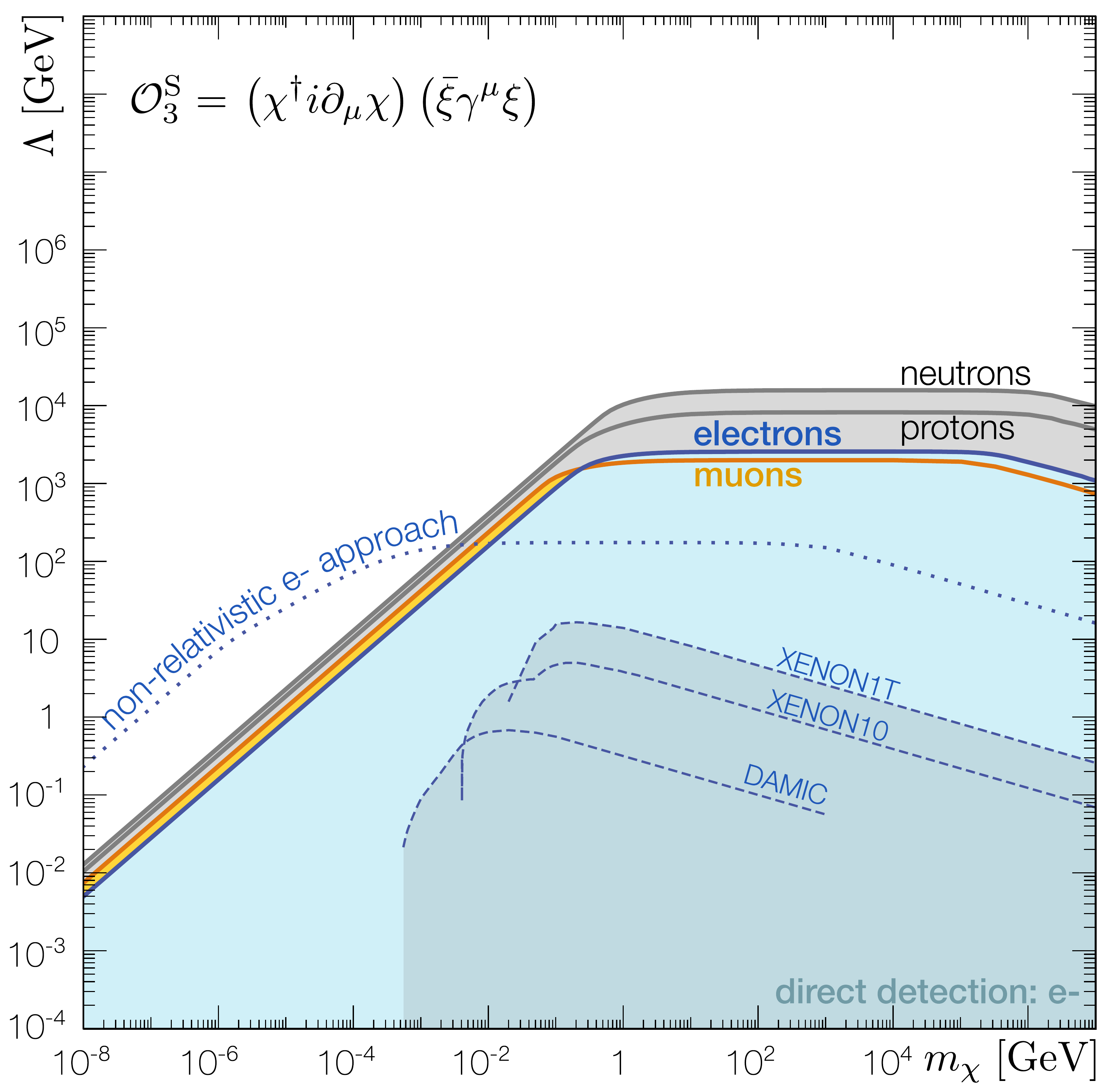}
  \includegraphics[width=0.48\textwidth]{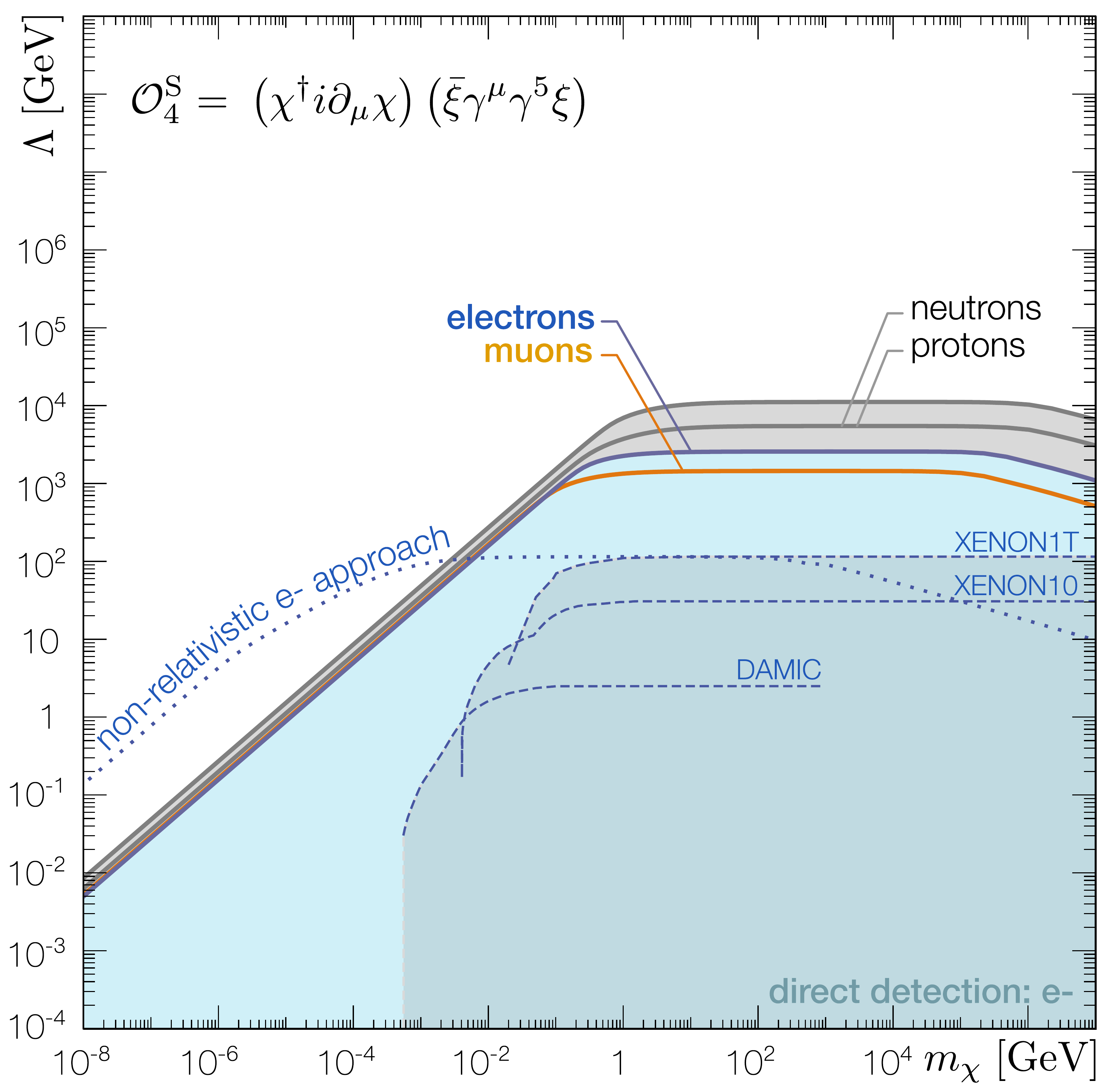}\\
  \caption{%
  Projected kinetic heating discovery reach for scalar dark matter interacting through the contact operators $\mathcal{O}_{1-4}^\text{F}$ in Table~\ref{tab:amplitudes12} for different targets: electrons (cyan), muons (yellow), nucleons (gray).
  Other features follow the caption in Figure~\ref{fig:FContactScalar}.
  }
  \label{fig:BContact}
\end{figure*}

Figures~\ref{fig:FContactScalar}--\ref{fig:BContact} present the results for each of the operators in Table~\ref{tab:amplitudes12}.
The numerical results follow the general behavior of relativistic and non-relativistic targets explained in Section~\ref{sec:RelScat}. We find that muons are semi-relativistic targets whose capture efficiency is accurately represented in the non-relativistic approximation. The muon reach typically tracks that of the nucleons. Note that this reflects the non-relativistic kinematics only; the dynamics generating the nucleon and lepton couplings are assumed to be completely independent. 

To highlight the importance of relativistic effects, we also plot the approximation where electron targets are treated non-relativistically. For light dark matter, the non-relativistic target approximation greatly overestimates the sensitivity of kinetic heating. This is because it is harder to transfer momentum to energetic ultra-relativistic targets rather than stationary targets. In the heavy dark matter regime, the non-relativistic target approximation significantly under-estimates the kinetic heating reach. 

Our results are particularly significant for models of leptophilic dark matter~\cite{Baltz:2002we, Chen:2008dh, Cirelli:2008pk, Fox:2008kb} for which the induced nucleon interactions are loop suppressed. Earlier work compared the kinetic heating reach of tree-level leptophilic interactions to the loop-induced nucleon interactions in the limit of non-relativistic leptons~\cite{Bell:2019pyc}. The result is that the loop-level coupling to nucleons is a weaker probe of leptophilic interactions than the tree-level coupling to non-relativistic muons. Our fully relativistic treatment revises these results and shows that electron scattering can be the dominant heating channel in these scenarios. In particular, we confirm that muons are accurately described by the non-relativistic target limit and show that the reach from relativistic electron scattering is generically stronger or comparable to that of non-relativistic muons.

The rightmost column of Table~\ref{tab:amplitudes12} shows how the squared amplitude of each contact operator depends on the masses and kinematics of the scattering.
Most of the operators contain a term proportional to $m_\chi^2 E_p^2$; this term dominates when it is present. 
Section~\ref{sec:CharFeatures} describes the scaling behavior of the capture efficiency $f$ with respect to $m_\chi$ assuming this $m_\chi^2 E_p^2$ term as a benchmark.
The exceptional cases are tabulated in Table~\ref{tab:contact:ops:M2:exceptions} of Appendix~\ref{sec:DominantContact}. 
For capture efficiencies that scale like $f\propto m_\chi^n$, the corresponding reach scales like $\Lambda\propto m_\chi^{n/4}$. From the scaling arguments summarized in Figure~\ref{fig:app:phaseblock:short:flow:chart}, the benchmark scenario gives:
\begin{align}
  \Lambda \sim 
    \begin{cases}
      \displaystyle
      m_\chi^{3/4}
      & \text{light and very light dark matter}
      \\
      \displaystyle
      m_\chi^{0}
      & \text{heavy (intermediate mass) dark matter}
      \\
      \displaystyle
      m_\chi^{-1/4}
      & \text{very heavy dark matter}
  \end{cases}
  \ ,
  \label{eq:lambda:mx:scaling:baseline}
\end{align}
where the dark matter mass regimes are defined in Figure~\ref{fig:app:phaseblock:phasespace} for relativistic and non-relativistic targets. Figures~\ref{fig:FContactVector}--\ref{fig:BContact} show that $\mathcal O^\text{F}_{5-10}$ and $\mathcal O^\text{S}_{3,4}$ follow this trend. One may deduce the scaling for operators with exceptional squared amplitudes following the analogous scaling arguments in Figures~\ref{fig:app:phaseblock:short:flow:chart:F:O1O2} and \ref{fig:app:phaseblock:short:flow:chart:F:O3O4} in Appendix~\ref{sec:DominantContact}.

\subsection{Comparison to Existing Bounds}

Our reach plots also show current and projected direct detection sensitivities to show how kinetic heating complements and may supersede these traditional searches. We also comment on collider and black hole bounds that we do not depict on these plots.

\paragraph{Direct Detection.}

We estimate the reach from direct detection experiments that probe non-relativistic dark matter scattering from non-degenerate targets~~\cite{Essig:2012yx, Agnes:2018ves, Aprile:2019xxb, Aguilar-Arevalo:2019wdi}.
We recast these direct detection bounds using the non-relativistic limit of the squared amplitudes in 
Table~\ref{tab:amplitudes12},
\begin{align}
s 
&\rightarrow 
\left(
\frac{1}{2}m_\chi v_\textnormal{halo}^2 
+ m_\text{T}
\right)^2
&
-t 
&\rightarrow 
\begin{cases}
      \displaystyle
      \alpha^2 m_e^2
      &
      \quad\text{(electron recoil)}
      \\
      \displaystyle
      2\mu^2_{\textnormal{T}\chi} v_\textnormal{halo}^2
      &
      \quad\text{(nuclear recoil)}
  \end{cases}
\ ,
\end{align}
where $\alpha$ is the electromagnetic fine structure constant and $t$ is estimated with the characteristic transfer momentum averaged over all scattering angles. For electron recoil, we set this characteristic momentum to the typical atomic ionization energy~\cite{Essig:2015cda}. We further assume that the electron recoil form factor is unity and a uniform dark matter velocity $v_\textnormal{halo}\sim 10^{-3}$. Observe that the $\mathcal O^\textnormal{F}_{2,5}$ operators highlight the powerful complementarity of kinetic heating to direct detection. Scattering through these operators is suppressed by the transfer momentum in the non-relativistic limit relevant for direct detection. When dark matter scatters in a neutron star, on the other hand, this suppression is lifted since the transfer momentum can be large.

\paragraph{Comparison to Collider Bounds.} We do not show bounds from collider searches for dark matter~\cite{Goodman:2010ku, Fox:2011fx}. While monojet and monophoton searches are sensitive to dark matter lighter than $\mathcal O(100~\text{GeV})$, the contact operator description may break down when the mediator mass is smaller than the center of momentum energy. Since dark matter capture is a $t$-channel process, there are regimes where the effective theory is valid for kinetic heating but not collider searches based on missing energy.

\paragraph{Black Holes.} Certain scenarios of dark matter capture in neutron stars are constrained simply from the existence of neutron stars~\cite{Kouvaris:2010jy,Bramante:2014zca}. The accumulation of non-annihilating dark matter in the interior of neutron star may produce a long-lived black hole. This process is subject to ongoing uncertainties about the stability of neutron star core against collapse~\cite{Gresham:2018rqo} and whether the scalar self-interactions generically prevent black hole formation~\cite{Bell:2013xk}. Our search is agnostic to whether or not dark matter annihilates and we do not include black hole bounds on our plots.

\paragraph{Neutrino experiments.} For cross sections much greater than electroweak cross sections, a small flux of dark matter at sub-GeV masses can scatter off cosmic rays, gain large kinetic energies, and trigger direct detection and neutrino detectors~\cite{Yin:2018yjn,Cappiello:2018hsu,Bringmann:2018cvk,Ema:2018bih,Cappiello:2019qsw}. Estimating the resultant operator-dependent constraints is beyond our scope and focus, and would constitute an interesting future study.

\subsection{Dependence on Energy Scales}
\label{sec:dependence:on:E:scales}

\begin{table}[t]
  \centering
  \begin{tabular}{p{1.6cm}p{2.6cm}p{2.7cm}p{1.9cm}p{1.9cm}p{1.9cm}p{1.9cm}}  
    \toprule
    $m_\text{T}$
    & 
    \multicolumn{2}{l}{Non-Relativistic}
    & 
    \multicolumn{3}{l}{Relativistic}
    \\
    \cmidrule(r){1-1}\cmidrule(r){2-3}\cmidrule(r){4-7}
    $m_\chi$
    & 
    {\small Heavy}
    & 
    {\small Light}
    & 
    {\small Heavy}
    & 
    {\small Light-ish}
    &
    {\small Med.~Light}
    & 
    {\small Very~Light}
    \\
    \midrule
    ${\rm Baseline}$ 
    & 
    $p_{\rm F}^{1/4}m_{\rm T}$
    & 
    $p_{\rm F}^{1/2}m_\chi^{3/4}$
    & 
    $p_{\rm F}^{5/4}$
    & 
    $p_{\rm F}^{1/2}m_\chi^{3/4}$
    & 
    $p_{\rm F}^{1/2}m_\chi^{3/4}$
    & 
    $p_{\rm F}^{1/2}m_\chi^{3/4}$
    \\
    \bottomrule
  \end{tabular}
  \caption{%
  	Experimental reach on 
  	$\Lambda R_\star^{-1/4}$
  	as a function of the target mass, $m_\textnormal{T}$, dark matter mass $m_\chi$, and the Fermi momentum $p_\textnormal{F}$ in the different regimes defined in Figure~\ref{fig:app:phaseblock:phasespace}. 
  	The baseline case corresponds to the behavior of the $\mathcal O^\textnormal{F}_{5-10}$ and $\mathcal O^\textnormal{S}_{3,4}$ operators in Table~\ref{tab:fullscalingLambda:allops}.
  	These follow the benchmark $|\mathcal M|^2\sim m_\chi E_p^2$ relation in \eqref{eq:app:dominant:terms:baseline:M2}.
	%
	The full set of scalings including the exceptional operators is presented in Table~\ref{tab:fullscalingLambda:allops}.
    }
  \label{tab:fullscalingLambda:baseline}
\end{table}

Table~\ref{tab:fullscalingLambda:baseline} shows the baseline scaling of the discovery reach, $\Lambda$, with respect to all relevant energy scales.
This matches the $m_\chi$ scaling in \eqref{eq:lambda:mx:scaling:baseline} and the behavior of the $\mathcal O^\text{F}_{5-10}$ and $\mathcal O^\text{S}_{3,4}$ operators in Figures~\ref{fig:FContactVector}--\ref{fig:BContact}.
Appendix~\ref{app:prob:scalings} derives the scaling of the discovery reach with respect to all relevant energy scales with the exceptional cases tabulated in Table~\ref{tab:fullscalingLambda:allops}.
The neutron star radius enters through the calculation of the kinetic heating effect in Section~\ref{sec:Kinetic:Heating}. It is notable that the target mass $m_\textnormal{T}$, the target Fermi momentum $p_\textnormal{F}$,  the dark matter mass $m_\chi$, the cutoff/coupling scale $\Lambda$, and the neutron star radius $R_\star$ saturate all relevant scales in the theory. The neutron star mass $M_\star$ is related to $R_\star$ and $p_\textnormal{F}$. 

The equilibrium densities in the BSk24 model of neutron star core lead to Fermi momenta $p_\textnormal{F}$ that are within a $\mathcal{O}(1)$ factor of each other for different targets.
This explains why the low-mass behavior of discovery reach plots appear as nearby parallel lines: in addition to the non-relativistic and relativistic targets having the same power-law scaling in $m_\chi$, the overall prefactor difference is quite small. 
This is a consequence of the equilibrium conditions between $\beta$-decay and the Urca processes in the neutron star core. 
Chemical equilibrium ties together the number densities, $n$, of the degenerate species, which in turn sets their Fermi momenta, $p_\text{F}\sim n^{1/3}$.
As a result, the electron Fermi momentum is approximately only a factor of ten smaller than the nucleon mass.
The closeness of these scales accounts for the closeness of the relativistic and non-relativistic target reach in light and very light dark matter regime that is sketched Figure~\ref{fig:app:phaseblock:phasespace:mesa} and appears numerically in our plots.
The $\mathcal O^\textnormal{F}_{1,2}$ and $\mathcal O^\textnormal{S}_1$ operators contradict the baseline scaling in the light-ish dark matter regime: the difference in the $m_\chi$-scaling between relativistic and non-relativistic targets leads to a pronounced suppression in the reach for electron targets relative to the non-relativistic targets in the light dark matter regimes.

\section{Sources of Uncertainties}
\label{sec:uncertainties}

We conservatively underestimate the heating rate since we do not include the effect of scatters that transfer energy but do not capture. Our multi-scatter bounds are conservative as well, as explained in Section~\ref{sec:multiple:scatter:Nhit}. The cooling of neutron star due to transfer of energy to the incident dark matter is prevented by the Pauli degeneracy of relativistic targets.

Our estimate of the $\Lambda$ reach shown in Figures~\ref{fig:FContactScalar}--\ref{fig:BContact} uses the volume-averaged target abundance and Fermi momentum for a benchmark functional~BSk24 with a specific mass-radius configuration of $M_\star=1.5\,M_\odot$ and $R_\star=12.6$~km, consistent with BSk24 functional. Choosing another functional such as BSk22, BSk25 or BSk26 or choosing a different mass-radius configuration for a given functional alters the radial profiles of stellar constituents and associated quantities such as target abundance and Fermi momentum. These choices in turn affect the Pauli blocking and subsequently the $\Lambda$ reaches shown in Figures~\ref{fig:FContactScalar}--\ref{fig:BContact}. The effect of nucleon density radial profile and corresponding radial variation in chemical potential on dark matter capture rate is studied in~\cite{Garani:2018kkd}. 

To estimate the maximum deviation due to change of functional as well as radial variation in density for a given functional, we follow the method outlined in~\cite{Joglekar:2019vzy}. A detailed analysis for individual functionals and radial density profiles is beyond the scope of this work. Here we consider the maximum range of baryon densities from $0.05$~fm$^{-3}$ to $0.95$~fm$^{-3}$ as allowed by viable mass-radius configurations corresponding to the functionals BSk22, Bsk24, Bsk25 and Bsk26~\cite{Bell:2019pyc,Pearson:2018tkr}. We take the values of target abundance and Fermi momentum corresponding to these extreme density values for estimation of respective kinetic heating cut-off reaches. The band defined by these extreme value estimates will contain all deviations in the kinetic heating reach due to functional change or radial density variation for a given functional. 

For operators $\mathcal{O}^F_1$, $\mathcal{O}^F_3$, $\mathcal{O}^F_5-\mathcal{O}^F_{10}$, $\mathcal{O}^S_3$ and $\mathcal{O}^S_{4}$, we obtain that the reach can at most be a factor~$2$ greater compared to the benchmark values shown in Figures~\ref{fig:FContactScalar}--\ref{fig:BContact} for neutron, proton and muon targets. For electron targets it can at most be a factor~$3$ higher. For neutron and electron targets it can at most be a factor $1.5$ lower, while for muon targets it can be lower by a large factor, if the central baryon density falls below $0.12$~fm$^{-3}$ due to muon abundance disappearing for very low baryon densities~\cite{Pearson:2018tkr}. For the remaining four operators, the maximum deviation factors are same as the above for non-electron targets. In the case of electrons targets, for light dark matter capture, we see negligible deviation from the benchmark values shown in Figures~\ref{fig:FContactScalar}~and~\ref{fig:BContact}. This occurs due to the fact that lowering of the target density is compensated by lesser Pauli Blocking. For heavy dark matter capture by electrons, the reach could be higher up to a factor of $4.5$ or lower up to a factor of $4$ for operators $\mathcal{O}^F_2$ and $\mathcal{O}^F_4$. These factors are $9$ and $7$ respectively in case of operators $\mathcal{O}^S_1$ and $\mathcal{O}^S_2$.

The relative shifts in the $\Lambda$ reaches due to change of nucleon density, among different targets for the same operator, are found to be qualitatively similar to those noted for $\mathcal{O}^F_5$ in~\cite{Joglekar:2019vzy}. This is true for all the operators with a few exceptions. In the light dark matter regime, there is a great overlap between the bands. The reach for all targets shifts while generally maintaining the relative ordering between them as shown in Figures~\ref{fig:FContactScalar}--\ref{fig:BContact}. In the case of muons, for the configurations with central baryon density below $0.12$~fm$^{-3}$, the muon reach rapidly falls and electron reach can dominate over muon reach even in the light dark matter regime for all operators, as found for the case of $\mathcal{O}^F_5$ in~\cite{Joglekar:2019vzy}. Exceptional behavior of electron reach dominating over neutron reach is possible for $\mathcal{O}^F_2$, $\mathcal{O}^F_4$, $\mathcal{O}^S_1$ and $\mathcal{O}^S_2$, where lowering of the baryon density does not have much effect on electron reach as noted above, but neutron reach is sufficiently lowered.

In the heavy dark matter regime, the electron band moves closer to nucleon band for higher baryon densities, as noted in~\cite{Joglekar:2019vzy}. This is because higher $p_\text{F}/m_\text{T}$ ratio for electrons compared to neutrons increases the capture of heavy dark matter much more than the corresponding increase in capture by neutrons. For the same reason, electron dominance over muons in this regime is enhanced even for higher baryon densities. 

We assume that all dark matter transits are diametric across the star, meaning the transit time is assumed to be $\Delta t=2R_\star$. This introduces another uncertainty in the capture rate by an $\mathcal{O}(1)$ factor. When translated to an uncertainty in the bound on cut-off for effective interactions, it gets diminished, since the cut-off goes as the fourth root of the capture rate. For example, in the case of a constant density sphere, according to special relativity, the transit time along the diameter is $\sim 3.2 R_\star$ for a falling object, with initial speed of $v_\textnormal{halo}$ far away from the star and relativistic speeds at the surface of star. Even the paths through shorter chords have $\Delta t>2R_\star$. The number of dark matter particles following these paths are an $\mathcal{O}(1)$ fraction of the total flux through the star. This $\mathcal{O}(1)$ factor uncertainty in the capture rate, when suppressed by a fourth root generates a deviation of $\mathcal{O}(10\%)$ at most in the $\Lambda$ reach. Thus, the cut-off scale bounds derived by assuming $\Delta t =2R_\star$ remain fairly robust to variations in transit time of different dark matter particles. 

We also neglect corrections due to the Schwarzschild metric in the stellar interior. Every dot product in the derivation of the capture rate is expected to carry a corresponding $\mathcal O(10\%)$ correction. This correction also appears in the relative velocity and the expressions for $|\mathcal M|^2$ in Table~\ref{tab:amplitudes12}. $\mathcal{O}(1)$ factor variations due to change of functional are dominant compared to $\mathcal{O}(10\%)$ corrections. Thus, an overall variation in the benchmark reach remains a small $\mathcal{O}(1)$ factor except when zero abundance of muons at low baryon densities leads to a significant reduction in the sensitivity for muons. Finally, we note that we have neglected exotic phases of matter that may occupy the core of a neutron star. In particular, a color-flavor-locked phase may have no electrons and may suppress dark matter scattering with nucleons~\cite{Bertoni:2013bsa}.


\section{Conclusions}

Neutron star kinetic heating is a novel way to discover dark matter--visible matter interactions. We present a new framework to calculate the kinetic heating from dark matter incident on Fermi-degenerate, ultra-relativistic targets in a neutron star, such as electrons. 
Our formalism accounts for the boost between the center of momentum frame where scattering kinematics are manifest and the neutron star frame where heating and Pauli blocking effects are defined.  

We classify the kinematics of dark matter capture in a neutron star according to the dark matter mass and whether or not the target is relativistic. Figure~\ref{fig:app:phaseblock:phasespace} summarizes this classification.
We apply our framework to determine the discovery reach of a kinetic heating observation for scalar and fermionic dark matter that interacts with visible matter through contact interactions. 
This extends the existing literature on kinetic heating to a complete set of effective interactions with Standard Model fermion up to dimension-6. 
Figures~\ref{fig:FContactScalar}--\ref{fig:BContact} present our numerical results. In large regions of parameter space, neutron star heating from electron scattering is a powerful and complementary technique to search for dark matter compared to terrestrial searches. This is especially true for leptophilic dark matter models where electron scattering is the primary discovery channel and for operators whose scattering is proportional to the transferred momentum. 

The neutron star kinetic heating program is especially exciting given the road map of upcoming radio (\acro{FAST}~\cite{FAST}, {\acro{SKA}~\cite{SKA}, and \acro{CHIME}~\cite{CHIME}}) and infrared telescopes (\acro{JWST}~\cite{Gardner:2006ky}, \acro{TMT}~\cite{Skidmore:2015lga}, \acro{ELT}~\cite{andersen2003euro50}). The detection of a single star sufficiently old and sufficiently nearby neutron stars may be sufficient to either discover dark matter through kinetic heating, or otherwise play an important role in model discrimination in concert with other experimental programs.

 \section*{Acknowledgements}

We thank
Nicole Bell, 
Joseph Bramante, 
Sekhar Chivukula, 
Lexi Costantino, 
James Dent, 
and Peter Denton 
for useful comments and discussions. 
\textsc{p.t.} and \textsc{n.r.} thank the Aspen Center for Physics (\acro{NSF} grant \#1066293) for its hospitality during a period where part of this work was completed. A part of this work was also completed at Kavli Institute for Theoretical Physics (\acro{A.J.}, \acro{H.B.Y.}), which is supported in part by the National Science Foundation under Grant No.~\acro{NSF} \acro{PHY}-1748958.
\textsc{a.j.}, \textsc{p.t.}, and \textsc{h.b.y.} are supported by the \acro{DOE} grant \textsc{de-sc}/0008541.
\textsc{n.r.} is supported by Natural Sciences and Engineering Research Council of Canada (\acro{NSERC}). 
\acro{TRIUMF} receives federal funding via a contribution agreement with the National Research Council Canada. 
\textsc{p.t.} thanks his mother for providing unsolicited copy editing for a draft of this paper on Christmas day.
%


\appendix

\section{Convention for Frame-Dependent Quantities}
\label{app:conventions}

The formalism in this paper involves quantities that are naturally defined in different frames. This can lead to ambiguous or cluttered notation. This appendix summarizes the conventions we use for labeling frame-dependent quantities.

  {Frames are specified by subscripts on individual quantities.} When there is ambiguity, the outermost subscript is the frame. Thus $d\sigma_\CM$ is the differential cross section in the center of momentum frame, $\left(d\sigma\, v_\text{rel}\, dn_\text{T}\, dn_\chi\right)_\text{T}$ is a product calculated in the target frame.
%
  {Quantities that are Lorentz invariant do not have frames specified.}
%
  Most of the calculations in this study are in the neutron star frame. For simplicity, we drop the neutron star frame subscript when there is no ambiguity. Thus {quantities that are Lorentz covariant and carry no specified frame are understood to be calculated in the neutron star frame.}

  Four-momenta and three-momenta: {$p^\mu = (E_p, \vec{p})$ and $k^\mu = (E_k, \vec{k})$ represent the four momenta of the target and the dark matter, respectively.} Without additional labeling, they are assumed to be in the neutron star frame. 
%
  {Contractions are written with dots}: $p\cdot k = p^\mu k_\mu = p^0 k^0 - \vec{p}\cdot\vec{k}$. 
%
  {An italicized four-momentum with no indices is understood to be the magnitude of the three-vector}: $p = |\vec p|$ and $k = |\vec k|$. This introduces no ambiguity with the norm of the four-vector since, for example~$p^\mu p_\mu = m_\text{T}^2$; in other words: we \emph{never} write $p^2$ to mean the Minkowski norm of a four-momentum. This slightly unconventional choice simplifies the visual interpretation of the expressions in these appendices.

  We use natural units throughout this document. {Physical velocities of particles are written as $v$, for example $v_\text{halo}$ is the asymptotic velocity of dark matter in the halo as measured in the neutron star frame.} The boost factor to the dark matter rest frame is $\gamma_\text{halo} = (1-v_\text{halo}^2)^{-1/2}$.

\section{Some Useful Derivations}

\subsection{Maximum Impact Parameter}
\label{app:bmax}

We derive $b_\text{max}$ in \eqref{eq:bmax}. Ref.~\cite{Baryakhtar:2017dbj} attributes this result to Ref.~\cite{Goldman:1989nd}, which in turn references a general relativity textbook. Our treatment is based on the textbook by Hartle~\cite{hartle2003gravity}. In the vicinity of the neutron star the space is described by the Schwarzschild metric, which in spherical coordinates is
\begin{align}
  ds^2 &= 
  \left( 1-\frac{2GM}{r}\right) dt^2
  -
  \left( 1-\frac{2GM}{r}\right)^{-1}dr^2
  - r^2 d\Omega^2 \ .
  \label{eq:Schwarzschild:metric}
\end{align}
This space has two constants of motion coming from invariance along translations in time and the polar direction:
\begin{align}
  \varepsilon
  &= \left( 1-\frac{2GM}{r}\right) \frac{dt}{d\tau}
  &
  \ell &= r^2 \sin^2\theta \frac{d\phi}{d\tau} \ ,
  \label{eq:Schwarzschild:energy}
\end{align}
where $\tau$ is the proper time of a test particle. These are simply energy per unit mass and angular momentum per unit mass. The normalization of a test particle‘s four-velocity, $u^\alpha u^\beta g_{\alpha\beta}$ gives
\begin{align}
  \frac{
  \varepsilon
  }{\ell^2} 
  &= 
  \frac{1}{\ell^2} \frac{dr}{d\tau} 
  + \frac{\left( 1-\frac{2GM}{r}\right)}{r^2}
  + \frac{\left( 1-\frac{2GM}{r}\right)}{\ell^2} \ .
  \label{eq:bmax:4velnorm}
\end{align}
The maximum impact parameter $b_\text{max}$ corresponds to the distance at which a dark matter particle approaching with some initial velocity $v_\text{halo}$ has a trajectory that is tangent to the neutron star:
\begin{figure}[tb]
	\centering 
	\includegraphics[width=0.65\textwidth]{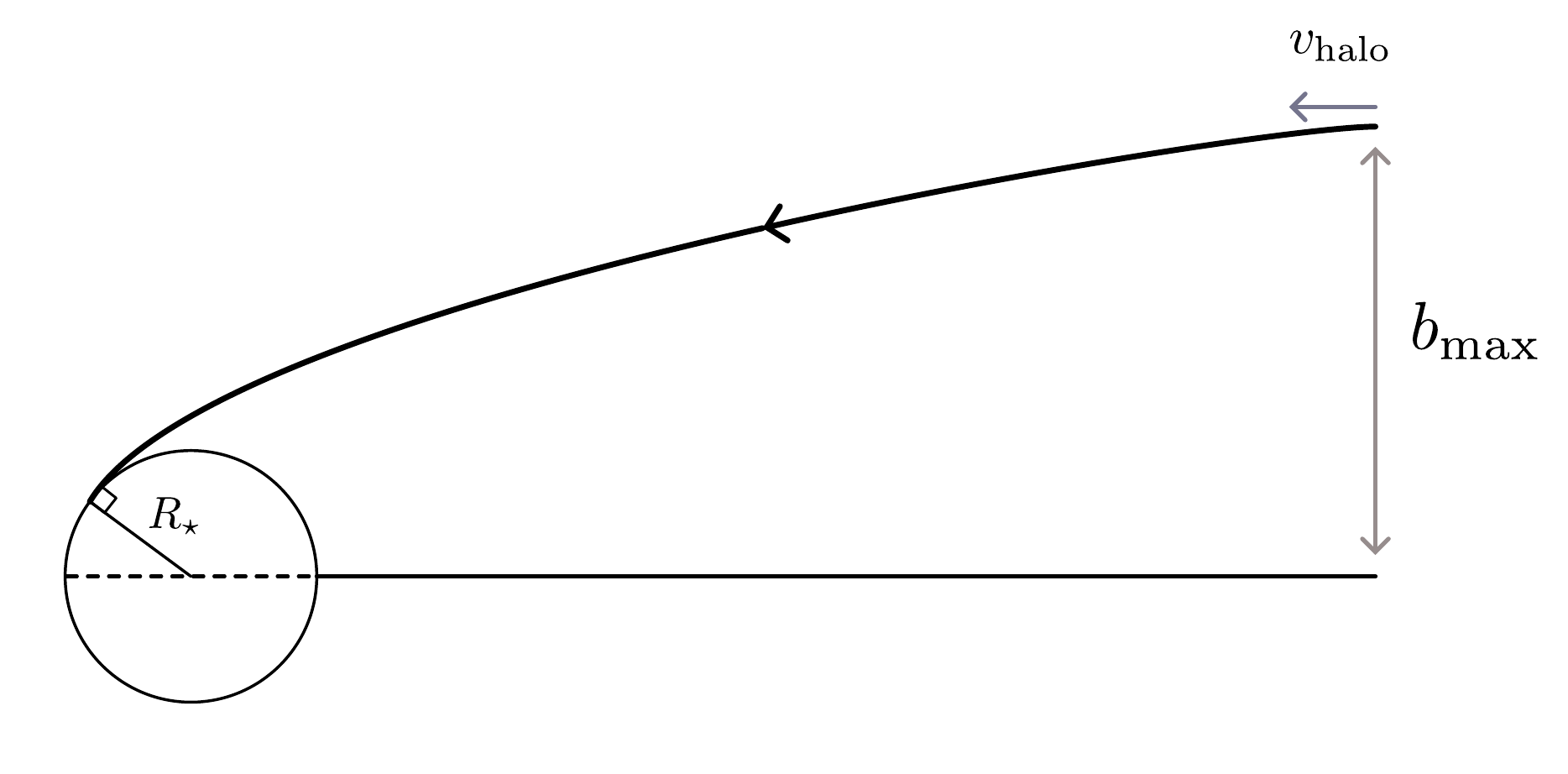}
	\caption{Schematic diagram showing $b_\text{max}$.}
	\label{fig:bmax}
\end{figure}
At the point of tangency, $dr/d\tau = 0$ and $r=R$. Thus \eqref{eq:bmax:4velnorm} gives 
\begin{align}
  \ell = R \sqrt{\frac{2GM}{R}} \left(1-\frac{2GM}{R}\right)^{-1/2} \ .
\end{align}
However, since $\ell$ is a constant of motion, we may set it to its initial value asymptotically far from the neutron star: $\ell = b_\text{max}v_\text{halo}$, from which we derive the expression for $b_\text{max}$.

\subsection{Energy Transfer for a Non-Relativistic Target}
\label{app:Delta:E:Non:Relativistic}

To assist in contrasting the relativistic and non-relativistic cases, we derive the energy transfer to a non-relativistic target that is stationary in the neutron star frame, equation (5) in Ref.~\cite{Baryakhtar:2017dbj}.
In the neutron star frame, the incident dark matter has four-momentum $k^\mu = (E_k, \vec{k})$ such that 
\begin{align}
  \vec{k}^2
  &= 
  E^2_k
  - m_\chi^2 
  = 
  \left(
    \gamma_\esc
    - 1\right)m_\chi^2
  = \frac{v_\esc^2}{1-v_\esc^2} m_\chi^2
  &
  \gamma_\esc^2 = \frac{1}{1-v_\esc^2} \ ,
\end{align}
where $v_\esc$ is the escape velocity at the surface of the neutron star. Similarly, let $p_\mu = (m_\text{T}, \vec{0})$ be the non-relativistic target four-momentum in the neutron star frame. Define $\bbeta$ and $\gamma^2 = (1-\beta^2)$ to be the boost parameter to the center of momentum frame. The center of momentum frame momenta are
\begin{align}
  k^\mu_\CM
  &
  =
  \begin{pmatrix}
    \gamma & -\gamma \boldsymbol{\beta}
    \\
    -\gamma \boldsymbol{\beta} & \gamma
  \end{pmatrix}
  \begin{pmatrix}
    E_k
    \\
    \vec{k}
  \end{pmatrix}
  &
  p^\mu_\CM
  &
  =
  \begin{pmatrix}
    \gamma & -\gamma \boldsymbol{\beta}
    \\
    -\gamma \boldsymbol{\beta} & \gamma
  \end{pmatrix}
  \begin{pmatrix}
    m_\text{T}
    \\
    \vec{0}
  \end{pmatrix}
  \ .
\end{align}
The total three-momentum vanishes in the center of momentum frame so that
\begin{align}
  \vec{p}_\CM + \vec{k}_\CM &= 0 
  &
  \text{which gives}
  &
  &
  \boldsymbol{\beta}
  &= \frac{\vec{k}}{E_\esc+ m_\text{T}} \ .
\end{align}
The transferred four-momentum is $q^\mu_\CM = k^\mu_\CM-k'^\mu_\CM = (0, \vec{q}_\CM)^T$ . In the neutron star frame, the energy transfer is
\begin{align}
  \Delta E &= q^0 
  = \gamma \boldsymbol{\beta}\cdot \vec{q}_\CM
  = \frac{\gamma \vec{k}\cdot\vec{q}_\CM}{E_k+m_\text{T}} 
  = \frac{%
        \gamma^2 m_\text{T} \vec{k}^2
        \left( 1-\cos\psi \right)
      }{
        (E_k+m_\text{T})^2
      }
  \ ,
\end{align}
where $\psi$ is the angle between the dark matter incoming and outgoing three-momenta in the center of momentum frame. We simplify this using
\begin{align}
  \frac{\gamma^2}{E_\esc + m_\text{T}} 
  &=
  \frac{
    E_\esc + m_\text{T}
  }{
    m_\chi^2 + m_\text{T}^2 + 2\gamma_\esc m_\chi m_\text{T}
  } 
  \ .
\end{align}
The energy transfer to a non-relativistic target in the neutron star frame is
\begin{align}
  \Delta E
  &= 
  \frac{
    m_\text{T} m_\chi^2
  }{
    m_\chi^2 + m_\text{T}^2 + 2\gamma_\esc m_\chi m_\text{T}
  }
  \frac{v_\esc^2}{1-v_\esc^2} 
  \left( 1-\cos\psi \right)
  \ ,
  \label{eq:app:NR:Delta:E}
\end{align}
where $v_\esc$ is the escape velocity so that in the $v_\esc^2\ll 1$ limit the second factor reduces to $v_\esc^2$.

\subsection{Flux Density and the M\o ller Velocity}
\label{sec:Moller}

The M\o ller velocity, $v_\Mol$, appears in the kinematics of non-collinear particle scattering such as dark matter annihilation~\cite{Gondolo:1990dk}. For colliding particles $T$ and $\chi$ with respective four-momenta $p^\mu=(E_p, \vec{p})$ and $k^\mu=(E_k,\vec{k})$, a convenient expression is
\begin{align}
  v_\Mol &= \frac{\sqrt{(p\cdot k)^2 - m_\text{T}^2 m_\chi^2}}{E_p E_k} \ ,
  \label{eq:Moller:velocity}
\end{align}
which is precisely the factor that shows up in the flux density of dark matter--target scattering rate: $n_\text{T} n_\chi v_\Mol$. The following discussion is a summary of the review by Cannoni~\cite{Cannoni:2016hro}, which in turn is based on \emph{The Classical Theory of Fields} by Landau \& Lifschitz~\cite{landau1975classical}. 

The relative velocity $v_\text{rel}$ between the target and dark matter is a Lorentz invariant~\cite{Cannoni:2013bza}. This can be seen, for example, by starting in the target frame where $v_\text{rel}$ is simply the dark matter velocity. One may subsequently write $v_\text{rel}$ in terms of Lorentz invariants:
\begin{align}
  v_\text{rel} 
  = \left. \frac{k}{E_k} \right|_\text{T}
  = \left. \frac{\sqrt{E_k^2 - m_\chi^2}}{E_k} \right|_\text{T}
  = \frac{\sqrt{(p\cdot k)^2 - m_\text{T}^2m_\chi^2}}{p\cdot k} \ .
  \label{eq:v:rel:Lorentz:Invariant}
\end{align}
For a given scattering process, the invariant rate density is 
  \begin{align}
    \mathcal{R} 
    &= \frac{d\nu}{\Delta V \Delta t}
    = \left( 
      d\sigma\, v_\text{rel} \, dn_\text{T} \, dn_\chi
      \right)_\text{T}
    \ ,
    \label{eq:R:dnu:dsig:dvrel:dnt:dnchi:in:T:frame}
  \end{align}
where the right-hand side we write the expression in the frame of the target particle because here the cross section $d\sigma$ and relative velocity $v_\text{rel} = \left. v_\chi\right|_\text{T}$ are unambiguously defined. By comparison, we do not simply plug in our expression for $d\nu$ from \eqref{eq:dnu:dsig:vrel:dnt:dnchi:Dv:Dt} because of the challenge of defining $d\sigma$ in an arbitrary frame due to Lorentz contraction.

In order to write the rate density in a general frame, $F$, we note that it must be proportional to the target and dark matter densities,
\begin{align}
  \mathcal{R}  &= 
  \left(
    A\, dn_\text{T} dn_\chi
  \right)_F  = 
  \left(A \frac{E_T E_\chi}{m_\text{T} m_\chi}\right)_F d\hat n_\text{T} d \hat n_\chi \ ,
  \label{eq:R:A:EtEx:dnt0:dnx0:div:mt:mx}
\end{align}
where $A$ is a proportionality factor and we factor out the target density in the target frame $d\hat n_\text{T} = (dn_\text{T})_\text{T}$ and the dark matter density in the dark matter frame $d\hat n_\chi = (dn_\chi)_\chi$. These densities are the zeroth components of four-currents so that boosting from their respective rest frames to a general frame rescales them by $\gamma = E/m$. 

Because $\mathcal R$ and $d\hat n_\text{T} d\hat n_\chi/m_Tm_\chi$ are invariant, the combination $AE_\text{T}E_\chi$ must also be invariant.
Comparing the right-hand sides of \eqref{eq:R:dnu:dsig:dvrel:dnt:dnchi:in:T:frame} and \eqref{eq:R:A:EtEx:dnt0:dnx0:div:mt:mx} gives the proportionality factor,
\begin{align}
  A &= \frac{p\cdot k}{E_TE_\chi} \left(d\sigma\, v_\text{rel}\right)_\text{T} \ ,
\end{align}
where the cross section $d\sigma$ is calculated in the target frame\footnote{Here one could equivalently replaced the target frame with the dark matter frame. Indeed, this is necessary if the target is formally massless. The cross section is invariant with respect to boosts along the collision axis so that 
$
  d\sigma_\text{T} 
  = 
  d\sigma_\chi
$
}. Using the invariance of $v_\text{rel}$ in \eqref{eq:v:rel:Lorentz:Invariant} and the invariance of $d\sigma_\text{T}$ along the collision axis, we have
$d\sigma_\text{T} 
  = 
  d\sigma_\CM$
so that the invariant rate density is conveniently expressed as
\begin{align}
  \mathcal{R} &= d\sigma_\CM 
  \, 
  \left(\frac{p\cdot k}{E_TE_\chi} v_\text{rel}\right)
   \, dn_\text{T} dn_\chi
   = 
   d\sigma_\CM 
    \, 
    v_\Mol
   \, dn_\text{T} dn_\chi \ .
\end{align}
Comparing this to the rate in \eqref{eq:R:dnu:dsig:dvrel:dnt:dnchi:in:T:frame}, we have a convenient transformation of $d \sigma\, v_\text{rel}$ into any frame:
\begin{align}
  \left(
    d\sigma\, v_\text{rel}
  \right)_F 
  &= 
  d\sigma_\CM 
    \, 
  \left(
    v_\Mol
  \right)_F \ .
\end{align}
The relative velocity $v_\text{rel}$ is invariant~\cite{Cannoni:2016hro}, but the M\o ller velocity, $v_\Mol$, is not. The above relation is a simple way to connect the cross section in a general frame to the center-of-momentum frame where it is most conveniently calculated. This proves \eqref{eq:dsig:vrel:dsig:CM:vMol}.

\subsection{Capture from Multiple Scattering}
\label{app:multiple:scatter:capture}

We discuss the treatment of multiple scattering, expanding on the presentation in Section~\ref{sec:multiple:scatter:Nhit}. For convenience, let us define a differential capture efficiency \emph{without} any kinematic conditions imposed:
\begin{align}
  d\hat f 
  \equiv \frac{d\nu}{dN_\chi} 
  = 
  d\sigma_\CM 
  \, v_\Mol 
  \, dn_\text{T} 
  \, \Delta t \ ,
\end{align}
this differs from $df$ in \eqref{eq:df:dsig:CM:v:dnt:dt} in that capture conditions are not imposed. The discussion of multiple scatters is unaffected by Pauli blocking, so for the purposes of this appendix we may assume that the targets are not degenerate.

Dark matter is captured if it loses its asymptotic kinetic energy $\Delta E_\text{min} = E_\text{halo}$ over its transit through the star. 
The capture efficiency $d\hat f$ is a measure of the total number of scatters that a transiting dark matter particle undergoes. 
We can restrict to cases where dark matter captures after only one capture ($N_\text{hit} =1$) by multiplying $d\hat f$ by a step function enforcing that any scatter must transfer $\Delta E \geq \Delta E_\text{min}$:
\begin{align}
  d\hat f_{N_\text{hit} = 1} 
  &= 
  d\hat f \, \Theta(\Delta E - E_\text{halo}) 
  \equiv  
  \left.d\hat f\right|_{\Delta E > E_\text{halo}} 
  \label{eq:dhat:f:1}
  \ .
\end{align}
$\Delta E$ is fixed for a given initial and final state kinematic configuration. The $\Theta$ function enforces that only scatters that satisfy the capture condition are counted when integrating \eqref{eq:dhat:f:1}.
When $d\hat f_{N_\text{hit} = 1}  \geq 1$, we assume that transiting dark matter scatters at least once and always captures. Otherwise, the probability for a given dark matter particle to capture is simply the capture efficiency,  $d\hat f_{N_\text{hit} = 1}$.

We extend this to the case where dark matter that captures from $N_\text{hit} =2$ scatters. In this case it is sufficient for each scatter to transfer energy $\Delta E > \Delta E/2$ but it must do so over each half of its total transit through the star.
Thus
\begin{align}
  d\hat f_{N_\text{hit} = 2} 
  &= 
  \frac 12 \left.d\hat f\right|_{\Delta E > \Delta E_\text{min}/2} \ .
\end{align}
This is clearly a conservative estimate as it undercounts configurations where the average energy loss is larger than $\Delta E_\text{min}/2$ but one scatter has $\Delta E <\Delta E_\text{min}/2$. 

When combining these results, one must be careful not to double count the configurations. If $\Delta E$ permits capture after one scatter, it should not be counted in the piece for two scatters. Thus the total number of scatters that ($i$) are part of a capturing transit and ($ii$) counting only captures that require up to $N_\text{hit}=2$ scatters is
\begin{align}
  d\hat f_{N_\text{hit} = 1\text{ or }2} 
  &= 
  \left.d\hat f\right|_{\Delta E > \Delta E_\text{min}} 
  +
  \frac{1}{2}
  \left.d\hat f\right|_{\Delta E_\text{min} > \Delta E > \Delta E_\text{min}/2}
  \ .\label{eqn:multiscat}
\end{align}
The generalization to larger $N_\text{hit}$ is straightforward. 

To analytically understand the dependence of capture efficiency $f$ on $m_\chi$, we make a further conservative estimate and assume that the expected number of capturing scatters $d\hat f$ is dominated by a single value of $N_\text{hit}$. 
\begin{align}
  f
  &\approx 
  \frac{1}{\hat N_\text{hit}}
  \int 
  \left.d\hat f\right|_{\Delta E > \Delta E_\text{min}/{\hat N_\text{hit}}}
  \ ,
\end{align}
where $\hat N_\text{hit}$ is the value of $N_\text{hit}$ that maximizes the integrand. Note that we also drop the upper limit on $\Delta E$ on the right-hand side, since this approximation would not over count those scatters. For the purpose of numerical results presented in Figures~\ref{fig:FContactScalar}--\ref{fig:BContact}, we use~\eqref{eqn:multiscat} generalized to very large $N_\text{hit}$.


\section{Kinematics}
\label{sec:Mandel+CMDM}

We present the expressions for the Mandelstam $s$ and $t$ parameters and the dark matter momentum in the center of momentum frame.
The boost from the neutron star frame to the center of momentum frame is
\begin{align}
  \boldsymbol{\beta} 
  &= 
  \frac{\vec{p}+\vec{k}}{E_p+E_k}
  &
  \gamma 
  = \frac{1}{\sqrt{1-\boldsymbol{\beta}^2}} 
  = \frac{E}{E_\CM}
  \label{eq:beta:p:k:gamma}
  \ .
\end{align}
The dilation factor $\gamma$ is simply the ratio of the total energy in the neutron star frame, $E=E_p+E_k$, to the total energy in the center of momentum frame, $E_\CM = \left(E_p\right)_\CM  + \left(E_k\right)_\CM $. This is readily seen by boosting the total energy in the center of momentum frame to the neutron star frame\footnote{This is this is the inverse transformation of \eqref{eq:beta:p:k:gamma}, but the $\gamma$ factors are the same. Since the total four-momentum in the center of momentum frame has no three-momentum component, the $\boldsymbol{\beta}$ term does not contribute.} so that $E = \gamma E_\CM$.
The center of momentum frame energies with respect to neutron star frame momenta are:
\begin{align}
  \left(E_p\right)_\CM 
  &=
  \gamma
  \left( E_p-\boldsymbol{\beta}\cdot\vec{p} \right)
  &
  \left(E_k\right)_\CM
  &=
  \gamma
  \left( E_k - \boldsymbol{\beta}\cdot\vec{k} \right) 
  \ .
\end{align}

\subsection{Expression for \texorpdfstring{$s$}{s}}
\label{sec:s}

The Mandelstam $s$ parameter is 
\begin{align}
  s 
  = (p+k)^2 
  = m_\text{T}^2 + m_\chi^2 + 2\left(E_pE_k-\vec{p}\cdot\vec{k}\right)
  =  E_\CM^2 
  \ .
  \label{eqn:s}
\end{align}

\subsection{Expression for \texorpdfstring{$t$}{t}}\label{sec:t}

In the center of momentum frame, the energy and magnitude of the three-momentum is conserved:
\begin{align}
  \left(E_{k}\right)_\CM &= \left(E_{k'}\right)_\CM
  &
  |\vec k_\CM| &= |\vec k_\CM'| \equiv k_\CM
  \ .
\end{align}
The Mandelstam $t$ parameter encodes the momentum transfer. It may be expressed with respect to the center-of-momentum frame polar angle, $\psi$ between $\vec{k}_\CM$ and $\vec{k}'_\CM$.
\begin{align}
  -t
  &=({k}_\CM-{k}'_\CM)^2
  = 2k_\CM^2\left(1-\cos\psi\right) 
  = 4 k_\CM^2 \sin^2\frac{\psi}{2}
  \ .
  \label{eqn:t}
\end{align}

\subsection{Dark Matter Three-Momentum}\label{sec:CMDM}

In \eqref{eqn:t} and in the appendices below, we require an expression for the dark matter momentum in the center of momentum frame, $k_\CM$, in terms of the neutron star frame kinematics.  $\vec{k}_\CM$ is related to its neutron star frame counterpart $\vec{k}$ by a boost. This boost only transforms the components of the three-momentum that are parallel to $\boldsymbol{\beta}$ in \eqref{eq:beta:p:k:gamma}. We thus separate the three momenta into pieces that are parallel, $\parallel$, and perpendicular, $\perp$, to $\boldsymbol\beta$:
\begin{align}
  \vec{k}_\CM
  &= 
    \left(\vec{k}_\perp\right)_\CM 
  + \left(\vec{k}_\parallel\right)_\CM
  = \left(\vec{k}_\perp\right)_\CM 
  + \gamma \vec{k}_\parallel - \gamma\boldsymbol\beta E_k 
  \ .
\end{align}
One may then write the parallel and perpendicular projections with respect to the neutron star frame dark matter momentum projected onto the boost direction, $\vec k\cdot \boldsymbol\beta$:
\begin{align}
  \vec{k}_\CM
  &=  \vec{k}
  + \left(\gamma-1\right) \frac{(\vec{k}\cdot\boldsymbol{\beta})\boldsymbol\beta}{\boldsymbol\beta^2} 
  - \gamma\boldsymbol\beta E_k 
  \ .
  \label{eq:kCM:k:gamma:beta}
\end{align}
We may express \eqref{eq:kCM:k:gamma:beta} in terms of the kinematic quantities in the neutron star frame:
\begin{align}
  E_\CM \vec{k}_\CM
  &= 
  E_\CM \vec{k}
  + 
  \left[
  \left( E-E_\CM \right)
  \frac{(\vec{k}\cdot\boldsymbol{\beta})}{\boldsymbol\beta^2} 
  - E E_k 
  \right] \boldsymbol\beta
  \ .
\end{align}
Use \eqref{eq:beta:p:k:gamma} to simplify the term in brackets. This gives:
\begin{align}
  1-\beta^2 &= \frac{E_\CM^2}{E^2}
  &
  \frac{1}{\beta^2}
  &= \frac{E^2}{\left(E+E_\CM\right)\left(E-E_\CM\right)} 
  \ ,
  \label{eq:beta:E:Ecm:trick}
\end{align}
which in turn yields:
\begin{align}
  E_\CM \vec{k}_\CM
  &= 
  E_\CM \vec{k}
  + 
  \left[ 
    \frac{\vec{k}\cdot(\vec{p}+\vec{k})}{E+E_\CM}
    - E_k
  \right](\vec{p}+\vec{k})
  \ .
\end{align}
To further simplify this expression, it is useful to separate a term $(E_p\vec{k} - E_k\vec{p})$:
\begin{align}
  E_\CM \vec{k}_\CM
  &\equiv
  \left(E_p\vec{k} - E_k\vec{p}\right)
  + A\vec{p}
  + B\vec{k} \ ,
  \label{eq:Ecm:kcm:AB}
\end{align}
where the coefficients are, using $E=E_k+E_p$,
\begin{align}
  A &= \frac{\vec{k}\cdot(\vec{p}+\vec{k})}{E+E_\CM}
  &
  B &=
  A
  + (E_\CM - E)
  = 
  A - \frac{(\vec{p}+\vec{k})^2}{E+E_\CM}
  = -\frac{\vec{p}\cdot (\vec{p}+\vec{k})}{E+E_\CM} 
  \ ,
\end{align}
where we simplified $B$ using \eqref{eq:beta:p:k:gamma} and \eqref{eq:beta:E:Ecm:trick}.
We may thus combine the $A$ and $B$ terms in \eqref{eq:Ecm:kcm:AB}:
\begin{align}
  E_\CM \vec{k}_\CM
  &\equiv
  \left(E_p\vec{k} - E_k\vec{p}\right)
  +
  \frac{(\vec{p}+\vec{k})\times\left(\vec{p}\times\vec{k}\right)}{E+E_\CM} 
  \ .
  \label{eqn:kcm}
\end{align}
We ultimately would like the norm of $\vec{k}_\CM$. A useful intermediate step is to write the cross products in terms of the angle $\theta$ between the neutron star frame momenta:
\begin{align}
  \vec{k}\cdot\left[ \vec{p} \times (\vec{p}+\vec{k}) \right]
  = -\vec{p}\cdot\left[ \vec{k} \times (\vec{p}+\vec{k}) \right]
  = -(\vec{p}\times\vec{k})^2
  = p^2 k^2\sin^2\theta
  \ .
\end{align}
We thus find that $k_\CM^2 = |\vec k_\CM|^2$ is
\begin{align}
  E_\CM^2 {k}_\CM^2 
  &= 
  E_p^2 k^2 + E_k^2 p^2
  - 2E_pE_k (\vec{p}\cdot \vec{k})
  - p^2k^2\sin^2\theta \ .
  \label{eq:kCM:2:Ecm}
\end{align}
Note that $\theta$ is the same angle defined in \eqref{eq:NS:angular:omega:F:phi:theta}.

\subsection{Energy Transfer in the Neutron Star Frame}
\label{sec:DeltaE}

We derive $\Delta E$, the energy transferred to the target by a dark matter scatter in the neutron star frame. It is convenient to relate this to quantities in the center of momentum frame:
\begin{align}
\Delta E 
= E_k - E_{k'}
= \gamma\left[(E_k)_\CM - (E_{k})_\CM \right]
  + \gamma \boldsymbol{\beta}\cdot \left(\vec{k}_\CM - \vec{k}_\CM'\right)
= 
  \gamma \boldsymbol{\beta}\cdot \vec{q}_\CM
  \ .
  \label{eqn:Ens}
\end{align}
Here $\vec{q}_\CM = \vec{k}_\CM - \vec{k}_\CM'$ is the transferred three-momentum in the center of momentum frame.
Note that this is the \emph{inverse} transformation of \eqref{eq:beta:p:k:gamma}, for which the boost parameter is $-\boldsymbol{\beta}$ rather than $\boldsymbol{\beta}$.

We may write the scattered dark matter three-momentum, $\vec{k}'_\CM$, with respect to the polar and azimuthal angles in the center of momentum frame. The polar angle $\psi$ is measured with respect to the $\vec{k}_\CM$ direction. The azimuthal angle $\alpha$ is further measured with respect to the component of $\boldsymbol{\beta}$ that is perpendicular to $\vec{k}_\CM$, which we call $\boldsymbol{\beta}_\perp$. In the center of momentum frame the length of the three-momentum is conserved, so that
\begin{align}
  \vec{k}'_\CM
  &= 
  k_\CM \, \sin\psi \, \cos\alpha \, \hat{\boldsymbol{\beta}}_\perp
  + 
  \sin\psi \, \sin\alpha \, \left(\vec{k}_\CM \times \hat{\boldsymbol{\beta}}_\perp\right)
  + 
  \cos\psi\, \vec{k}_\CM \ ,
  \label{eq:kpCM:beta:projection}
\end{align}
where $\hat{\boldsymbol{\beta}}_\perp$ is a unit vector in the direction of $\boldsymbol{\beta}_\perp$. Plugging \eqref{eq:kpCM:beta:projection} into \eqref{eqn:Ens} and using the orthogonality of $\bbeta$ and $(\vec{k}_\CM\times \hat{\boldsymbol{\beta}}_\perp)$, we have:
\begin{align}
  \Delta E 
  & = 
  \gamma\boldsymbol{\beta}
  \cdot
  \left[
    \vec{k}_\CM \left(1-\cos\psi\right)
    - k_\CM \sin\psi\, \cos\alpha\, \hat{\boldsymbol{\beta}}_\perp
  \right]
  \\
  &= \gamma(\boldsymbol{\beta}\cdot \vec{k}_\CM)\left(1-\cos\psi\right)
  - \gamma \sqrt{\bbeta^2 \, \vec{k}_\CM^2 - \left(\bbeta\cdot\vec{k}_\CM\right)^2} \, \sin\psi\,\cos\alpha \ .
  \label{eqn:recoil}
\end{align}
In Appendix~\ref{app:PhaseBlock} we reduce this expression to special cases that illuminate the qualitative features of relativistic capture.

\section{Scaling Relations}
\label{app:PhaseBlock}

This appendix first shows how kinematic conditions on $\Delta E$ conditions impose constraints on the phase space variables $\cos\theta$, $\cos \psi$, $\alpha$, and $p$ as a function of the dark matter mass $m_\chi$.
We use the energy transfer expression \eqref{eqn:recoil} to develop a qualitative understanding of the capture rate as a function of dark matter mass. We establish a set of necessary conditions to diagnose the size of the phase space accessible to capture. 
The main result of this appendix is Table~\ref{fig:app:phaseblock:flow:chart}, which systematically determines the $m_\chi$ scaling of the capture efficiency, $f$. It extends and clarifies the Table~\ref{fig:app:phaseblock:short:flow:chart} according to the detailed treatment in this appendix.

\subsection{%
Energy Transfer and \texorpdfstring{$\cos\delta$}{Cos Delta}
}

The center of momentum frame momentum, from \eqref{eq:beta:p:k:gamma} and \eqref{eqn:kcm}, is
\begin{align}
  E_\CM \vec{k}_\CM
  =
  \left(E_p\vec{k} - E_k\vec{p}\right)
  +
  \frac{E\boldsymbol{\beta}\times\left(\vec{p}\times\vec{k}\right)}{E+E_\CM} 
  \ ,
\end{align}
where the second term is orthogonal to the boost from the neutron star to center of momentum frame, $\bbeta$. We define the angle $\cos\delta$ between the boost parameter from the neutron star frame to the center of momentum frame, $\bbeta$, and the dark matter three-momentum in the center of momentum frame $\kcm$:
\begin{align}
  \bbeta \cdot \kcm 
  &\equiv
  \beta k_\CM\cos \delta 
  =
  \frac{E_p k^2 - E_k p^2 + (E_p-E_k)\vec{p}\cdot\vec{k}}{EE_\CM} 
  \ .
  \label{eq:beta.kcm}
\end{align}
With respect to this variable, the energy transfer expression \eqref{eqn:recoil} is
\begin{align}
  \frac{\Delta E}{\gamma\beta k_\CM}
  &= 
  \cos\delta\, \left(1-\cos\psi\right) 
  - 
  \left|\sin\delta\right|
  \cos\alpha\, \sin\psi 
  \ ,
  \label{eq:Delta:E:delta:psi:alpha}
\end{align}
where we identify $\sqrt{1-\cos^2\delta} = \left|\sin\delta\right|$.
The quantity $\cos\delta$ is proportional to the projection of the total three-momentum in the neutron star frame, $\vec{p}+\vec{k}$, onto the dark matter three-momentum in the center of momentum frame, $\vec{k}_\textnormal{CM}$. It plays a key role in determining the scaling of the phase space volume.

\subsection{Rules of Thumb for Phase Space Scaling}

We establish a set of heuristics to diagnose the volume of phase space.

\begin{thumb}[Independent Integration Assumption]
We assume that the phase space integrals are independent of one another. For simplicity, we ignore the dependence on phase space integrals in the differential cross section, $d\sigma/d\Omega_\CM$. This is sufficient to understand the scaling behavior with respect to the dark matter mass.
\label{thumb:independent:integration}
\end{thumb}

\begin{thumb}[Positive Energy Transfer Condition]
A necessary---but not sufficient---condition for dark matter to capture is that the dark matter transfers energy to the target, $\Delta E >0$. 
\label{thumb:positive:Delta:E}
\end{thumb} 
\begin{corollary}[Easy Condition]
A sufficient condition for $\Delta E> 0$ (Rule~\ref{thumb:positive:Delta:E}) is that $\cos\delta > 0$ for an unsuppressed volume of phase space.
\label{thumb:positive:cos:delta}
\end{corollary}
\begin{proof}
This follows from the positivity of the right-hand side of \eqref{eq:Delta:E:delta:psi:alpha}. Over the range of the polar angle $0\leq \psi \leq \pi$ the first term is non-negative. The second term is non-negative for $\cos\alpha <0$ which is available for half of the scattering phase space and so it is unsuppressed.
\end{proof}
The Easy Condition is a simpler diagnostic than the Positive Energy Transfer Condition. One only needs to check the latter condition when the former fails.
The following heuristic accounts for the possibility of positive energy transfer subject to Pauli blocking:

\begin{thumb}[Hard Condition]
The phase space for the initial target momentum must be large enough that after scattering, the outgoing target has momentum larger than its Fermi momentum. The necessary condition to diagnose this is based on the maximum kinematically allowed energy transfer, $\Delta E_\text{max}$:
\begin{align}
  p + \Delta E_\text{max} > p_\text{F} \ .
  \label{thumb:strong:p:DeltaEmax:pF}
\end{align}
\label{thumb:strong:condition:pauli:block}
\end{thumb}

\begin{figure}[tb]
  \centering
  \includegraphics[width=\textwidth]{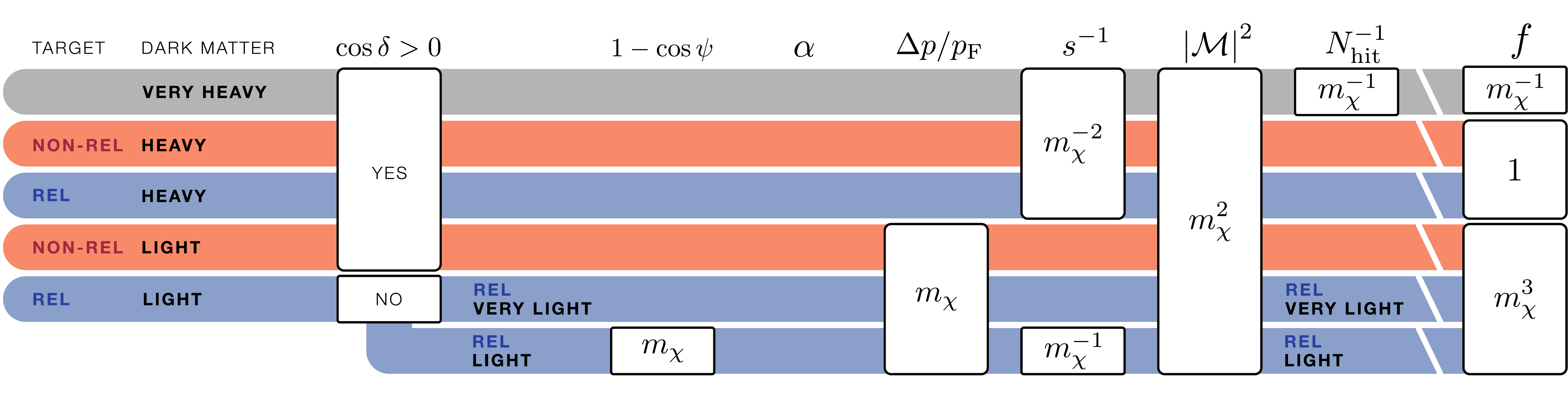}
  \caption{
  Flow chart showing the $m_\chi$ suppression of the capture efficiency, $f$,
  for target and dark matter cases defined in Table~\ref{fig:app:phaseblock:phasespace}.
  These scalings follow from the $\Delta E$ conditions defined in this appendix.
  The $\cos\delta>0$ diagnoses the Easy Condition, wheras the $\Delta p/p_\text{F}$ factors come from the Hard Condition. 
  We assume the baseline scaling of $|\mathcal M|^2$ in \eqref{eq:app:dominant:terms:baseline:M2}. Exceptional scalings are shown in Appendix~\ref{sec:DominantContact}.
  This table extends Table~\ref{fig:app:phaseblock:short:flow:chart} according to the detailed analysis in this appendix. 
  }
  \label{fig:app:phaseblock:flow:chart}
\end{figure}

We use the Independent Integration Assumption (Rule~\ref{thumb:independent:integration}) to determine the $m_\chi$ scaling of phase space based on the Easy and the Hard Conditions. We use the Easy Condition to determine the scaling of the angular phase space variables and the Hard Condition ot determine the scaling of the radial phase space variable. We proceed as follows:
\begin{enumerate}
\item Check if the Easy Condition holds; use this to determine the $m_\chi$ suppression of the angular phase space variables. Below we show that passing the Easy Condition gives no $m_\chi$ suppression. Failing the Easy Condition requires one to check the Positive Energy Transfer Condition; this imposes $m_\chi$ suppression through the $\cos\psi$ integration.

\item Check the conditions for which the Hard Condition holds; use this to determine the $m_\chi$ suppression from the integration of the target three-momentum magnitude.
\end{enumerate}
This process is shown in the flow chart in Figure~\ref{fig:app:phaseblock:flow:chart}. 
Though these heuristic arguments are approximations, they accurately capture the scaling behavior of our numerical results.

\subsection{\texorpdfstring{$\cos\delta$}{cosdelta} and the Easy Condition}

Factoring out overall positive factors in \eqref{eq:beta.kcm}, we find that the condition for $\cos\delta>0$ is
\begin{align}
  E_p(k^2 + \vec{p}\cdot\vec{k})
  >
  E_k(p^2 + \vec{p}\cdot\vec{k}) \ .
\end{align}
We square both sides and use the kinematic relations (the energy and momentum of dark matter at the point of impact):
\begin{align}
  E_p^2 &= m_\text{T}^2 + p^2
  &
  E_k^2 &= \gamma_\esc^2 m_\chi^2
  &
  k^2 &= \gamma_\esc^2 v_\esc^2 m_\chi^2 
  \label{eq:app:phasespace:Ep:Ek:k:expressions}
\end{align}
to distill Corollary~\ref{thumb:positive:cos:delta} to the following: phase space is unsuppressed when
  \begin{align}
    \left(m_\text{T}^2 + p^2\right)
    \left( 
      \gamma_\esc^2 v_\esc^2 m_\chi^2 
      + p\gamma_\esc v_\esc m_\chi\cos\theta
    \right)^2
    >
    \gamma_\esc^2m_\chi^2
    \left(
      p^2 
      + p \gamma_\esc v_\esc m_\chi \cos\theta
    \right)^2 \ .
    \label{eq:cos:delta:positive:cos:theta}
  \end{align}
  When this condition is not satisfied, we diagnose the phase space effects in more detail.
We examine \eqref{eq:cos:delta:positive:cos:theta} in each of the regimes in Figure~\ref{fig:app:phaseblock:phasespace}.

\subsection{\texorpdfstring{$\cos\theta$}{Cos Theta} Volume}
\label{app:phasespace:cos:theta}

The expression \eqref{eq:cos:delta:positive:cos:theta} for the Easy Condition, Corollary~\ref{thumb:positive:cos:delta}, is an inequality that must be satisfied by $\cos\theta$. When this range overlaps with the constraint $|\cos\theta| \leq 1$, we assume that $\cos\theta$ is unconstrained. 

\begin{figure}[tb]
  \centering 
  \includegraphics[width=\textwidth]{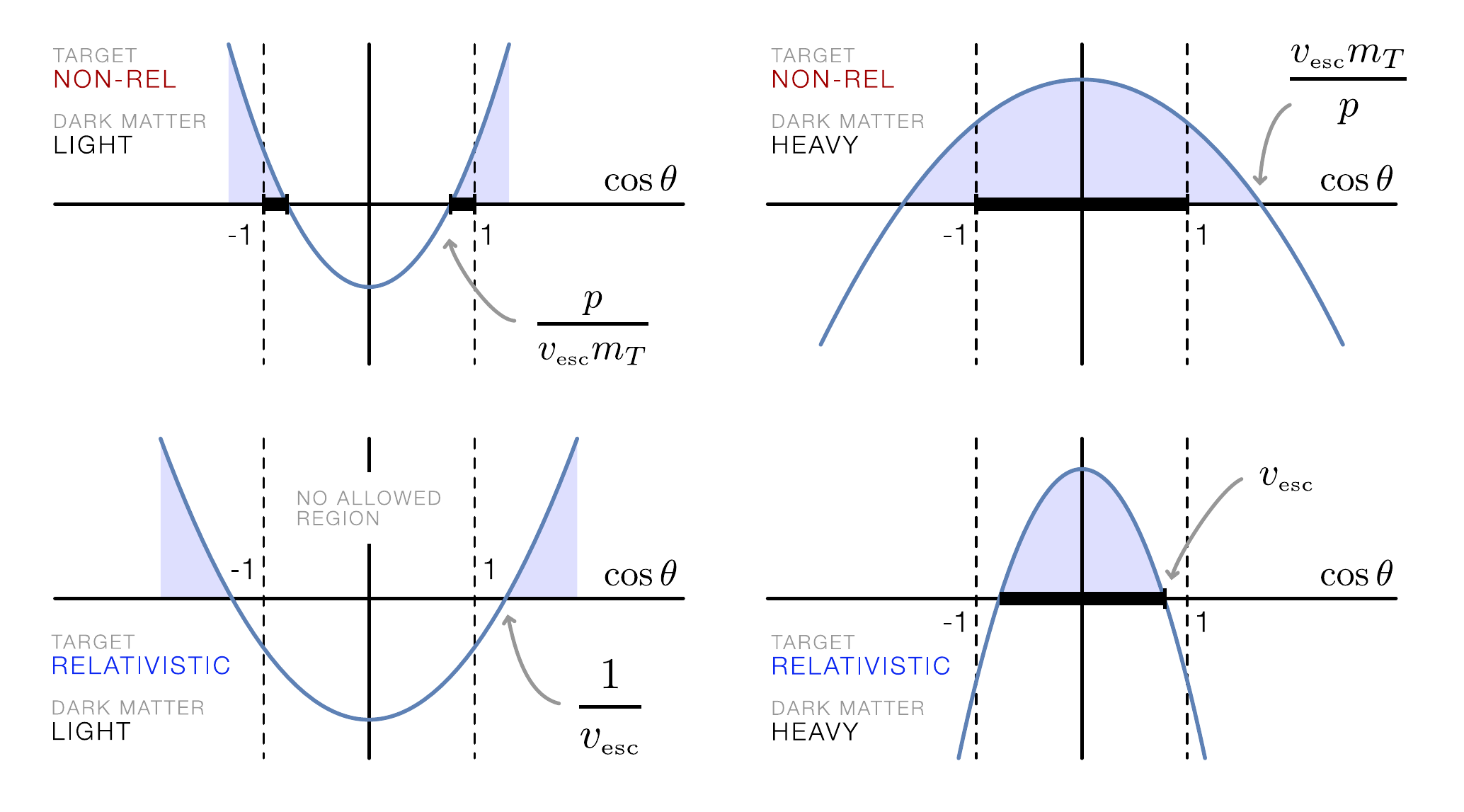}
  \caption{
  The Easy Condition \eqref{eq:cos:delta:positive:cos:theta} applied to the cases in Figure~\ref{fig:app:phaseblock:phasespace}. Parabolas correspond to \eqref{eq:theta:blocking:rel:target:light:DM:costheta:ineq}, \eqref{eq:theta:blocking:rel:target:heavy:DM:costheta:ineq}, \eqref{eq:theta:blocking:NR:target:light:DM:costheta:ineq}, and \eqref{eq:theta:blocking:NR:target:heavy:DM:costheta:ineq} as appropriate. Shaded regions are allowed by the condition, and thick black lines contained in these regions indicate the allowed range subject to $\cos\theta \in [-1,1]$. The case of a relativistic target with light dark matter (lower left) is {seen to be incompatible} with the Easy Condition. 
  }
  \label{fig:app:phaseblock:costheta}
\end{figure}

We show that relativistic targets with light and very light dark matter do not simultaneously satisfy \eqref{eq:cos:delta:positive:cos:theta} and $|\cos\theta|\leq 1$. This implies that the Easy Condition is not satisfied in that case and $\cos\delta < 0$. This constraint, in turn, feeds into the $\cos\psi$ conditions require to satisfy the Positive Energy Transfer Condition when $\cos\delta <0$. All other cases are unconstrained. These results are visualized in Figure~\ref{fig:app:phaseblock:costheta}.

\subsubsection*{Non-Relativistic Target, Heavy Dark Matter}

In this regime, $m_\text{T} > p_\text{F}$ and $m_\chi\gg  m_\text{T}$. 
The Easy Condition, \eqref{eq:cos:delta:positive:cos:theta}, reduces to
\begin{align}
  -  \left(\gamma_\esc^2 p^2 m_\chi^2\right) \cos^2\theta
  + \left(2 \gamma_\esc v_\esc p\, m_\text{T}^2 m_\chi\right) \cos\theta
  + \gamma_\esc^2 v_\esc^2 m_\text{T}^2 m_\chi^2
  & > 0 \ .
  \label{eq:theta:blocking:NR:target:heavy:DM:costheta:ineq}
\end{align}
Solving for the critical points of the inequality in this limit gives
\begin{align}
  -\frac{m_\text{T}v_\esc}{p} 
  &\lesssim
  \cos\theta
  \lesssim
  \frac{m_\text{T}v_\esc}{p}
  \ .
\end{align}
For the non-relativistic targets we consider, $m_\text{T}v_\esc \gtrsim p_\text{F}$, so that the Easy Condition implies that the entire range $-1\leq\cos\theta\leq 1$ admits scattering with $\Delta E >0$. Thus there is no suppression in the $\cos\theta$ volume that admits capture.

\subsubsection*{Non-Relativistic Target, Light Dark Matter}

In this regime, $m_\text{T} > p_\text{F}$ and $m_\chi \ll  m_\text{T}$. 
The Easy Condition, \eqref{eq:cos:delta:positive:cos:theta}, reduces to
\begin{align}
  \left(v_\esc^2 p^2 m_\text{T}^2\right) \cos^2\theta
  + \left(2 \gamma_\esc v_\esc^3 p\, m_\text{T}^2 m_\chi\right) \cos\theta
  - p^4
  & > 0 \ .
  \label{eq:theta:blocking:NR:target:light:DM:costheta:ineq}
\end{align}
Solving for the critical points of the inequality in this limit gives
\begin{align}
  -\frac{p}{v_\esc m_\text{T}} 
  \lesssim
    \cos\theta + \frac{\gamma_\esc v_\esc m_\chi}{p}
  & 
  \lesssim
  \frac{p}{v_\esc m_\text{T}} 
  \ .
\end{align}
For light dark matter with non-relativistic targets, the $m_\chi/p$ term is negligible compared to $p/m_\text{T}$. This means that the Easy Condition (Corollary~\ref{thumb:positive:cos:delta}) implies that almost entire range $-1\leq\cos\theta\leq 1$ admits scattering with $\Delta E >0$. Thus, there is almost no suppression in the $\cos\theta$ volume that admits capture due to $v_\esc m_\text{T}/p$ being a small $\mathcal{O}(1)$ factor.

\subsubsection*{Relativistic Target, Heavy Dark Matter}

In this regime, $m_\text{T} < p_\text{F}$ and $m_\chi \gg  p_\text{F}$. 
The Easy Condition, \eqref{eq:cos:delta:positive:cos:theta}, reduces to
\begin{align}
  -\left(\gamma_\esc^3 v_\esc p^2 m_\chi^2\right) \cos^2\theta
  - \left(2 p^3 m_\chi\right) \cos\theta
  - \gamma_\esc^3 v_\esc^3 p^2 m_\chi^2
  & > 0 \ .
  \label{eq:theta:blocking:rel:target:heavy:DM:costheta:ineq}
\end{align}
Solving for the critical points of the inequality in this limit, neglecting the powers of $p/m_\chi$, gives
\begin{align}
  -v_\esc
  &
  \lesssim \cos\theta 
  \lesssim
  v_\esc
  \ ,
\end{align}
Because $v_\esc < 1$, the $\Delta E >0$ condition shrinks the allowed range of $\cos\theta$. However, the neutron star’s gravitational acceleration is so large that this is only a modest suppression of the $\cos\theta$ volume by a factor of $v_\esc^{-1}\sim 1.7$.
For the purposes of understanding the mass scaling of the capture phase space, this suppression is negligible.

\subsubsection*{Relativistic Target, Light or Very Light Dark Matter}
\label{app:phasespace:cos:theta:relativistic:light:dark:matter}

In this regime, $m_\text{T} < p_\text{F}$ and $m_\chi \ll  m_\text{T}$. 
The Easy Condition, \eqref{eq:cos:delta:positive:cos:theta}, reduces to
\begin{align}
  \left(\gamma_\esc v_\esc^4 p^4 \right) \cos^2\theta
  - \left(2 v_\esc p^3 m_\chi\right) \cos\theta
  - \gamma_\esc p^4
  & > 0 \ .
  \label{eq:theta:blocking:rel:target:light:DM:costheta:ineq}
\end{align}
Solving for the critical points of the inequality in this limit gives
\begin{align}
  \cos\theta<-\frac{1}{v_\esc} \quad
  \text{or}\quad
    \frac{1}{v_\esc} <\cos\theta,
\end{align}
where we have used $p\approx p_\text{F}$ to neglect powers of $m_\chi/p$.
This is qualitatively different from the above cases. Because $v_\esc^{-1}>1$, there is \emph{no value of $\cos\theta$ that satisfies $\cos\delta>0$}.
For positive energy transfer in this case, \eqref{eq:Delta:E:delta:psi:alpha} requires a configuration with $\cos\delta < 0$. Diagnosing the phase space suppression requires further diagnostics to understand the allowed phase space for capture. Below we show that this maps onto a bound on the $\cos\psi$ volume.

\subsection{\texorpdfstring{$\cos\psi$}{CosPsi} Volume}
\label{app:phasespace:PsiBlock}

\subsubsection*{Cases Satisfying Easy Condition}

The cases in Appendix~\ref{app:phasespace:cos:theta} for which the Easy Condition ($\cos\delta > 0$) is compatible with $-1\leq \cos\theta\leq 1$ have no additional suppression from the integration over center of momentum scattering angle, $\cos\psi$.
This is clear from examining \eqref{eq:Delta:E:delta:psi:alpha} and recalling that $\psi$ is a polar angle with range $0\leq \psi \leq \pi$. Observe that every term on the right-hand side is positive when $\cos\delta>0$ and $\cos\alpha <1$. Thus there is at least an $\mathcal O(1)$ part of phase space that remains unsuppressed by the Easy Condition.

\subsubsection*{Relativistic Target, Light or Very Light Dark Matter}
\label{app:phase:block:rel:light}

For \emph{relativistic targets with dark matter lighter than the Fermi momentum}, on the other hand, Appendix~\ref{app:phasespace:cos:theta:relativistic:light:dark:matter} shows that $\cos \delta$ must be negative.
The Positive Energy Transfer Condition is not obviously satisfied and may impose further phase space suppression through \eqref{eq:Delta:E:delta:psi:alpha}. Applying trigonometric half-angle formulas, requiring $\Delta E >0$ imposes
\begin{align}
  \cos\delta \sin^2\frac{\psi}{2} > |\sin\delta| \cos\alpha\, \cos\frac{\psi}{2} \, \sin \frac{\psi}{2} \ .
\end{align}
This can be written---minding the sign of $\cos\delta$---as a condition on $\tan\psi/2$:
\begin{align}
  \tan \frac{\psi}{2} < \frac{\sqrt{1-\cos^2\delta}}{\cos\delta} 
  \cos\alpha
  \ .
  \label{eq:app:phasespace:psi:tan:psi:2:lessthan:cos:deltas}
\end{align}
Note that $\cos\frac{\psi}{2}\geq 0$, because $0\leq \psi\leq\pi$ by virtue of being a polar angle.

In this way the center of momentum polar angle $\psi$ is constrained by $\cos\delta = \bbeta\cdot\kcm/\beta k_\CM$, the projection of the neutron star-to-center of momentum boost $\bbeta$ and the center of momentum dark matter momentum $\kcm$. Note that both $\cos\alpha$ and $\cos\delta$ are negative so that the right-hand side is positive.

We may directly relate this to a bound on the phase space integral over $\cos\psi$. We make the simplifying approximation that the cross section $d\sigma/d\Omega_\CM$ does not introduce additional $\cos\psi$ dependence, as per Rule of Thumb~\ref{thumb:independent:integration}. The $\cos\psi$ phase space integral is thus
\begin{align}
  \int_{\cos\psi_\text{max}}^1 d\cos\psi 
  = 1 - \cos\psi_\text{max} 
  \approx 
  2\tan^2\frac{\psi_\text{max}}{2}
  \ ,
  \label{eq:app:phasespace:psi:tan:2:psi}
\end{align}
where we use the assumption that $\psi$ is small from \eqref{eq:app:phasespace:psi:tan:psi:2:lessthan:cos:deltas} so that $(1-\cos\psi) = {2} \sin^2\psi/2 \approx {2} \tan^2\psi/2$. Combining \eqref{eq:app:phasespace:psi:tan:psi:2:lessthan:cos:deltas} and \eqref{eq:app:phasespace:psi:tan:2:psi} gives a constraint on the volume of $d\cos\psi$ phase space:
\begin{align}
  \tan^2 \frac{\psi}{2} < \frac{{1-\cos^2\delta}}{\cos^2\delta} 
  \cos^2\alpha
  \ .
  \label{eq:phasespace:psi:tan2:bound}
\end{align}
This expression evaluates to
\begin{align}
  \tan^2 \frac{\psi}{2}
  &<
  \frac{
    v_\esc^2 \sin^2\theta \cos^2\alpha
  }{
    \left(
      1- 
      v_\esc \cos\theta
    \right)^2
  }
  \left[
      \frac{m_T^2}{p^2}
      +
      \frac{m_\chi}{p}
      \left(
        2\frac{m_\text{T}^2}{p^2}X
        + 2\gamma_\esc (1-v_\esc \cos\theta)
      \right)
      +\mathcal O\left(\frac{m_\chi^2}{p^2}\right)
    \right]
  \ .
  \label{eq:app:phasespace:psi:tan2psi:2:upper:limit:approx}
\end{align}
The key here is that the upper bound on $\tan^2\psi/2$ scales either independently of $m_\chi$ or linearly with $m_\chi$ depending on which term in the square brackets dominates. 
This, in turn, defines two sub-regimes:
\begin{itemize}

  \item \textbf{Very light dark matter}: When $m_\chi \lesssim m_\text{T}^2/p_\text{F}$, the $\mathcal O(m_\chi^0)$ term sets the bound in \eqref{eq:app:phasespace:psi:tan2psi:2:upper:limit:approx}. In this case the phase space suppression is independent of $m_\chi$.

  \item \textbf{Light dark matter}: When $m_\text{T}^2/p_\text{F} \lesssim m_\chi \lesssim p_F$, the $\mathcal O(m_\chi^1)$ term sets the bound in \eqref{eq:app:phasespace:psi:tan2psi:2:upper:limit:approx}. In this case the volume of the $\cos\psi$ phase space scales with $m_\chi$.
\end{itemize}
In the above, we have made use of the fact that $p\approx p_\text{F}$ for light and very light dark matter due to Pauli blocking. The remainder of this sub-section derives \eqref{eq:app:phasespace:psi:tan2psi:2:upper:limit:approx}.
%
%
\begin{proof}
We evaluate the right-hand side of \eqref{eq:phasespace:psi:tan2:bound}. As an intermediate step, write $\cos\delta$ in terms of a ratio using \eqref{eq:beta.kcm}:
\begin{align}
  \cos\delta 
  = 
  \frac{E_p k^2 - E_k p^2 + (E_p-E_k)\vec{p}\cdot\vec{k}}{E\beta E_\CM k_\CM} 
  \equiv 
  \frac{B}{A}
  \ .
  \label{eq:phasespace:psi:cos:delta:A:B:def}
\end{align}
Then the right-hand side of \eqref{eq:phasespace:psi:tan2:bound} is 
\begin{align}
  \frac{{1-\cos^2\delta}}{\cos^2\delta} 
  &=
  \frac{A^2 - B^2}{B^2}
  \ .
  \label{eq:app:phasespace:psi:cos:delta:A2-B2:B2}
\end{align}
$A^2$ is written using the expression for $E_\CM^2 k_\CM^2$ from \eqref{eq:kCM:2:Ecm} and $E^2 \beta^2 = (\vec{p}+\vec{k})^2$ from \eqref{eq:beta:p:k:gamma}.
The numerator of \eqref{eq:app:phasespace:psi:cos:delta:A2-B2:B2} greatly simplifies to 
\begin{align}
  A^2 - B^2 = k^2 p^2 E_\CM^2 \sin^2\theta \ ,
\end{align}
where $E_\CM^2$ is simply the Mandelstam $s$ parameter, \eqref{eqn:s}. 
The full expression for the upper bound in \eqref{eq:app:phasespace:psi:cos:delta:A2-B2:B2} is
\begin{align}
  \tan^2 \frac{\psi}{2}
  &<
  \frac{
    k^2p^2 \sin^2 \theta 
    \left(m_\text{T}^2 + m_\chi^2 + 2E_pE_k - 2pk\cos\theta\right)
  }{
    \left[
      E_p k^2 - E_k p^2 + (E_p-E_k)
      pk\cos\theta
    \right]^2
  }
  \cos^2\alpha
  \ .
  \label{eq:app:phasespace:psi:tan2psi:2:upper:limit:full}
\end{align}

Assuming that the characteristic target momentum is the Fermi momentum, $p\sim p_\text{F}$, the denominator can be expanded with respect to the $m_\chi \ll p_\text{F}$ regime:
\begin{align}
  B 
  &= 
  -\gamma_\esc m_\chi
  \left[
    \left(p^2 - E_p p v_\esc \cos\theta \right)
    -
    \gamma_\esc v_\esc m_\chi
    \left(
      E_p v_\esc - p \cos\theta
    \right)
  \right]
  \\
  &=
  -\gamma_\esc m_\chi p^2 \left[
   \left(1-\frac{E_pv_\esc}{p}\cos\theta\right)
   - \frac{m_\chi}{p}X
   \right]
  \ ,
\end{align}
where the $X$ term is higher order in $m_\chi/p$.
Plugging in the quantities \eqref{eq:app:phasespace:Ep:Ek:k:expressions}, using $E_p\approx p$ as the target is relativistic, and expanding to $\mathcal O(m_\chi/p)$ gives \eqref{eq:app:phasespace:psi:tan2psi:2:upper:limit:approx}.
\end{proof}

\subsection{\texorpdfstring{$\alpha$}{alpha} Volume}\label{app:phasespace:AlphaBlock}

The center of momentum frame azimuthal angle, $\alpha$, does not affect the $m_\chi$ scaling of the capture efficiency.

\subsubsection*{Cases Satisfying Easy Condition}

The cases in Appendix~\ref{app:phasespace:cos:theta} that pass the Easy Condition ($\cos\delta > 0$) carry no additional suppression from the center of momentum frame azimuthal angle phase space, $\alpha$.
This is clear from \eqref{eq:Delta:E:delta:psi:alpha} where every term on the right-hand side is positive when $\cos\alpha <1$. Thus there is at least an half of the $\alpha$ phase space that remains unsuppressed by the Easy Condition, Rule~\ref{thumb:positive:Delta:E}.

\subsubsection*{Relativistic Target, Light or Very Light Dark Matter}

\emph{Relativistic targets with light or very light dark matter} do not satisfy the Easy Condition. This leads to a bound on the $\cos\psi$ phase space \eqref{eq:app:phasespace:psi:tan2psi:2:upper:limit:approx} that depends on $\cos^2\alpha$. However, because the dependence is an overall prefactor that is independent of the dark matter mass, there is no additional $m_\chi$-dependent suppression in the $\alpha$ phase space.

\subsection{\texorpdfstring{$p$}{p} Volume and Maximum Energy Transfer}
\label{sec:app:max:DeltaE:wrt:psi:alpha}

For the initial momentum phase space we invoke Rule~\ref{thumb:strong:p:DeltaEmax:pF} and account for the Pauli blocking.
The phase space volume is
\begin{align}
  \int_{p_\text{min}}^{p_\text{F}}
  \frac{p^2 dp}{\frac{1}{3}p_\text{F}^3}
  =
  \frac{p^3_\text{F} - p^3_\textnormal{min}}{p^3_{\rm F}}
  &\approx
  \frac{{3}\Delta p}{p_\text{F}}\sim \frac{3E_\text{F}\Delta E}{p_\text{F}^2}
  \approx \begin{cases}
  \frac{{3}\Delta E}{p_\text{F}}\left(1+\frac{m_{\rm T}^2}{2 p_{\rm F}^2}\right) & m_{\rm T} \ll p_\text{F}
  \\
  \frac{{3}m_{\rm T}\Delta E}{p_\text{F}^2}\left(1+\frac{p_{\rm F}^2}{2 m_{\rm T}^2}\right)    &  m_{\rm T} \gg p_\text{F}
  \end{cases}
  \label{eqn:p:phase}
\end{align}
Thus, the scaling of this allowed phase space volume is determined by the scaling of the maximum allowed $\Delta E$.
The results are shown in the relevant column of Figure~\ref{fig:app:phaseblock:flow:chart}.
One may carry over intuition from non-relativistic scattering where the kinematics depends on the reduced mass. In the relativistic case, one may replace the the target mass with its characteristic Fermi momentum.

We maximize $\Delta E$ with respect to the center of momentum frame scattering polar angle $\psi$ and azimuthal angle $\alpha$. For simplicity we define a positive rescaling of $\Delta E$ that shares the same extrema,
\begin{align}
	\Delta \mathcal E = \frac{\Delta E}{\gamma \beta k_\CM} \ .
\end{align}
The expression for $\Delta E$ in \eqref{eq:Delta:E:delta:psi:alpha} is thus
\begin{align}
  \Delta \mathcal E &= \cos \delta (1-\cos \psi) - |\sin\delta| \cos\alpha \sin \psi \ .
  \label{eq:app:max:Delta:E:Delta:Curly:E}
\end{align}
Because $|\sin\delta|\sin\psi \geq 0$, the term in $\Delta \mathcal E $ that depends on $\alpha$ is maximized when $\cos\alpha$ is as negative as possible. 
Thus it is clear that the maximum of $\Delta \mathcal E$ occurs for $\cos\alpha = -1$.
Using $\partial\Delta \mathcal E/\partial \psi=0$, the critical point for $\Delta \mathcal E$ with respect to $\psi$ is
\begin{align}
  \tan \psi &=
  \frac{|\sin\delta|}{\cos\delta} 
  \cos\alpha
   = - \frac{|\sin\delta|}{\cos\delta}
  \ .
\end{align}
This is a maximum because $\Delta \mathcal E$ is written as a linear combination of eigenfunctions of $\partial^2/\partial\psi^2$ with non-positive eigenvalues (constants and trigonometric functions). Thus we are guaranteed to have $\partial^2\Delta \mathcal E/\partial\psi^2 \leq 0$.

We may succinctly write the conditions for the maximum energy transfer as
\begin{align}
  \cos\alpha &= -1
  &
  \cos\psi &= -\cos\delta
  &
  \sin\psi &= |\sin\delta| = \sqrt{1-\cos^2\delta} \ ,
\end{align}
{where} we have used the range $0\leq \psi \leq \pi$ to assign the negative sign to $\cos\psi$.
With this result, the maximum energy transfer, $\Delta E_\text{max}$, is
\begin{align}
  \frac{\Delta E_\text{max}}{\gamma\beta k_\CM} 
  &= 
  \cos\delta (1+\cos\delta) + \sin^2\delta
  = \cos\delta + 1 \ .
  \label{eq:app:max:Delta:E:wrt:cos:delta}
\end{align}
The expression for $\cos\delta$ is presented in \eqref{eq:phasespace:psi:cos:delta:A:B:def}. 
We present approximations for this expression for the limiting cases of interest.

\subsubsection*{Non-Relativistic Targets, Heavy Dark Matter}

We assume $m_\text{T} \gg p$ and $m_\chi \gg m_\text{T}$. In this limit, \eqref{eq:app:max:Delta:E:wrt:cos:delta} gives
\begin{align}
  \Delta E_\text{max}
  &\approx
  \frac{
    4\gamma_\esc^2\left(m_\text{T} v_\esc - p \cos\theta\right)^2
    + p^2 \sin^2 \theta
  }{
    2\left(m_\text{T} - \frac{p}{v_\esc}\cos\theta\right)
  } \ .
  \label{eqn:DeltaE:nonrel:heavy}
\end{align}
By substituting this into~\eqref{thumb:strong:p:DeltaEmax:pF}, one can solve for minimum allowed target momentum by Pauli exclusion principle. This, in turn, gives the maximum fraction of the $p$ volume that is not Pauli blocked. This process is independent of $m_\chi$ because $\Delta E_\text{max}$ is independent of $m_\chi$.

As a check, in the non-relativistic limit, $p\to 0$ and $E_p\to m_\text{T}$, this expression reduces to the well known result \eqref{eq:app:NR:Delta:E}, maximized over $\psi$, $\Delta E_\text{max} \to 2m_\text{T}\gamma_\esc^2 v_\esc^2$.

\subsubsection*{Non-Relativistic Targets, Light Dark Matter}

We assume $m_\text{T} \gg p$ and $m_\chi \ll m_\text{T}$. In this limit, \eqref{eq:app:max:Delta:E:wrt:cos:delta} gives
\begin{align}
  \Delta E_\text{max}
  &\approx
  \frac{p\gamma_\esc  m_\chi}{m^2_\text{T}}
  \left[
    (m_\text{T} v_\esc - p)
    (1+\cos\theta)
    + 
    \frac{p^2 \sin^2\theta}{2\gamma_\esc^2 (m_\text{T}v_\esc - p \cos\theta)}
  \right] 
  \ .
  \label{eqn:DeltaE:nonrel:light}
\end{align}
Maximizing over the target orientation $\theta$ gives
\begin{align}
  \Delta E_\text{max}
  &\approx
  \frac{2p v_\esc}{m_\text{T}}
  \left(1 - \frac{p}{m_\text{T}v_\esc}\right) \gamma_\esc m_\chi
  \ .
\end{align}
We thus have $\Delta E_\text{max}\propto m_\chi$.

\subsubsection*{Relativistic Targets, Heavy Dark Matter}

We assume $m_\text{T} \ll p_\text{F}$ and $m_\chi \gg p_\text{F}$. In this limit, \eqref{eq:app:max:Delta:E:wrt:cos:delta} gives
\begin{align}
  \Delta E_\text{max}
  &\approx
  p\gamma_\esc^2 v_\esc (1-v_\esc \cos\theta)(v_\esc+1)
  \ .
  \label{eqn:DeltaE:rel:heavy}
\end{align}
Thus, $\Delta E_\text{max}$ is independent of $m_\chi$.

\subsubsection*{Relativistic Targets, Light Dark Matter}

We assume that the target is light: $p_\text{F} \gg m_\chi \gg m_\text{T}^2/p_\text{F}$. In this limit, \eqref{eq:app:max:Delta:E:wrt:cos:delta} gives
\begin{align}
  \Delta E_\text{max}
  &\approx
  \frac{v_{esc}^2\sin^2\theta}{2}\cdot\frac{\gamma_{\rm esc}m_\chi}{1-v_{\rm esc}\cos\theta}
  \ .
  \label{eqn:DeltaE:rel:light}
\end{align}
We thus have $\Delta E_\text{max}\propto m_\chi$ for $\vec{p}$ near the Fermi surface.

\subsubsection*{Relativistic Targets, Very Light Dark Matter}

We assume a very light target, $p_\text{F} \gg m_\text{T} \gg m_\text{T}^2/p_\text{F} \gg m_\chi$, which implies $m_\chi/p_{\rm F}\ll m_{\rm T}^2/p_{\rm F}^2$. Since in this case all of the interactions occur very close to Fermi surface, $E_{\rm F}\sim p_{\rm F}$ is still true, but the leading order difference between $E_{\rm F}$ and $p_{\rm F}$ is $\mathcal{O}\left(m_{\rm T}^2/p_{\rm F}^2\right)$, which is significant with respect to $m_\chi/p_{\rm F}$. Taylor expanding~\eqref{eq:app:max:Delta:E:wrt:cos:delta} with respect to both $m_\chi/p_{\rm F}$ and $m_{\rm T}^2/p_{\rm F}^2$ gives
\begin{align}
  \Delta E_\text{max}
  &\approx
  \frac{v_{\rm esc}^2\sin^2\theta}{2}\cdot\frac{\gamma_{\rm esc}m_\chi}{1-v_{\rm esc}\cos\theta}
  \ .
  \label{eqn:DeltaE:rel:vlight}
\end{align}
We thus have $\Delta E_\text{max}\propto m_\chi$ for $\vec{p}$ near the Fermi surface.

\subsection{Dominant Terms in Contact Operators}
\label{sec:DominantContact}

This appendix derives the scaling of the squared amplitude, $|\mathcal M|^2$ for the operators in Table~\ref{tab:amplitudes12}. 
We derive the benchmark scaling \eqref{eq:app:dominant:terms:baseline:M2} followed by most of the operators as well as the scaling for the exceptional operators. For the latter, we show how the flow chart in Figure~\ref{fig:app:phaseblock:flow:chart} is modified according to the dynamics of each case.

We tabulate the $m_\chi$ and $E_p$ scalings of Mandelstam variables $s$ and $t$ in Table~\ref{tab:sandt}. Use the limiting behavior of the Mandelstam $s$ variable in~\eqref{eq:qualitative:s:scaling}. The scalings of $t$ with respect to $m_\chi$ can be derived from~\eqref{eqn:t} in conjuction with~\eqref{eq:kCM:2:Ecm} and the results of Appendix~\ref{app:phasespace:PsiBlock}, which derive the scaling of $(1-\cos\psi)$ with $m_\chi/p_{\rm F}$.

\begin{table}[H]
  \centering
  \begin{tabular}{p{1.6cm}p{2.6cm}p{2.7cm}p{1.9cm}p{1.9cm}p{1.9cm}p{1.9cm}}  
    \toprule
    $m_\text{T}$
    & 
    \multicolumn{2}{l}{Non-Relativistic}
    & 
    \multicolumn{3}{l}{Relativistic}
    \\
    \cmidrule(r){1-1}\cmidrule(r){2-3}\cmidrule(r){4-7}
    $m_\chi$
    & 
    {\small Heavy}
    & 
    {\small Light}
    & 
    {\small Heavy}
    & 
    {\small Light-ish}
    &
    {\small Med.~Light}
    & 
    {\small Very~Light}
    \\
    \midrule
    $s$
    &
    $m_\chi^2$
    &
    $m_\text{T}^2$
    &
    $m_\chi^2$
    &
    $m_\chi p_\text{F}$
    &
    $m_\chi p_\text{F}$
    &
    $m_\text{T}^2$
    \\
    $-t$
    &
    $m_\text{T}^2$
    &
    $m_\chi^2$
    &
    $p_\text{F}^2$
    &
    $m_\chi^2$
    &
    $m_\chi^2$
    &
    $
    m_\chi^2
    $
    \\
    \bottomrule
  \end{tabular}
  \caption{
    Mandelstam variable $s$ and $t$ scalings with respect to $m_\chi$, $p_{\rm F}$ and $m_{\rm T}$. The scaling corresponds to each kinematic regime defined in Section~\ref{sec:kinematic:regimes}.}
  \label{tab:sandt}
\end{table}

The $t$ scaling for very light dark matter case requires special attention. In this regime, $m_\chi/p_{\rm F}\ll m_{\rm T}^2/p_{\rm F}^2$. 
One may then Taylor expand right-hand side of~\eqref{eq:kCM:2:Ecm} in both $m_\chi/p$ and $m_{\rm T}^2/p^2$ for targets close to Fermi surface.
The leading order term is proportional to $m_\chi^2 m_{\rm T}^2$, using $s\sim m_{\rm T}^2$ as per~\eqref{eq:qualitative:s:scaling}. This gives $-t\propto m_\chi^2$.

\subsection{Fermionic Dark Matter Operators}

The operators $\mathcal O_{5-10}^\text{F}$ follow the same benchmark behavior. One may use~\eqref{eqn:s} to show
\begin{align}
	(m_{\rm T}^2+m_\chi^2)^2-2s(m_{\rm T}^2+m_\chi^2)+s^2=4\gamma_{\rm esc}^2m_\chi^2E_p^2\left(1-\frac{p}{E_p}v_{\rm esc}\cos\theta\right)^2
	\ .
	\label{eqn:operator_reduction}
\end{align}
Substituting this identity and the results of Table~\ref{tab:sandt} into the expressions for $|\mathcal M|^2$ in Table~\ref{tab:amplitudes12} shows that the $m_\chi^2E_p^2$ term dominates for each of the $\mathcal O_{5-10}^\text{F}$ operators, deriving the baseline behavior in~\eqref{eq:app:dominant:terms:baseline:M2}.

\begin{table}[H]
  \centering
  \begin{tabular}{p{1.6cm}p{2.6cm}p{2.7cm}p{1.9cm}p{1.9cm}p{1.9cm}p{1.9cm}}  
    \toprule
    $m_\text{T}$
    & 
    \multicolumn{2}{l}{Non-Relativistic}
    & 
    \multicolumn{3}{l}{Relativistic}
    \\
    \cmidrule(r){1-1}\cmidrule(r){2-3}\cmidrule(r){4-7}
    $m_\chi$
    & 
    {\small Heavy}
    & 
    {\small Light}
    & 
    {\small Heavy}
    & 
    {\small Light-ish}
    &
    {\small Med.~Light}
    & 
    {\small Very~Light}
    \\
    \midrule
    $\mathcal O_1^\text{F}$ 
    & 
    & 
    & 
    & 
    $m_\chi^4$
    & 
    & 
    \\
    $\mathcal O_2^\text{F}$ 
    & 
    $m_\text{T}^4$ 
    & 
    & 
    $p_\text{F}^4$
    & 
    $m_\chi^4$ 
    & 
    & 
    \\
    $\mathcal O_3^\text{F}$  
    & 
    & 
    $m_\chi^4$
    & 
    & 
    $m_\chi^4$
    & 
    $m_\chi^4$
    & 
    $\displaystyle
    m_\chi^4$
    \\
    $\mathcal O_4^\text{F}$ 
    & 
    $m_\text{T}^4$
    & 
    $m_\chi^4$
    & 
    $p_\text{F}^4$
    & 
    $m_\chi^4$
    & 
    $m_\chi^4$
    & 
    $\displaystyle
    m_\chi^4$
    \\
    \bottomrule
  \end{tabular}
  \caption{
    Fermionic operators with exceptional scaling in $\Lambda^4 | \mathcal M |^2$ compared to the baseline case $\Lambda^4 | \mathcal M |^2\propto m_\chi^2 E_p^2$, \eqref{eq:app:dominant:terms:baseline:M2}. The scaling corresponds to each kinematic regime defined in Section~\ref{sec:kinematic:regimes}. Blank entries correspond to the baseline scaling, $m_\chi^2 E_p^2$.}
  \label{tab:contact:ops:M2:exceptions}
\end{table}

\begin{figure}[tb]
  \centering
  \includegraphics[width=\textwidth]{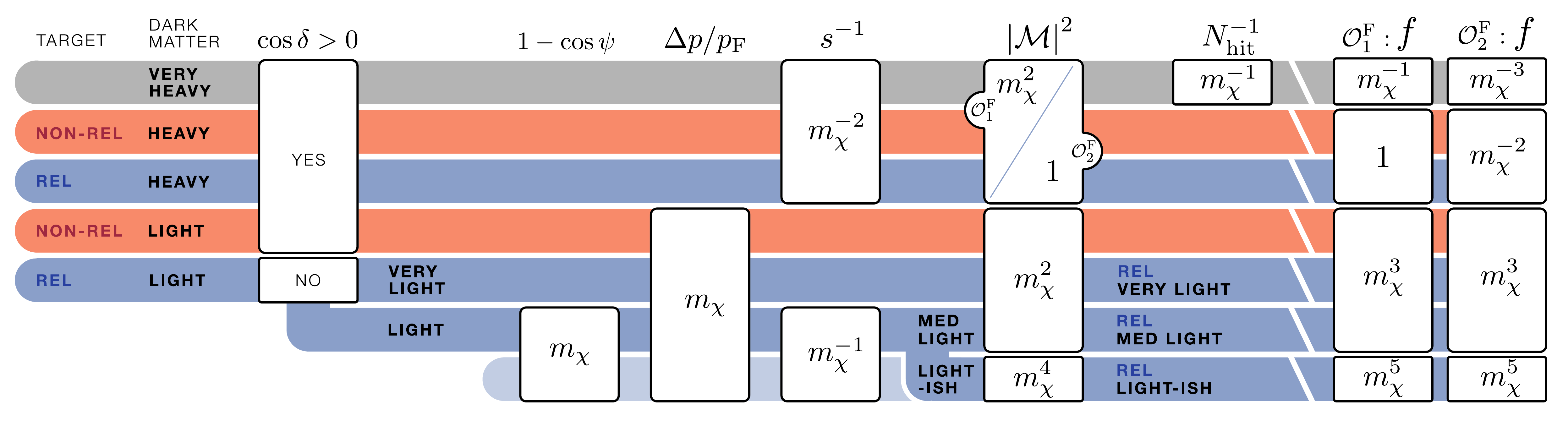}
  \caption{
  Extends Figure~\ref{fig:app:phaseblock:flow:chart} to account for the exceptional $m_\chi$ scaling of $\mathcal O^\textnormal{F}_{1,2}$. 
  }
  \label{fig:app:phaseblock:short:flow:chart:F:O1O2}
\end{figure}

\begin{figure}[tb]
  \centering
  \includegraphics[width=\textwidth]{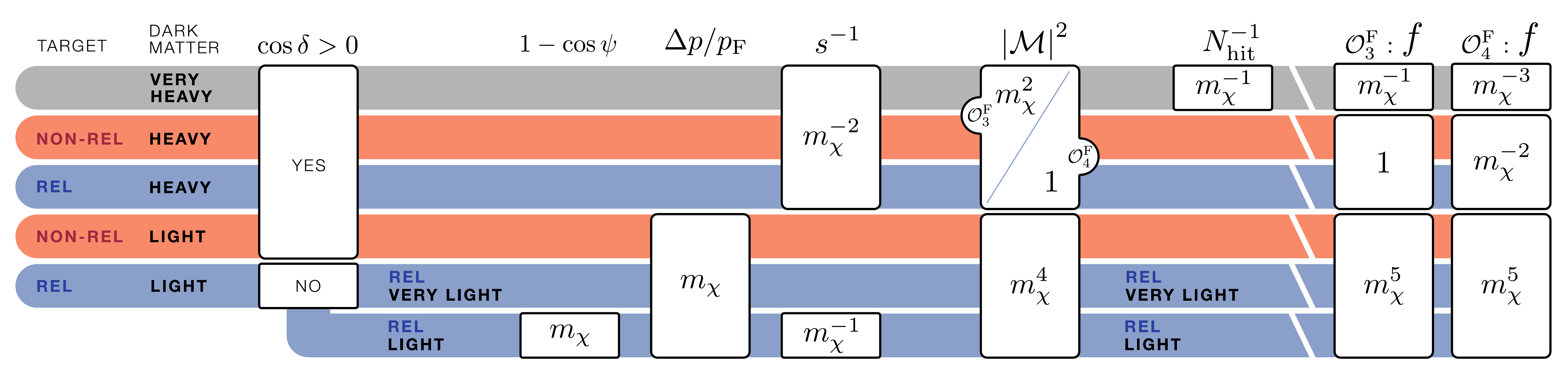}
  \caption{
  Extends Figure~\ref{fig:app:phaseblock:flow:chart} to account for the exceptional $m_\chi$ scaling of $\mathcal O^\textnormal{F}_{3,4}$.
  } 
  \label{fig:app:phaseblock:short:flow:chart:F:O3O4}
\end{figure}

The $\mathcal O_{1-4}^\text{F}$ operators deviate from~\eqref{eq:app:dominant:terms:baseline:M2} for certain dark matter masses. Table~\ref{tab:contact:ops:M2:exceptions} shows the scaling of $|\mathcal M|^2$ with respect to $m_\chi$ for these fermionic operators. These are derived from the standard results for $|\mathcal M|^2$ in 
Table~\ref{tab:amplitudes12} 
using the behavior of $s$ and $t$ in various DM regimes as tabulated in Table~\ref{tab:sandt}. This scaling can then be complete flow charts for the $m_\chi$ scaling of the capture efficiency, $f$, which we present in Figures~\ref{fig:app:phaseblock:short:flow:chart:F:O1O2} and \ref{fig:app:phaseblock:short:flow:chart:F:O3O4}.

\subsection{Scalar Dark Matter Operators}

Following the same analysis that we use for the fermionic dark matter operators, one observes that the scalar dark matter operators $\mathcal O^\text{S}_{3,4}$ follow the benchmark scaling $|\mathcal M|^2 \sim m_\chi^2 E_p^2$ for all dark matter masses.

\begin{figure}[tb]
  \centering
  \includegraphics[width=\textwidth]{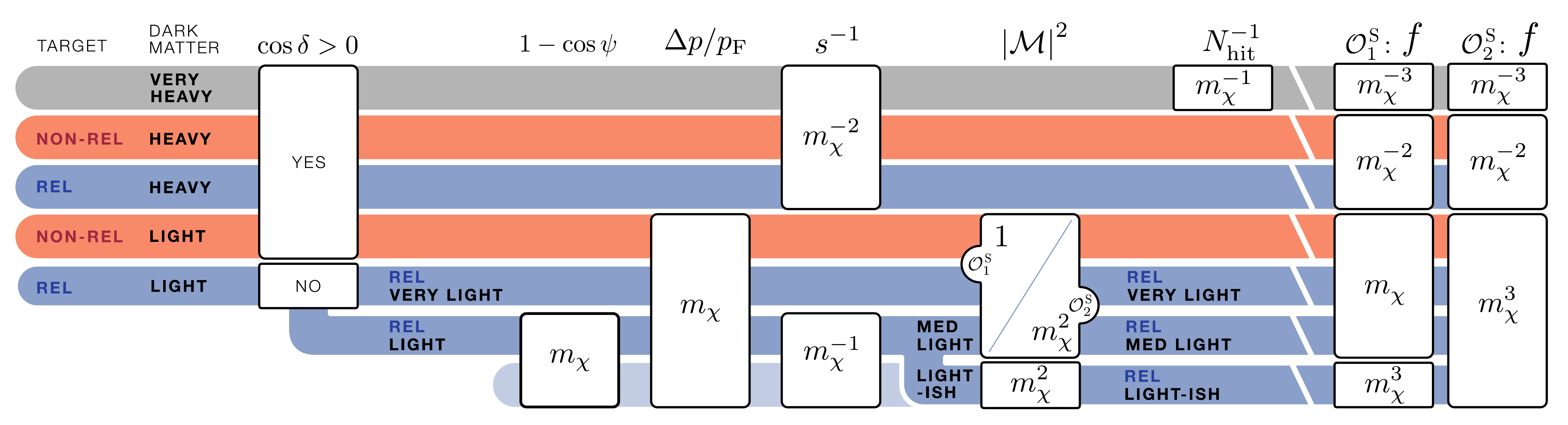}
  \caption{
  	Extends Figure~\ref{fig:app:phaseblock:flow:chart} to account for the exceptional $m_\chi$ scaling of $\mathcal O^\textnormal{S}_{1,2}$.} 
  \label{fig:app:phaseblock:short:flow:chart:B:O1O2}  
\end{figure}

The $\mathcal O^\text{S}_{1,2}$ scalar operators, on the other hand, demonstrate exceptional scaling. These are tabulated in Table~\ref{tab:contact:ops:M2:exceptions:bosons} and produce a flow chart of capture efficiency scaling shown in Figure~\ref{fig:app:phaseblock:short:flow:chart:B:O1O2}.

\begin{table}[H]
  \centering
  \begin{tabular}{p{1.6cm}p{2.6cm}p{2.7cm}p{1.9cm}p{1.9cm}p{1.9cm}p{1.9cm}}  
    \toprule
    $m_\text{T}$
    & 
    \multicolumn{2}{l}{Non-Relativistic}
    & 
    \multicolumn{3}{l}{Relativistic}
    \\
    \cmidrule(r){1-1}\cmidrule(r){2-3}\cmidrule(r){4-7}
    $m_\chi$
    & 
    {\small Heavy}
    & 
    {\small Light}
    & 
    {\small Heavy}
    & 
    {\small Light-ish}
    &
    {\small Med.~Light}
    & 
    {\small Very~Light}
    \\
    \midrule
    $\mathcal O^\textnormal{S}_1$ 
    & 
    $m_{\rm T}^2\Lambda^2$
    & 
    $m_{\rm T}^2\Lambda^2$
    & 
    $p_{\rm F}^2\Lambda^2$
    & 
    $m_\chi^2\Lambda^2$
    & 
    $m_{\rm T}^2\Lambda^2$
    & 
    $m_{\rm T}^2\Lambda^2$
    \\
    $\mathcal O^\textnormal{S}_2$ 
    & 
    $m_{\rm T}^2\Lambda^2$
    & 
    $m_\chi^2\Lambda^2$
    & 
    $p_\text{F}^2\Lambda^2$
    & 
    $m_\chi^2\Lambda^2$
    & 
    $m_\chi^2\Lambda^2$
    & 
    $m_\chi^2\Lambda^2$
    \\
    \bottomrule
  \end{tabular}
  \caption{
    Bosonic operators with exceptional scaling in $\Lambda^4 | \mathcal M |^2$ compared to the baseline case $\Lambda^4 | \mathcal M |^2\propto m_\chi^2 E_p^2$, \eqref{eq:app:dominant:terms:baseline:M2}. The scaling corresponds to each kinematic regime defined in Section~\ref{sec:kinematic:regimes}.}
  \label{tab:contact:ops:M2:exceptions:bosons}
\end{table}

\subsection{Capture Efficiency Scalings with Energy Scales}
\label{app:prob:scalings}

In order to determine the relative reach of scattering from different targets, 
we derive scaling of the capture efficiency with respect to all independent energy scales involved in dark matter capture by a neutron star: $m_\chi$, $m_{\rm T}$, $p_{\rm F}$ and $R_\star^{-1}$. This appendix extends the discussion in Section~\ref{sec:dependence:on:E:scales}.

The formula for the capture efficiency, $f$, is~\eqref{eq:f:dependence:on:mx}.
Following the discussion in Section~\ref{sec:CharFeatures}, we highlight the dependence of $f$ on the key energy scales of the system:
\begin{align}
  f &\propto 
  \frac{M_\star Y_{\rm T}}{m_nR_\star^2} 
  |\mathcal{M}|^2
  \int\limits_{p_{\rm min}}^{p_\text{F}}
  \frac{p^2 dp }{p_{\rm F}^3}
  \left(\frac{1}{s}\int\limits_{\cos\psi_0}^1 d\cos\psi\right)
  \label{eq:f:full:reduced}
\end{align}
where $Y_{\rm T}$ is the abundance and $p_{\rm F}$ is the Fermi momentum of a target species $T$. 
We may relate the total number of target particles $N$ to the Fermi momentum $p_\text{T}$ and the neutron star volume $V$ by
\begin{align}
  p_{F}&\propto\left(\frac{N}{V}\right)^{1/3}\quad\implies\quad N\propto p_{\rm F}^3R_\star^3,
\end{align}
where the volume of the neutron star is $V=\frac{4}{3}\pi R_\star^3$. From the definition of $Y_{\rm T}$, the total number of targets is $N=M_\star Y_{\rm T}/m_n$. This implies
\begin{align}
  Y_{\rm T}
  &\propto 
  \frac{p_{\rm F}^3R_\star^3\, m_n}{M_\star}
  \ .
  \label{eqn:yt}
\end{align}
Observe that this highlights that $M_\star$ is not an independent dimensionful parameter in the system.
Substituting \eqref{eqn:yt} into~\eqref{eq:f:full:reduced} gives
\begin{align}
  f&\propto \,p_{\rm F}^3R_\star
  |\mathcal{M}|^2
  \int\limits_{p_{\rm min}}^{p_\text{F}}
  \frac{p^2 dp }{p_{\rm F}^3}
  \left(\frac{1}{s}\int\limits_{\cos\psi_0}^1 d\cos\psi\right).\label{eqn:fullscaling}
\end{align}
We make the approximation where the integrals may be performed independently with trivial integrand; this captures the dominant scaling. 
The $p$ integral is then proportional to $\Delta p/p_{\rm F}$, as shown in~\eqref{eqn:p:phase}, and its scaling with respect to the energy scales can be evaluated in various phases of scattering kinematics of Figure~\ref{fig:app:phaseblock:phasespace} using the results of Appendix~\ref{sec:app:max:DeltaE:wrt:psi:alpha}.
Similarly, \eqref{eq:app:phasespace:psi:tan2psi:2:upper:limit:approx} gives the scaling of the $d\psi$ integral. We take the approximation that $E_p\sim m_{\rm T}$ in nonrelativistic limit and $E_p\sim p_{\rm F}$ in relativistic limit and tabulate the results in Table~\ref{tab:intpieces}.

\begin{table}[H]
  \centering
  \begin{tabular}{p{2.2cm}p{2cm}p{2.7cm}p{1.9cm}p{1.9cm}p{1.9cm}p{1.9cm}}  
    \toprule
    $m_\text{T}$
    & 
    \multicolumn{2}{l}{Non-Relativistic}
    & 
    \multicolumn{3}{l}{Relativistic}
    \\
    \cmidrule(r){1-1}\cmidrule(r){2-3}\cmidrule(r){4-7}
    $m_\chi$
    & 
    {\small Heavy}
    & 
    {\small Light}
    & 
    {\small Heavy}
    & 
    {\small Light-ish}
    &
    {\small Med.~Light}
    & 
    {\small Very~Light}
    \\
    \midrule
    $\Delta p/p_{\rm F}$
    &
    $m_{\rm T}^2/p_{\rm F}^2$
    &
    $m_\chi/p_{\rm F}$
    &
    $1$
    &
    $m_\chi/p_\text{F}$
    &
    $m_\chi/p_\text{F}$
    &
    $m_\chi/p_\text{F}$
    \\
    $\frac{1}{s}\left(1-\cos\psi_0\right)$
    &
    $m_\chi^{-2}$
    &
    $m_{\rm T}^{-2}$
    &
    $m_\chi^{-2}$
    &
    $p_{\rm F}^{-2}$
    &
    $p_{\rm F}^{-2}$
    &
    $p_{\rm F}^{-2}$
    \\
    \bottomrule
  \end{tabular}
  \caption{Scalings of phase space pieces of~\eqref{eqn:fullscaling} with respect to microscopic energy scales.}
  \label{tab:intpieces}
\end{table}

The scaling of the discovery reach on the cutoff $\Lambda$
follows from~\eqref{eqn:fullscaling}:
\begin{align}
\Lambda_{\rm max}&\propto \,\left(p_{F_{\rm T}}^3R_\star\right)^{1/4}\left(\Lambda^4|\mathcal{M}|^2\right)^{1/4}
\left(\,\,\int\limits_{p_{\rm min}}^{p_\text{F}}
\frac{p^2 dp }{p_{\rm F_{\rm T}}^3}\right)^{1/4}
\left(\frac{1}{s}\int\limits_{\cos\psi_0}^1 d\cos\psi\right)^{1/4}
\ ,
\label{eqn:fullLambdamxscaling}
\end{align}
where the factor $\Lambda^4|\mathcal M|^2$ is independent of $\Lambda$.
Table~\ref{tab:fullscalingLambda:allops} collects the discovery reach scaling for the baseline behavior of $\Lambda^4|\mathcal{M}|^2\sim m_\chi^2E_p^2$ and  the exceptional behaviors of Table~\ref{tab:contact:ops:M2:exceptions} and~\ref{tab:contact:ops:M2:exceptions:bosons}.

\begin{table}[H]
  \centering
  \begin{tabular}{p{1.6cm}p{2.6cm}p{2.7cm}p{1.9cm}p{1.9cm}p{1.9cm}p{1.9cm}}  
    \toprule
    $m_\text{T}$
    & 
    \multicolumn{2}{l}{Non-Relativistic}
    & 
    \multicolumn{3}{l}{Relativistic}
    \\
    \cmidrule(r){1-1}\cmidrule(r){2-3}\cmidrule(r){4-7}
    $m_\chi$
    & 
    {\small Heavy}
    & 
    {\small Light}
    & 
    {\small Heavy}
    & 
    {\small Light-ish}
    &
    {\small Med.~Light}
    & 
    {\small Very~Light}
    \\
    \midrule
    ${\rm Baseline}$ 
    & 
    $p_{\rm F}^{1/4}m_{\rm T}$
    & 
    $p_{\rm F}^{1/2}m_\chi^{3/4}$
    & 
    $p_{\rm F}^{5/4}$
    & 
    $p_{\rm F}^{1/2}m_\chi^{3/4}$
    & 
    $p_{\rm F}^{1/2}m_\chi^{3/4}$
    & 
    $p_{\rm F}^{1/2}m_\chi^{3/4}$\\
    \midrule
    $\mathcal O_1^\text{F}$ 
    & 
    & 
    & 
    & 
    $m_\chi^{5/4}$
    & 
    & 
    \\
    $\mathcal O_2^\text{F}$ 
    & 
    $p_{\rm F}^{1/4}m_\chi^{-1/2}m_{\rm T}^{3/2}$
    & 
    & 
    $p_\text{F}^{7/4}m_\chi^{-1/2}$
    & 
    $m_\chi^{5/4}$
    & 
    & 
    \\
    $\mathcal O_3^\text{F}$ 
    & 
    & 
    $p_{\rm F}^{1/2}m_\chi^{5/4}m_{\rm T}^{-1/2}$
    & 
    & 
    $m_\chi^{5/4}$
    & 
    $m_\chi^{5/4}$
    & 
    $m_\chi^{5/4}$
    \\
    $\mathcal O_4^\text{F}$ 
    & 
    $p_{\rm F}^{1/4}m_\chi^{-1/2}m_{\rm T}^{3/2}$
    & 
    $p_{\rm F}^{1/2}m_\chi^{5/4}m_{\rm T}^{-1/2}$
    & 
    $p_\text{F}^{7/4}m_\chi^{-1/2}$
    & 
    $m_\chi^{5/4}$
    & 
    $m_\chi^{5/4}$
    & 
    $m_\chi^{5/4}$
    \\
    \midrule
    $\mathcal O^\textnormal{S}_1$ 
    & 
    $p_{\rm F}^{1/2}m_\chi^{-1}m_{\rm T}^2$
    & 
    $p_{\rm F}\,m_\chi^{1/2}$
    & 
    $p_{\rm F}^{5/2}m_\chi^{-1}$
    & 
    $m_\chi^{3/2}$
    & 
    $m_\chi^{1/2}m_{\rm T}$
    & 
    $m_\chi^{1/2}m_{\rm T}$
    \\
    $\mathcal O^\textnormal{S}_2$ 
    & 
    $p_{\rm F}^{1/2}m_\chi^{-1}m_{\rm T}^2$
    & 
    $p_{\rm F}\,m_\chi^{3/2}m_{\rm T}^{-1}$
    & 
    $p_{\rm F}^{5/2}m_\chi^{-1}$
    & 
    $m_\chi^{3/2}$
    & 
    $m_\chi^{3/2}$
    & 
    $m_\chi^{3/2}$
    \\
    \bottomrule
  \end{tabular}
  \caption{%
  	Experimental reach on $\Lambda$ as a function of the target mass, $m_\textnormal{T}$, dark matter mass $m_\chi$, and the Fermi momentum $p_\textnormal{F}$ in the different regimes defined in Figure~\ref{fig:app:phaseblock:phasespace}. 
  	The baseline case corresponds to the behavior of the $\mathcal O^\textnormal{F}_{5-10}$ and $\mathcal O^\textnormal{S}_{3,4}$ operators in Table~\ref{tab:fullscalingLambda:allops}.
	Powers of the neutron star radius $R_\star$ account for the additional dimensional dependence so that for dimension-6 operators the scaling applies to $\Lambda R_\star^{-1/4}$ while for dimension-5 operators the scaling applies to $\Lambda R_\star^{-1/2}$.
	%
	Blank entries correspond to the baseline scaling behavior.
	}
  \label{tab:fullscalingLambda:allops}
\end{table}

\bibliographystyle{utcaps}  
\bibliography{neutronstar}

\end{document}